\documentclass[11pt,a4paper]{article}
\usepackage{jheppub}
\usepackage{amsmath}
\usepackage{xcolor}
\usepackage{bbm}
\usepackage[utf8]{inputenc}
\usepackage{graphicx}
\usepackage{multirow}
\usepackage{hyperref}
\usepackage{comment}
\usepackage{amssymb,amsfonts}
\usepackage{mathrsfs}

\usepackage{array}   
\newcolumntype{L}{>{$}l<{$}} 
\usepackage{float} 
\usepackage{chngpage} 

\title{Fermionic Rational Conformal Field Theories and Modular Linear Differential Equations}

\author[1]{Jin-Beom Bae,}
\author[2]{Zhihao Duan,}
\author[2]{Kimyeong Lee,}
\author[2]{Sungjay Lee,}
\author[2]{and Matthieu Sarkis}

\affiliation[1]{Mathematical Institute, University of Oxford \\
$~~ $Andrew Wiles Building, Radcliffe Observatory Quarter \\
$~~ $Woodstock Road, Oxford, OX2 6GG, U.K.}
\affiliation[2]{Korea Institute for Advanced Study \\
$~~ $85 Hoegiro, Dongdaemun-Gu, Seoul 02455, Korea}

\preprint{KIAS-P20056}

\abstract{We define Modular Linear Differential Equations (MLDE) for the level-two congruence subgroups $\Gamma_\theta$, $\Gamma^0(2)$ and $\Gamma_0(2)$ of $\text{SL}_2(\mathbb Z)$. Each subgroup corresponds to one of the spin structures on the torus. The pole structures of the fermionic MLDEs are investigated by exploiting the valence formula for the level-two congruence subgroups. We focus on the first and second order holomorphic MLDEs without poles and use them to find a large class of `Fermionic Rational Conformal Field Theories', which have non-negative integer coefficients in the $q$-series expansion of their characters. We study the detailed properties of these fermionic RCFTs, some of which are supersymmetric. This work also provides a starting point for the classification of the fermionic Modular Tensor Category.}

\begin{document}

\maketitle

\parskip 0.2 cm

\section{Introduction and Concluding Remarks}


The classification of all unitary conformal field theories in two dimensions certainly plays a key role in our understanding of the critical phenomena. Utilizing the conformal and modular bootstraps, the unitary conformal field theories with $c<1$ in particular can be solved and classified completely \cite{Belavin:1984vu}. They are called minimal models obeying the so-called ADE classification. One of the important features of minimal models is that they have finitely many conformal primaries. It admits a generalization that leads to a very rich class of two-dimensional conformal field theories, namely rational conformal field theories. A rational conformal field theory (RCFT) refers to a conformal field theory whose torus partition function can be expressed as a finite sum of products of holomorphic and anti-holomorphic functions. Such holomorphic functions can be understood as characters with respect to an extended chiral algebra that includes Virasoro algebra. It is well-known that, in a given RCFT, the central charge as well as conformal weights are rational numbers \cite{Anderson:1987ge,Vafa:1988ag}. One prominent example of RCFTs is the Wess-Zumino-Witten (WZW) model where the extended chiral algebra is the current algebra. 

A few approaches based on chiral algebras and lattices have been proposed to solve a more tractable problem of the classification of RCFTs \cite{Knizhnik:1984nr, Ginsparg:1987eb, Dijkgraaf:1987jta, Goddard:1983at}. Since there are RCFTs that fit into neither of those approaches, the classification however remains incomplete. On the other hand, the authors of \cite{Mathur:1988na} have proposed a rather different approach based on the modular invariant linear differential equations (MLDE).

Let us first briefly explain how an MLDE can provide a systematic procedure to classify the RCFTs. Suppose that a given RCFT has $N$ independent characters. From  the modular invariance of the torus partition function, one can see that the holomorphic characters transform as a vector-valued modular form under  SL$_2(\mathbb{Z})$. It implies that the $N$ characters can be regarded as independent solutions to an $N$-th order differential equation invariant under SL$_2(\mathbb{Z})$. This method is particularly useful to obtain a complete list  for RCFTs with small numbers of characters, which was studied extensively in \cite{Mathur:1988na}. The classification of the super-characters in $\mathcal{N}=1$ superconformal field theory has been considered in \cite{Durganandini:1989es}. Readers can also find a recent status on this programme in \cite{Mukhi:2019xjy} and the references therein. 

Rational conformal field theories are also closely related to the modular tensor categories (MTC) that have extensive applications to the study of anyonic systems and topological quantum computation
\cite{Moore:1991ks, Wen:2016lrc, Schoutens:2015uia}. When two different RCFTs are related to two MTCs conjugate to each other, it turns out that they satisfy a certain bilinear relation studied recently in \cite{Gaberdiel:2016zke,Hampapura:2016mmz,Bae:2018qfh,Bae:2020pvv}.  Given that the MTCs of low rank are only classified in \cite{Gepner:1994bb, Rowell:2007dge}, we expect that the classification of RCFTs based on the MLDE sheds new light on the problem of the classification of MTCs. Recently the classification of fermionic MTCs was considered a view toward the three-dimensional topological field theories\cite{Cho:2020ljj, Gaiotto:2020iye}. 

In the present work, we extend the MLDE method to classify the fermionic rational conformal field theories, some of which appear to be supersymmetric. A fermionic CFT refers to a conformal field theory which contains operators of half-integer spin. To define a fermionic CFT on a manifold, one has to choose a spin structure. On a torus, there are four different spin structures, (NS,NS), (R,NS), (NS,R) and (R,R). For later convenience, we use the shorthand notation NS, $\widetilde{\rm NS}$, R and $\widetilde{\rm R}$ for those spin structures. A fermionic CFT in our study is further restricted to have a certain extended chiral  algebra that includes conserved currents of half-integer weight. In other words, there are half-integer spin descendants of the vacuum in the NS sector. 
 
It is well-known that the Jordan-Wigner transformation maps the critical Ising model to a theory of free Majorana fermion. Later on, the Jordan-Wigner transformation has been revisited to  fermionize a given bosonic CFT with a non-anomalous $\mathbb{Z}_2$ to a fermionic CFT \cite{Gaiotto:2015zta, Kapustin:2017jrc, Karch:2019lnn}. Along the way, the ``Beauty and the Beast'' ${\cal N}=1$ superconformal theory \cite{Dixon:1988qd} can be reinterpreted as a fermionization of the Monster CFT \cite{Lin:2019hks}, and the fermionic minimal models are constructed in \cite{Runkel:2020zgg, Hsieh:2020uwb, Kulp:2020iet}. The goal of this paper is to classify such fermionic RCFTs with small numbers of conformal characters systematically via the MLDE method. The classification of fermionic RCFTs will provide the classification  of the fermionic modular tensor category that characterizes the fermionic topological phases of matter, modulo the fact that two different fermionic RCFTs sharing the same fusion rule algebra are related to a single fermionic MTC. 

In order to discuss the extension of the MLDE method, let us note that the characters of a given fermionic RCFT in NS,  $\widetilde{\rm NS}$, and R sectors become a vector-valued modular functions  for the level-two congruence subgroups $\Gamma_\theta$, $\Gamma^0(2)$ and $\Gamma_0(2)$ of the modular group. This is because each of these congruence subgroups is associated to a specific spin structure on the torus, i.e., NS,  $\widetilde{\rm NS}$, and R sectors. Hence each of the `   
]fermionic' MLDEs associated to them, transform into each other under $\text{SL}_2(\mathbb Z)$ transformations.

The first step consists of understanding the relation between the pole structure of the coefficients of the MLDE and the set of zeros of the characters which are solutions to the MLDE. This was neatly understood in the case of $\text{SL}_2(\mathbb Z)$, see for instance \cite{Mathur:1988na}, in terms of the zeros of the Wronskian associated to the MLDE. Having automorphic properties, the Wronskian happens to be subject to the so-called valence formula, a classical result in the theory of modular forms constraining the possible set of zeros (together with their multiplicity) in terms of the weight of the automorphic form. This allows to reduce the choice of the pole structure characterizing the order $N$ MLDE down to the choice of a single integer $\ell$. In this paper, we introduce the equivalent of the valence formula for various relevant level-two congruence subgroups of $\text{SL}_2(\mathbb Z)$. Equipped with this tool, we generalize the MLDE to the case of level-two congruence subgroups $\Gamma_\theta$, $\Gamma^0(2)$ and $\Gamma_0(2)$ of the modular group. The level-two valence formula boils down to the following relation between the central charge $c$, the conformal weights in the NS and R sectors $h^{\text{NS}}$ and $h^{\text{R}}$, $N$ and $\ell$:
\begin{equation}
   -\frac{Nc}{8}+2\sum_j h_j^{\rm NS} +\sum_j h_j^{R} + \frac{\ell}{2}= \frac{N(N-1)}{4}\,.
\end{equation}

After discussing the possible poles structures for a fermionic MLDE in terms of the level-two congruence subgroups valence formula, we focus on the second order fermionic MLDE in the simplest case, that of a trivial pole structure. In this context, we classify the possible values of the central charge and conformal weights corresponding to some fermionic RCFT for which there exist character-like solutions to the fermionic MLDE. More precisely, we consider the  solutions with the property that all the coefficients are non-negative integers in $q$-series. We classify the solutions by six classes, which can be found in table \ref{tab:2mldesol}. In particular, we also derive a closed-form expression of the $S$-matrix of these fermionic RCFTs and investigate their fusion coefficients. We discard the solutions of the second order fermionic MLDE when they do not yield a consistent fusion rule algebra. For the solutions with consistent fusion rule algebra, we find the identifications in terms of the $\mathcal{N}=1$ supersymmetric minimal models or the WZW models. 

One can construct the partition function of individual spin structures using the solutions of the second order fermonic MLDE. The sum of four partition functions defines an SL$_2(\mathbb{Z})$ invariant partition function of a certain bosonic CFT and this procedure is often referred to as the GSO projection \cite{Gliozzi:1976qd}, or equivalently bosonization. In some cases we study below, we find that the bosonization works with the assumption that the torus partition function for the spin structure $\widetilde{R}$ becomes constant, $Z_{\widetilde{\text{R}}} = \text{const}$. Especially when the following three conditions are satisfied, 
 \begin{enumerate}
     \item a vacuum descendant of weight $3/2$ is present,
     \item the supersymmetic unitarity bound $h^\text{R}\geq c/24$ is obeyed, 
     \item $Z_{\widetilde{\text{R}}} = \text{constant}$,
 \end{enumerate}
then we suggest interpreting the corresponding solutions as the characters of unitary supersymmetric RCFT. For instance, we will show that the fermionization of $su(2)_6$ and $(\mathfrak{e}_6)_3$ WZW model could potentially be understood as the supersymmetric RCFT, as they satisfy the above three conditions. More examples will be discussed in the main context.

We also notice that some solutions can be combined into known partition functions, e.g., the Conway extremal CFT \cite{FLM85} or the $\mathcal{N}=1$ extension of the $(\mathfrak{e}_8)_1$ WZW model. This will be referred to as a bilinear relation in what follows. When such bilinear relation is satisfied for the characters of two different RCFTs, these two theories ought to share the same fusion rule algebra. Therefore, we expect that our classification will provide new insight into fermionic MTCs. On the one hand, some bilinear relations are known to appear as an evidence of deconstruction of the Monster group \cite{Bae:2018qfh, Bae:2020pvv}. In a similar way, we test the splitting of the supersymmetric VOA for the Conway group Co$_0$ with $c=12$. Specifically, the solution with $c=11$ exhibits moonshine phenomena for the Suzuki group as shown in \cite{Johnson-Freyd:2019wgb}. Further examples of fermionic deconstructions will be discussed in an upcoming paper \cite{ONP20}.

In a separate upcoming paper \cite{ONP20third}, we generalize the work of the present paper to the case of a third order fermionic MLDE. There, we focus on a subfamily of solutions for which the BPS bound is saturated in the Ramond sector. We provide a closed-form expression of the characters and the $S$-matrix for these solutions.

This article is organized as follows. In Section \ref{sec:SL2Z}, we review some mathematical facts concerning the modular group $\text{SL}_2(\mathbb{Z})$ and the well-known classification of second order MLDEs by \cite{Mathur:1988na}. In Section \ref{sec:subgroup}, we introduce some standard results of three level-two congruence subgroups and propose MLDEs with holomorphic coefficients, to which we also refer as (holomorphic) fermionic MLDEs. Then we move on to finding solutions that can possibly be identified as characters of fermionic RCFTs. In Section \ref{sec:1stMLDE}, we classify all possible solutions of fermionic first order MLDEs, which turn out to consist of products of Majorana-Weyl fermions. In Section \ref{sec:2ndMLDE}, we study fermionic second order MLDEs with trivial pole structure and find six families of solutions, listed in Table \ref{tab:2mldesol}. For all the consistent solutions we are able to express them in terms of the characters of some known RCFT.

We would like to dedicate this article to the memory of Professor Tohru Eguchi who passed away last year.  K.L. recalls a personal meeting with him about 20 years ago during his visit to Tokyo University and many more wonderful interactions later on. Many of our works  got influenced by   Prof. Eguchi's works. In relation to this article, we note that Eguchi and Ooguri derived in \cite{Eguchi:1987qd} the third order MLDE for conformal characters of the Ising model as well as the exact form of the characters. It would be the first place where the MLDE made its appearance in the study of conformal field theories.  Anderson and Moore wrote a general MLDE and used it show the rationality of $c$ and $h$ for rational conformal field theories \cite{Anderson:1987ge}.  Afterwards, Mathur, Mukhi and Sen \cite{Mathur:1988na} expanded and established the modular linear differential equation and utilized it as a tool for the classification of RCFTs. They found the famous MMS series of theories discussed in Section \ref{sec:SL2ZMLDE}.


\section{Modular Linear Differential Equation   for ${\rm SL}_2({\mathbb Z})$}\label{sec:SL2Z}


\subsection{${\rm SL}_2({\mathbb Z})$ Group and its Valence Formula}

It is natural to consider RCFTs on a torus. The partition function of the theory is required to be modular invariant. Thus, the  modular group  ${\rm SL}_2({\mathbb Z})$ plays an essential role in our understanding of the conformal field theory. Here we review its important features  relevant to our analysis. 

\begin{figure}[h!]
\centering
\includegraphics[width=6cm]{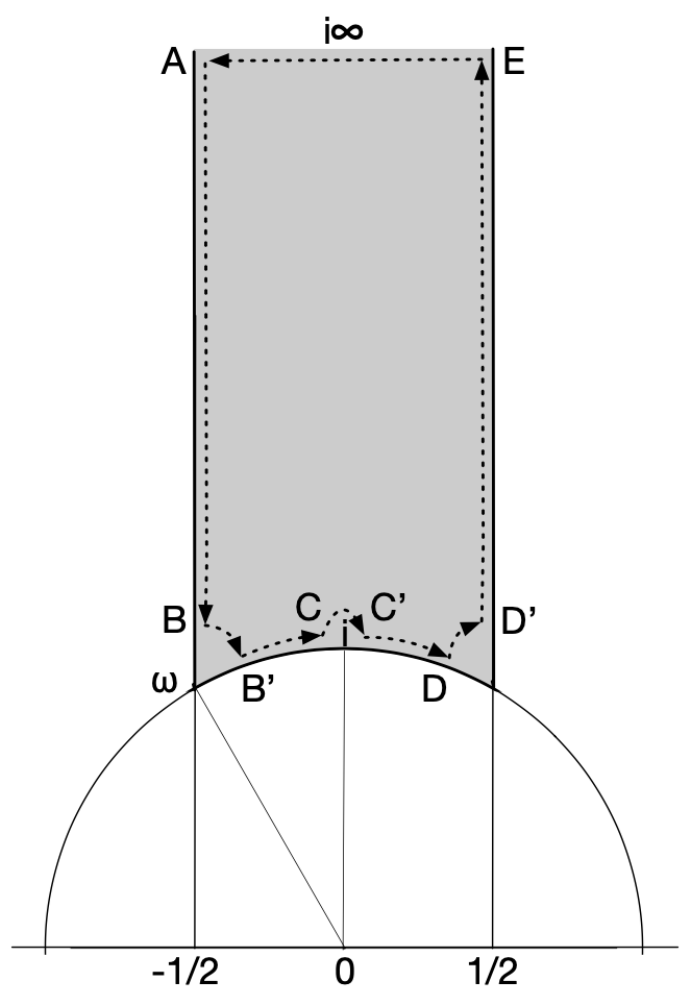}
\caption{ Fundamental Domain of ${\rm SL}_2(\mathbb{Z})$  and the contour for the valence formula}
\label{fig:g1fund}
\end{figure}

The modular group ${\rm SL}_2({\mathbb Z})$ is the group if invertible $2\times 2$ matrices with integer coefficients and unit determinant:
\begin{align}
   {\rm SL}_2({\mathbb Z})=\Big\{ \gamma=\big({_a \ _b \atop ^c\ ^d}\big) \Big|\  a,b,c,d \in {\mathbb Z}, \ \ ad-bc=1 \Big\}.
\end{align}
The modular group is generated by $S=\big({_0\atop^1} {_{-1}\atop ^0}\big)$ and $T=\big({_1\atop^0} {_{1}\atop ^1} \big)$. The modular parameter $\tau \in {\mathbb H}=\{  \tau\in {\mathbb C}| {\rm Im}(\tau)>0\}$  transforms by fractional linear transformations as  $\gamma\tau= (a\tau+b)/(c\tau+d)$.  The fundamental domain  $\mathscr{D}={\rm SL}_2({\mathbb Z})\backslash {\mathbb H}  $  for the modular group ${\rm SL}_2({\mathbb Z}) $ is given as  
\begin{align}
    {\mathscr D}({\rm SL}_2({\mathbb Z}))=\Big\{ \tau \in {\mathbb H}\ \Big|\  |{\rm Re}(\tau)|\le \frac12, |\tau|\ge 1\Big\},
\end{align}
with additional identification of the boundaries by $T:\tau\rightarrow \tau+1$ and $S:\tau\rightarrow -1/\tau$. The fundamental domain is drawn in Figure~\ref{fig:g1fund}. The $\tau=i$ is an orbifold point of order two and $\tau=\omega\equiv e^{2\pi i/3}$ is an orbifold point of order three. The topology of the fundamental domain is a sphere with a single puncture, the {\it cusp} at $i\infty$.

Given an element $\gamma\in {\rm SL}_2({\mathbb Z})$, one defines the following action on functions $f(\tau)$ from $\mathbb H$ to $\mathbb{C}$:
\begin{align}
\label{eq:mtrans}
    (f|_k\gamma)(\tau)\equiv \rho(\gamma)^{-1} (c\tau+d)^{-k}f(\gamma \tau),
\end{align}
where $\rho(\gamma)$ is a possibly non-trivial $\gamma$-dependent phase, and $k$ an integer called the weight of $f$. If the function $f$ is periodic under $T: \tau\rightarrow\tau+1$ so that $\rho(T)=1$, we can have a Fourier expansion:
\begin{align}
    f(\tau) = \sum_{k=-\infty}^\infty a_k q^k, \ \ q\equiv e^{2\pi i \tau}.
\end{align}
The function $f$ is {\it meromorphic} at $i\infty$ if only a finite number of negative powers of $q$ appears in the above expansion, {\it holomorphic} at $i\infty$ if there is no negative power of $q$, and {\it vanishes} at $i\infty$ if only positive powers of $q$ appear. 

Let us first focus on the  ${\rm SL}_2({\mathbb Z})$ weight $k$ {\rm forms} which have the following property: 
\begin{align}
   ( f|_k\gamma)(\tau)= f(\tau)  \ {\rm with} \ \rho(\gamma)=1,\  {\rm for \ all }\  \gamma\in 
   {\rm SL}_2({\mathbb Z}) \ {\rm and}\   {\rm for\ all} \ \tau \in {\mathbb H}.
\end{align}
These functions $f$ can be extended to maps $f: \overline{\mathbb H}\rightarrow {\mathbb C}$ where $\overline{\mathbb H}=  {\mathbb H}\ \cup \ \{i\infty \}$. Depending on the pole structures of these weight $k$ forms, we call them {\it automorphic} forms  if they have  poles in $\mathbb{H}$, {\it weakly-holomorphic} forms if they are holomorphic on $\mathbb{H}$ and have poles at $i\infty$, {\it modular} forms if they are holomorphic on $\overline{\mathbb H}$, {\it cusp} forms if they are holomorphic in $\mathbb{H}$ and vanish at $i\infty$. The Klein $j(\tau)$-invariant is a weakly-holomorphic function, that is, a form of  weight zero with simple pole $\tau = i\infty$.  Eisenstein series $E_4, E_6$ are the only modular forms of weight 4 and 6, respectively.  The modular discriminant $\Delta=E_4^3-E_6^2=1728 \eta^{24}$ is the only cusp form of weight 12.  The detail definition of these functions are given in appendix  \ref{sec:conventions}.
 
In this work we are interested in the  function space  $ M_k$ of modular forms of weight $k$ mainly.  It is generated by $E_4,E_6$ Eisenstein series as follows: 
    \begin{equation}
        \mathcal{M}_k({\rm SL}_2({\mathbb Z})) = \bigoplus_{4a+6b=k \atop a,b\ge 0} {\mathbb C} E_4^a E_6^b\ . \nonumber
    \end{equation}
    The dimension of $M_k$ for $k\ge 2$ are well-known to satisfy
    \begin{align} \sum_{k\in \mathbb{Z}}   {\rm dim}( M_k)  t^k = \frac{1}{(1-t^4)(1-t^6)}.
    \end{align} 
The space  $\mathcal{S}_k$ of   cusp forms   of weight $k$   is related to the space of the modular forms as follows:   \begin{equation}
        \mathcal{S}_k({\rm SL}_2({\mathbb Z})) = \Delta(\tau)\mathcal{M}_{k-12}({\rm SL}_2({\mathbb Z})).
\end{equation}

For the class of functions $f:{\mathbb H}\rightarrow {\mathbb C}$ which satisfy  $f|_k\gamma =f$  for any $\gamma\in{\rm SL}_2(\mathbb{Z})$ of given weight $k$ \eqref{eq:mtrans}, holomorphic in ${\mathbb H}$ and possibly meromorphic at $i\infty$,  there exists a so-called {\rm valence formula} relating the number of zeros in the fundamental domain $ {\rm SL}_2(\mathbb{Z})\backslash \mathbb{H}$ to   the weight $k$. Although it is a standard textbook material \cite{Serre:1993,ranestad20081}, let us review the derivation briefly as we want to extend it to the case of level-two congruent subgroups later.

To get the valence formula, we use the Cauchy formula on the integration of $\frac{\partial\tau f }{f}$ along the contour given in Figure~\ref{fig:g1fund}, which is inside   the fundamental domain $  {\rm SL}_2(\mathbb{Z})\backslash \mathbb{H}$. The resulting integral can be evaluated in two ways,
\begin{align}
\label{eq:g1balance}
-\frac{1}{2\pi i} \oint d\tau \frac{\frac{\text{d}}{\text{d}\tau} f(\tau)}{f(\tau)} & = - \sum_{p\in{\rm SL}_2(\mathbb{Z})\backslash \mathbb{H}
\atop p\neq i, \omega, i\infty} 
 \nu_{ p}(f)  \nonumber \\ 
&= \nu_{\infty}(f)+ \frac12 \nu_i(f)+\frac13\nu_{\omega}(f)    - \frac{k}{12} , 
\end{align}
where $\omega=e^{2\pi i /3}$. For the first equality, the contour integration is split into integrals surrounding all the zeros in the interior of $\mathbb{H}$ and becomes $-\sum_{{\rm interior}\  p} \nu_{p}(f)$. The second equality comes from the contour integration along the boundary. One can read off the origin of each terms in the RHS of the above equation from the dashed lines in Figure~\ref{fig:g1fund}. The contribution from each segment is:
\begin{align}
   & C_{EA}: \ \nu_\infty(f), \ \ \ C_{CC'}:\ \frac12 \nu_i(f), \ \ \         C_{BB'}+C_{DD'}:\ \frac13 \nu_\omega(f),    \nonumber \\
   &  C_{AB}+C_{D'E}: \ 0,\ \ \   C_{B'C}+C_{C'D}:\   -\frac{k}{12}. \ \ \  
\end{align}
The order $\nu_{p}(f)=n$ if $f(\tau)$ near $\tau=p$ has a Laurent expansion with leading term $f(\tau)\sim(\tau-p)^n$. For $f\sim q^n$ near $i\infty$, $\log(f) \sim 2\pi i n \log \tau$ and so $\nu_\infty(f)=n$. At orbifold points $i$ and $\omega$, the contribution gets only $\frac12$ and $\frac13$ fractions, respectively as one integrate over one-half or one-third of $2\pi$ integration. On the arc $C_{B'C}$ and $C_{C'D}$, a point $\tau$ and its S-dual points  $-\tau^{-1}$ are matched. One changes the variable $\tau\rightarrow -1/\tau$ on $C_{B'C}$ and use the fact $f(-1/\tau)=\rho(S)\,\tau^k f(\tau)$ to
get that the contribution from the contour $C_{B'C}$ is minus of that from the contour $C_{C'D}$ together with $-k/12$. The factor $1/12$ is due to the fact that the angle between $\omega=e^{2\pi i /3}$and $i$ is $2\pi/12$. 
 
In short the valence formula for weight $k$ form of  $ {\rm SL}_2(\mathbb{Z} )$ is
\begin{align}
\label{eq:valenceg0}
   \nu_{\infty}(f)+ \frac12 \nu_i(f)+\frac13\nu_{\omega}(f) +
    \sum_{p\in{\rm SL}_2(\mathbb{Z})\backslash \mathbb{H}
\atop p\neq i, \omega, i\infty} 
 \nu_{ p}(f)  = \frac{k}{12}  .
\end{align}
Table \ref{tab:valence1} shows the validity of the valence formula for well-known modular forms which have no zeros or poles besides  $i\infty, i$ and $\omega$.

\begin{table}[]
    \centering
     \begin{tabular}{c|c|ccc}
    $f(\tau)$ & $\frac{k}{12}$ & $\nu_\infty$ & $\frac12\nu_i$ & $\frac13\nu_\omega$   \\
    \hline
   $E_4$  &  $\frac13$ & 0 & 0 & $\frac13$   \\
   $E_6$ & $\frac12$ & 0 & $\frac12$ & 0   \\
   $j$ & 0 & $-1$   & 0  & 1 \\
   $\Delta$ & 1 & $1$ & 0 & 0   
\end{tabular}
    \caption{Weight $k$ (weak) modular forms of ${\rm SL}_2(\mathbb{Z})$}
    \label{tab:valence1}
\end{table}

\subsection{Modular Linear Differential Equation}\label{sec:SL2ZMLDE}
A two-dimensional rational conformal field theory of central charge $c$ has a finite number  of primary operators.  Its modular invariant partition function on a torus can thus be expressed as 
\begin{align}\label{toruspartition}
    Z(\tau,\bar\tau)= \sum_{a,b=0}^{N-1} {\mathcal M}_{ab} \bar\chi_a(\bar \tau) \chi_b(\tau),
\end{align}
where $\chi_a$ ($\bar \chi_a$) denote left-moving (right-moving) characters with respect to an extended chiral algebra including the Virasoro algebra. The modular invariance of \eqref{toruspartition} implies that the characters $\chi_a(\tau)$ transform as a finite-dimensional representation of $\text{SL}_2(\mathbb{Z})$, i.e, they transform under $S$ and $T$ as follows: 
\begin{align}\label{modularmatrices}
    \chi_a(-1/\tau)  = \sum_{b=0}^{N-1}S_{ab}\chi_b(\tau),  \ \ 
    \chi_a(\tau+1)  = \sum_{b=0}^{N-1}T_{ab}\chi_b(\tau) , 
\end{align}
where the modular matrices $S_{ab}$ and $T_{ab}$ are symmetric and satisfy the relations below
\begin{align}
  \big( S^\dagger \mathcal{M}   S\big)_{ab}=\mathcal{M}_{ab}, \quad 
  T_{ab}=q^{2\pi i h_a} \delta_{ab}, 
\end{align}
where $h_a$ is the conformal weight for $\chi_a$. Note also that $S_{ab}$ and $T_{ab}$ should satisfy 
\begin{align}
\label{eq:modularRelations}
    S^2 = \big(ST\big)^3 = C,
\end{align}
where $C$ is the charge conjugation matrix. 

One can see from \eqref{modularmatrices} that the characters of the RCFT constitute a vector-valued modular forms of weight zero and thus satisfy a modular-invariant linear differential equation (MLDE) of order $N$ where $N$ is the number of linearly independent characters.  Let us review the argument for the derivation of the modular linear differential equation in  \cite{Mathur:1988na,Chandra:2018pjq}. We start with an $(N+1)$-dimensional square matrix made of $\chi_0,\chi_1,...,\chi_{N-1}, f$ and their derivatives with the Ramanujan-Serre covariant derivative, 
\begin{equation}
    {\cal D} = \frac{1}{2\pi i } \frac{\text{d}}{\text{d}\tau} -\frac{r}{12}\, E_2(\tau) 
\end{equation}
acting on a weight $r$ modular form, up to the $N$-th power. This covariant derivative transforms a weight $r$ modular form to weight $r+2$ modular form. (See the Appendix \ref{App:A} for the details.) If the function $f$ is a linear combination of $N$ characters, the determinant of this $(N+1)$-dimensional matrix vanishes, implying that
\begin{align}
\label{eq:mlde0}
    \sum_{k=0}^N (-1)^k W_k {\cal D}^k f=0,
\end{align}  
where each coefficient $W_k$ is given by
\begin{align}
\label{eq:wronskian0}
    W_k = {\rm det} \left( \begin{array}{ccc}
    \chi_0 & \cdots &\chi_{N-1}  \\
    {\cal D} \chi_0 & \cdots & {\cal D} \chi_{N-1} \\
    \vdots & & \vdots \\
    {\cal D}^{k-1}\chi_0 & \cdots &D^{k-1}\chi_{N-1}  \\
    {\cal D}^{k+1}\chi_0 & \cdots & D^{k+1}\chi_{N-1} \\
    \vdots & & \vdots \\
    {\cal D}^N \chi_0 & \cdots & {\cal D}^n \chi_{N-1}  \\
\end{array}\right).
\end{align}
One can recast \eqref{eq:mlde0} into
\begin{align}\label{MLDE:wronsk}
    \left[ {\cal D}^N  + \sum_{k=0}^{N-1} \phi_k(\tau) {\cal D}^k \right] f(\tau) =0 , 
\end{align}
where the coefficients $\phi_k(\tau)= (-1)^{N-k} W_k/W_N$ are automorphic forms of weight $2N-2k$ for ${\rm SL}_2(\mathbb{Z})$. These forms $\phi_k(\tau )$  could have poles at zeros of the Wronskian $W_N(\tau)$.  

Let us first apply the valence formula \eqref{eq:valenceg0} to the Wronskian $W_N(\tau)$, which transforms under $S$ as a modular form of weight $N(N-1)$ and is invariant under $T$ up to a constant phase $\rho(T)$. At $\tau =i\infty$, the Wronskian has the asymptotic expansion:  
\begin{align}
    W_N(\tau)\sim q^{-\frac{Nc}{24}+\sum_a h_a} \Big(1+ {\mathcal O}(q) \Big) , 
\end{align}
because each characters $\chi_a$ is asymptotic to 
\begin{align}
\label{eq:nearinfty0}
    \chi_a \sim q^{-\frac{c}{24}+h_a}\Big(1+ {\mathcal O}(q)\Big) \quad \text{at} \quad \tau =i\infty.  
\end{align} 
This implies that  
\begin{align}
    \nu_{i\infty}= -\frac{Nc}{24}+\sum_{i=a}^{N-1} h_a ,
\end{align}
and the valence formula \eqref{eq:valenceg0}   becomes   
\begin{align}\label{sl2zvalence}
    -\frac{Nc}{24}+\sum_a h_a + \frac{\ell}{6} = \frac{N(N-1)}{12} ,
\end{align}
where
\begin{align}
\label{eq:ellsl2z}
    \frac{\ell}{6}= \frac12 \nu_i+\frac13 \nu_\omega +\sum_{\rm interior} \nu_\tau\ . 
 \end{align}
The zeros of the Wronskian $W_N$ become the poles of the coefficient functions $\phi_k$. As the modular forms are covariant under the modular transformation, the coefficient $\phi_k$ of weight $2N-2k$ can be expressed as rational functions of  $E_4$ and $E_6$, whose denominator is constrained by the parameter $\ell$.   

The second order MLDE with $n=2$ has been studied extensively in  \cite{Mathur:1988na,Chandra:2018pjq}. For the simple case without poles $\ell=0$, the MLDE becomes 
\begin{align} 
\label{eq:2omlde}
   \Big[{\mathcal D}^2 +\mu E_4\Big]f(\tau) = 0.  
\end{align}
Note that the coefficient of the first derivative vanishes as there exists no modular form of weight two and $E_4$ is the unique modular form of weight four up to a constant factor. Two independent solutions to the above equation can be regarded as two characters that can be expanded in powers of $q$ as follows 
\begin{align}\label{sl2zMLDE}
    \chi_0\sim q^{-\frac{c}{24}}, \quad 
    \chi_1\sim q^{-\frac{c}{24}+h}. 
\end{align}
Since the valence formula \eqref{sl2zvalence} says $h=\frac{c+2}{12}$, the free parameter $\mu$ of \eqref{eq:2omlde} can be determined as 
$\mu=-\frac{c(c+4)}{576}=\frac{(1+6h)(1-6h)}{144}$. It has been shown in   \cite{Mathur:1988na}  that there exist only   
ten values of allowed central charges,  
\begin{align}
    c = \left\{ \frac{2}{5}, 1, 2, \frac{14}{5}, 4, \frac{26}{5}, 6, 7, \frac{38}{5}, 8 \right\},
\end{align}
such that the solution of \eqref{sl2zMLDE}  has  all $q$-expansion coefficients given by  non-negative integers.  Although the first case with $c=2/5$ and $h=1/5$ appears to be consistent, it has negative fusion coefficients. To resolve this, one interchanges the two characters and obtain a theory with $c=-22/5$ and $h=-1/5$. As discussed in \cite{Mathur:1988na,Chandra:2018pjq}, the above ten solutions provide the characters for the Lee-Yang edge singularity and level-one WZW models for $ {\mathfrak a}_1 $, $ {\mathfrak a}_2$,   $ {\mathfrak g}_2 $, $ {\mathfrak d}_4 $, $ {\mathfrak f}_4 $,  $ {\mathfrak e}_6 $, $ {\mathfrak e}_7 $, $ {\mathfrak e}_{7\frac12}  $, ${\mathfrak e}_8 $, respectively. This series of Lie groups is known as the Deligne-Cvitanovic series \cite{cvitanovic2008group}. The explicit solutions to \eqref{MLDE:wronsk} as well as their $S$-matrix for each central charge $c$ can be found in \cite{Mathur:1988gt}. One can also show that the modular invariant partition function on the torus is diagonal, i.e., 
\begin{align}
    Z = |\chi_0|^2 + \mathcal{M} |\chi_1|^2, 
\end{align}
where  $\cal M$ is a certain non-negative integer. We present the central charge $c$, and corresponding values of $h$, and multiplicity $\mathcal{M}$ in Table \ref{Deligne}.
\begin{table}[t]
\begin{center}
\begin{tabular}{c|cccccccccc}
 & LY & $ {\mathfrak a}_1 $ & $ {\mathfrak a}_2 $ & $ {\mathfrak g}_2 $ & $ {\mathfrak d}_4 $ & $ {\mathfrak f}_4 $  &  $ {\mathfrak e}_6 $ & $ {\mathfrak e}_7 $ & $ {\mathfrak e}_{7\frac12}  $   & 
 $ {\mathfrak e}_8 $ \\ \hline
$c$ &  $- \frac{22}{5}$   &   1 &  2  &  
$\frac{14}{5}$ &  4&   $\frac{26}{5} $ &   6 & 
7 &   $\frac{38}{5} $&  8  \\
$h $ & $-\frac15$ & $\frac14$ & $\frac13$ & $\frac25$ &  $\frac12$  & $\frac35$ & $\frac23$ & $\frac34$ & $\frac45$ & $\left[\frac56\right]$ \\
$\mathcal{M}$ & 1 & 1  & 2 & 1 & 3 & 1 & 2 & 1 & 1 & $[0]$ 
\end{tabular}
\caption{
\label{Deligne} $c$ and $h$ for Mathur-Mukhi-Sen series.}
\end{center}
\end{table}%

As a remark, the WZW model for $\mathfrak{e}_8$ at level one appears in Table \ref{Deligne}, but in fact it is a single-character RCFT. The vacuum character which is its unique character is given by
\begin{align} \chi^{{\mathfrak e}_8}_0 = \frac{E_4(\tau)}{\eta(\tau)^8} =j(\tau)^\frac13. 
\end{align}
The characters of the above series satisfy also the bilinear relations
\begin{align}\label{e8bilinear}
 \chi^{{\mathfrak e}_8}_0=\chi^{c}_0\chi^{8-c}_0 + \mathcal{M}\chi^{c}_1 \chi^{8-c}_1.
\end{align}
This shows multiple ways to divide the $\mathfrak{e}_8$ level one WZW model into two pairs of complementary CFTs. 
 
In the following sections, we generalize MLDEs to the fermionic theories including supersymmetric theories by considering the modular forms and the valence formula for the related congruence groups $\Gamma_\theta,\Gamma^0(2)$, and $\Gamma_0(2)$.


\section{Congruence Subgroups $\Gamma_\theta,\Gamma^0(2)$ and $\Gamma_0(2)$} \label{sec:subgroup}


\subsection{$\Gamma_\theta,\Gamma^0(2),\Gamma_0(2), \Gamma(2)$ modular subgroups and forms} \label{subsec:3.1}

The level-two congruence subgroups $\Gamma_\theta, \Gamma^0(2)$ and $\Gamma_0(2)$ are related to the modular symmetry group of the fermionic conformal field theories in the NS-NS, R-NS and NS-R spin structure sectors of the partition functions, respectively. We will also refer to these three spin structures as NS, $\widetilde{\text{NS}}$ and R sectors respectively. Their study provides a good starting point for building up fermionic modular linear differential equations.   

The principal congruence subgroup of $\text{SL}_2(\mathbb Z)$ of level two, $\Gamma(2)$, is defined as follows:
\begin{align}
     \Gamma(2) = \left.\Big\{ \gamma\in {\rm SL}_2(\mathbb{Z})\right| \gamma\equiv \big({_1 \ _0 \atop ^0 \ ^1 }\big)  \   {\rm mod} \ 2 \Big\}.
\end{align}
It is generated by $S^2=-1, T^2,ST^2S$. Note that $S^2=(ST)^2=-1$. The  weight two $\Gamma(2)$ modular forms,    $\vartheta_a^4(\tau),a=2,3,4$, are built out of the Jacobi theta functions  and transform under $S$ and $T$ as
\begin{align}
    (\vartheta_2^4,\vartheta_3^4,\vartheta_4^4)|_2 S \ = \ & -(\vartheta_4^4,\vartheta_3^4,\vartheta_2^4) ,\nonumber \\
    (\vartheta_2^4,\vartheta_3^4,\vartheta_4^4)|_2 T \ = \ &  (-\vartheta_2^4,\vartheta_4^4,\vartheta_3^4)  .
\end{align}
We refer the reader to Appendix \ref{sec:conventions} for the definition and properties of various modular objects used in the core of the article.

\begin{figure}[t]
	\centering
    \includegraphics[width=13 cm]{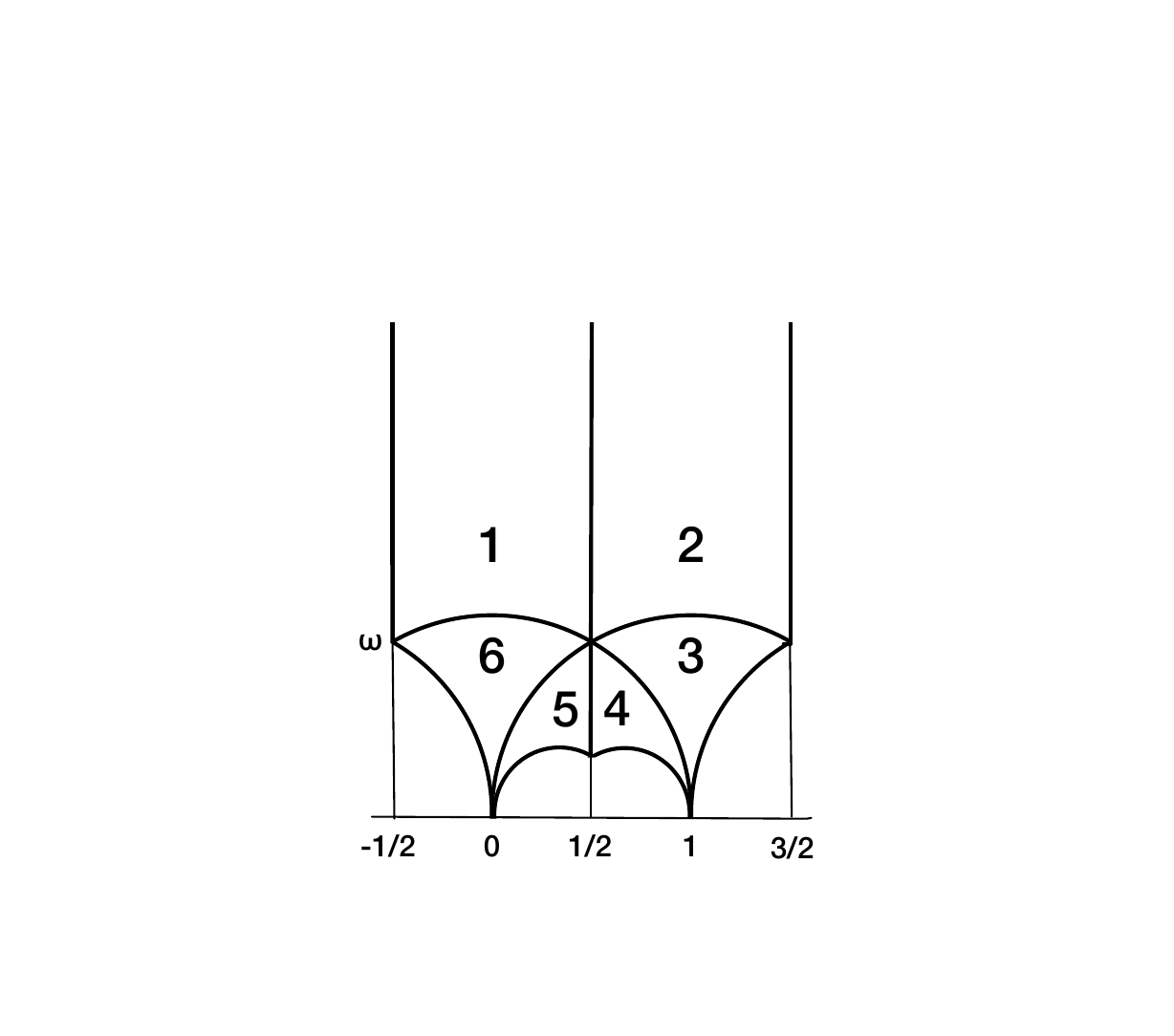}
    \caption{ Fundamental Domain of $\Gamma(2)$ }
    \label{fig:gamma2}
    \end{figure}

\begin{figure}[t]
	\centering
    \includegraphics[width=15cm]{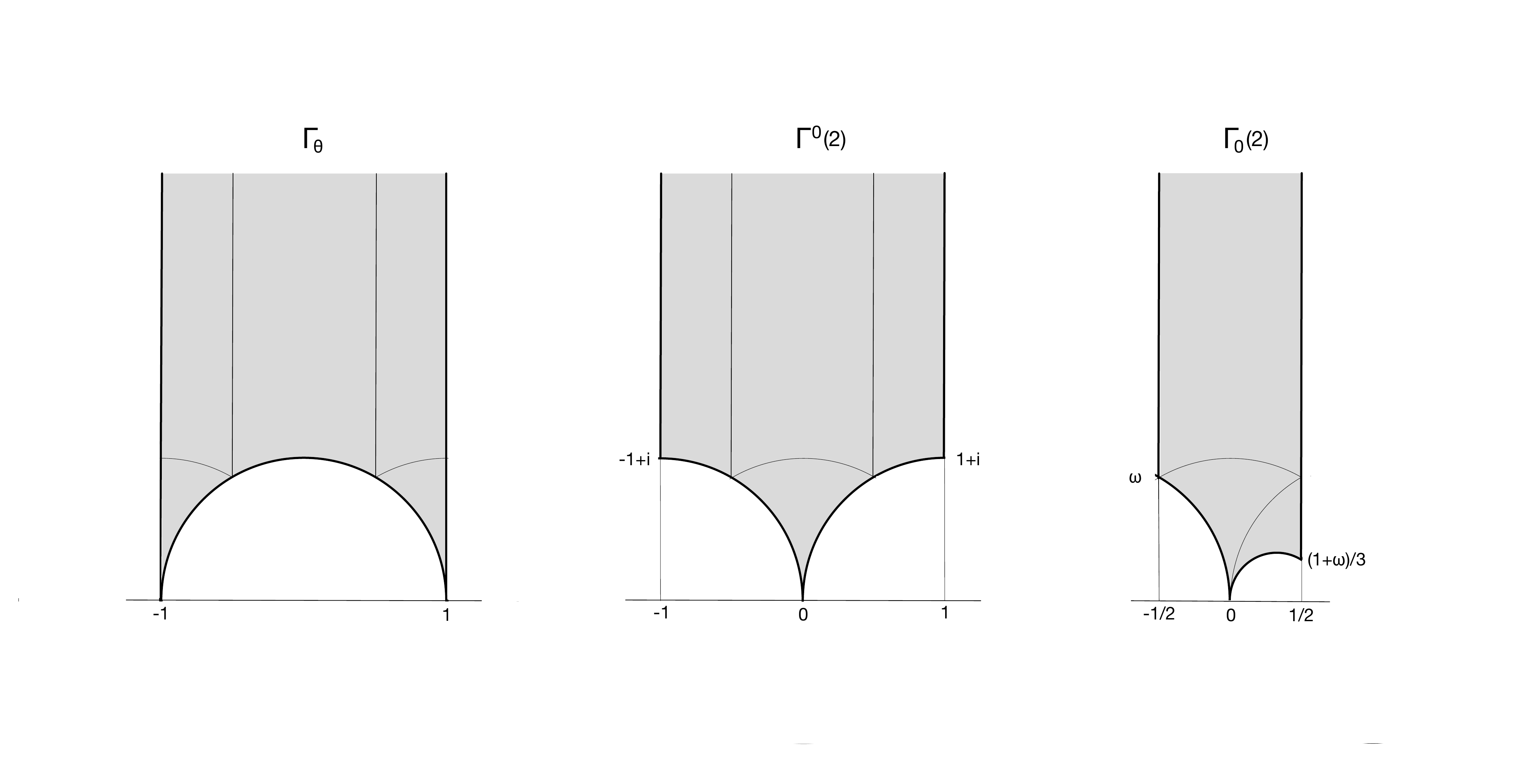}
    \caption{ Fundamental Domains of $\Gamma_\theta,\Gamma^0(2), \Gamma_0(2)$ }
    \label{fig:threeg}
\end{figure}

The quotient group ${\rm SL}_2({\mathbb Z})/\Gamma(2)$ is isomorphic to the permutation group $S_3$, which acting on modular forms permutes $\vartheta_2^8$, $\vartheta_3^8$ and $\vartheta_4^8$. The hauptmodul of $\Gamma(2)$, the Picard lambda function (which has of weight zero), is given in terms of the Jacobi theta functions by: 
\begin{align}
    \lambda(\tau)=\frac{\vartheta_2^4(\tau)}{\vartheta_3^4(\tau)}\,.
\end{align}
The lambda function $\lambda(\tau)$ transforms under $\gamma\in {\rm SL}_2({\mathbb Z})$ as follows:
\begin{align}
\label{eq:behaviorLambda}
    \begin{array}{c|cccccc}
  \gamma   & \ \ 1 \ \  & \ \ S \ \  &\ \  T \ \ & \ \ ST \ \ & \   (ST)^2 \ \ & \  TST\ \ \\ \hline 
    \lambda(\gamma\tau)    & \lambda  & 1-\lambda &  \frac{\lambda}{\lambda-1} &  \frac{\lambda-1}{\lambda} &
  \frac{1}{1-\lambda} & 
  \frac{1}{\lambda}
\end{array}.
\end{align}
The above transformations of the $\lambda$ functions indeed show that the congruence group  $\Gamma(2)$ decomposes ${\rm SL}_2({\mathbb Z})$ group to the permutation group $S_3$:
\begin{align}
     {\rm SL}_2({\mathbb Z})/\Gamma(2)=  \langle 1, S, T, ST, (ST)^2, TST\rangle\,.
\end{align}
The fundamental domain $ \Gamma(2)\backslash {\mathbb H} $   is made of the sum of the $S_3$ transformations of the fundamental domain ${\mathscr F}_{{\rm SL}_2({\mathbb Z})}= {\rm SL}_2({\mathbb Z})\backslash \mathbb{H} $:  
\begin{align}
    {\mathscr F}_{\Gamma(2)}  = \langle1, S, T, ST, (ST)^2, TS T \rangle {\mathscr F}_{{\rm SL}_2({\mathbb Z})}  
\end{align}
A representative of the six regions of the fundamental domain $\mathscr{F}_{\Gamma(2)}$ are given in 
Figure~\ref{fig:gamma2} where they are labelled as $(1,6,2,3,5,4)$.
The theta subgroup  $\Gamma_\theta$ of ${\rm SL}_2({\mathbb Z})$ is defined as follows:
\begin{equation} 
    \Gamma_\theta=  \left.\Big\{ \gamma\in {\rm SL}_2(\mathbbm{Z})\right.| \gamma\equiv \big({_1 \ _0 \atop ^0 \ ^1}\big) \ {\rm or}\   \big({_0 \ _1  \atop ^1 \ ^0} \big) \ {\rm mod} \ 2 \Big\}
\end{equation}
The Hecke congruence subgroups of level two are defined as
\begin{align}
    \Gamma_0(2)&=  \left.\Big\{ \gamma\in {\rm SL}_2(\mathbb{Z})\right.| \gamma\equiv \big({_\star \ _\star \atop ^0 \ ^\star}\big)   \   {\rm mod}\ 2 \Big\}\,,\\
    \Gamma^0(2)&=  \left.\Big\{ \gamma\in {\rm SL}_2(\mathbb{Z})\right.| \gamma\equiv \big({_\star \ _0 \atop ^\star \ ^\star }\big)  \   {\rm mod} \ 2\Big\}\,.
\end{align}
One can easily see that at level two, these coincide with the subgroups of unipotent elements modulo 2, namely $\Gamma_1(2)$ and $\Gamma^1(2)$. $\Gamma(2)$ is obviously a subgroup of $\Gamma_\theta,\Gamma^0(2),\Gamma_0(2)$. The index of $\Gamma_\theta$, $\Gamma_0(2)$ and $\Gamma^0(2)$ in $\text{SL}_2(\mathbb Z)$ is 3. Moreover, one has the following indices:
\begin{align}
    \Big[{\rm SL}_2(\mathbb{Z}):\Gamma_\theta \Big]=3,\ \Big[\Gamma_\theta:\Gamma(2)\Big]=2, \ 
    \Big[{\rm SL}_2(\mathbb{Z}):\Gamma(2)\Big]=6 . 
\end{align}
We have the following quotients:
\begin{align}
   &  {\rm SL}_2(\mathbb{Z})/\Gamma_\theta=\langle 1,T,ST  \rangle,     \ \ \Gamma_\theta/\Gamma(2) = \langle 1, S \rangle ,  \nonumber \\
   &  {\rm SL}_2(\mathbb{Z})/\Gamma^0(2)=\langle 1,T,S  \rangle,    \ \ \Gamma^0(2)/\Gamma(2) = \langle 1, TST \rangle ,  \nonumber \\
   &  {\rm SL}_2(\mathbb{Z})/\Gamma_0(2)=\langle 1,S,(ST)^2  \rangle ,   \ \ \Gamma_0(2)/\Gamma(2)= \langle 1, T \rangle .
\end{align}
The fundamental domains for these congruence subgroups are given by acting with the generators of the quotient spaces on the fundamental domain of $\text{SL}_2(\mathbb Z)$, up to  a proper identification of boundaries.  A connected representative of these fundamental domains is drawn in Figure~\ref{fig:threeg}.  
The spaces of the modular forms for these groups are simply spanned by combinations of Jacobi theta functions:
\begin{align}
\label{modular form basis}
    \mathcal{M}_{2k}(\Gamma_\theta) &={\rm span} \{  (-\vartheta_2^4)^r\vartheta_4^{4s} +  (-\vartheta_2^4)^s\vartheta_4^{4r}, \ r\leq s , \ r+s=k \} , \nonumber\\
    \mathcal{M}_{2k}(\Gamma^0(2)  ) &={\rm span} \{  (-1)^{r+s}  ( \vartheta_3^{4r}\vartheta_4^{4s} +  \vartheta_3^{4s}\vartheta_4^{4r} , \ r\leq s \ r+s=k \} , \nonumber\\
  \mathcal{M}_{2k}(\Gamma_0(2)) &={\rm span} \{ \vartheta_2^{4r}\vartheta_3^{4s} +   \vartheta_2^{4s}\vartheta_3^{4r} ,\ r\leq s,  \ r+s=k \}  . 
\end{align}
From the $\Gamma(2)$ weight 0 modular function $\lambda(\tau)$, one can form the  $\Gamma_\theta$ hauptmodul:
\begin{align}
    K(\tau)&=\frac{16}{\lambda(1-\lambda)}-24=   \frac{\vartheta_2^{12}+ \vartheta_3^{12}+ \vartheta_4^{12}}{2\eta^{12}} =\frac{\vartheta_3^{12}}{\eta^{12}}-24 \nonumber \\
    &= q^{-\frac12}  + 276 q^\frac12 + 2048 q  + 11202 q^\frac32 + 49152 q^2 + 
 184024 q^\frac52 +  614400 q^3+\cdots\ . 
\end{align}
 It is indeed invariant under $S$ and $T^2$ and has played a fundamental role in the moonshine phenomena for the sporadic group ${\rm Co}_0$ as the vacuum character of a  $\mathcal{N}=1$ supersymmetric conformal field theory (super VOA) \cite{FLM85}. 

The various modular functions $\lambda, j, K$ satisfy the following simple identities:
\begin{align} 
\label{eq:dlambda}
\frac{1}{2\pi i}\frac{\text{d}}{\text{d}\tau} \lambda(\tau)  &= \frac12 \vartheta_4^4\lambda , \nonumber\\
\frac{1}{2\pi i}\frac{\text{d}}{\text{d}\tau} K(\tau) & =
-\frac12 (\vartheta_4^4-\vartheta_2^4)(K(\tau)+24) ,\nonumber\\
\frac{1}{2\pi i }\frac{\text{d}}{\text{d}\tau} j(\tau) &=
 \frac12 (\vartheta_4^4-\vartheta_2^4)  \Big( 1-\frac{3}{1-\lambda(1-\lambda)}  \Big)j(\tau).
\end{align}
We will use the $\lambda$ function to re-express the MLDEs in terms of the new variable $\tau\to\lambda(\tau)$. It so happens that in some nice situations, doing so will allow us to explicitly solve the ODE, hence providing a closed form expression of the characters. For instance, it is known that the MLDE \eqref{eq:2omlde} can be written in terms of $\lambda$ and take the form:
\begin{align}
    \left[\frac{\text{d}^2}{\text{d}\lambda^2}+\frac{2(1-2\lambda)}{3\lambda(1-\lambda)}\frac{\text{d}}{\text{d}\lambda}+\frac{4\mu (1-\lambda+\lambda^2)}{\lambda^2(1-\lambda)^2}\right]f=0\ .
\end{align}
which then can explicitly be solved in term of the hypergeometric function $\left._2F_1\right.$.

    
\subsection{Valence Formula for $\Gamma_\theta$}

Let us consider the $\Gamma_\theta$ valence formula  for definiteness. The valence formula for $\Gamma_0(2)$ and $\Gamma^0(2)$ could be found similarly. The fundamental domain for $\Gamma_\theta$ is depicted above in Figure \ref{fig:threeg}. With the $S, T^2$ generators leave the fundamental domain invariant, one has to fold along ${\rm \tau}=0$ vertical line, identifying ${\rm Re}(\tau)=\pm 1$ vertical line by $T^2$ transformation and the left and right half of the semicircle by $S$ transformation.	 The fundamental domain  $\mathscr{D}_{\Gamma_\theta}$ has two cusp points at $i\infty$ and $1$,  and a single orbifold point   $\tau=i$ of order two.

\begin{figure}[t]
	\centering
   \includegraphics[width=7cm]{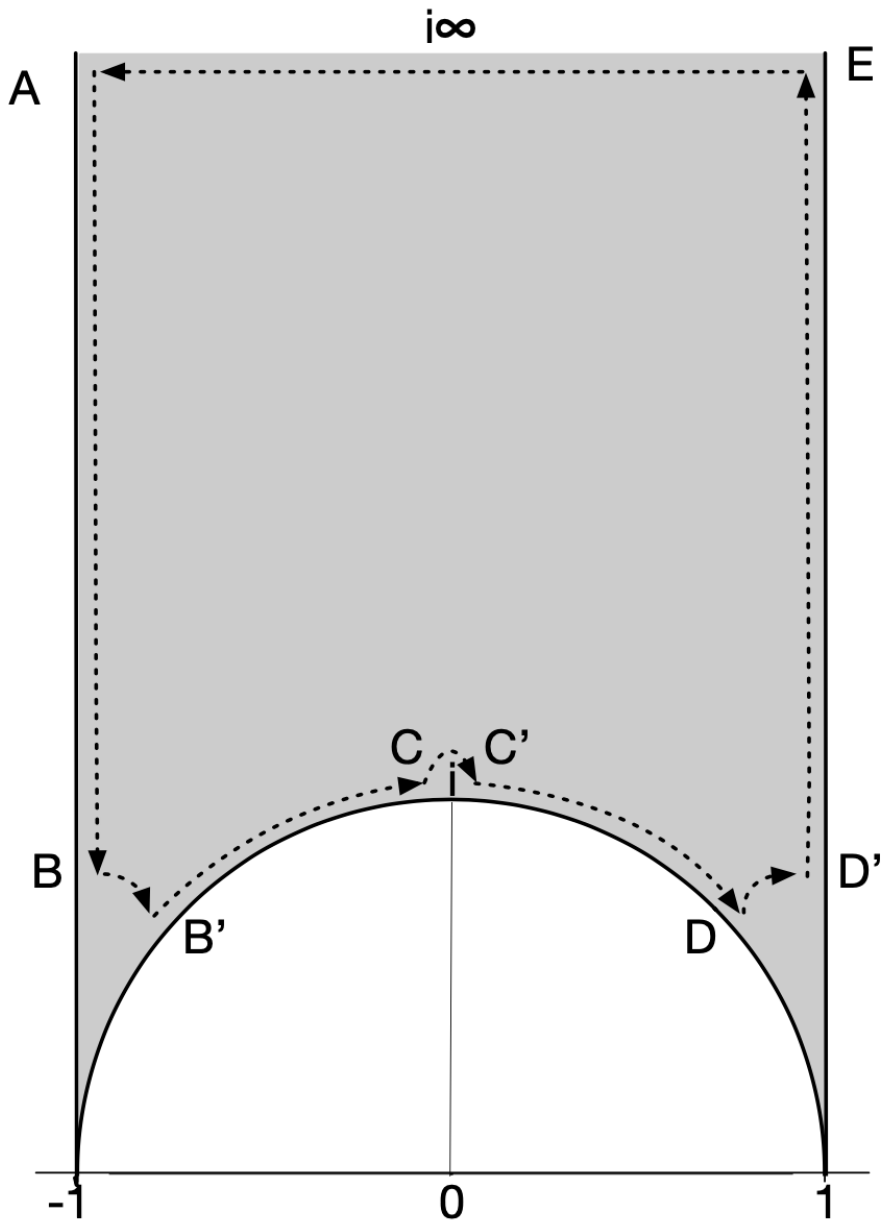}
    \caption{Integration contour along the boundary of the fundamental domain of $\Gamma_\theta$}
    \label{fig:contour}
    \end{figure}

Analogous to the $\text{SL}_2(\mathbb{Z})$ case, we consider a class of the functions $f:\mathbb H\to\mathbb C$ that $f$ are holomorphic inside $\mathbb H$, at worst meromorphic at the cusps $i\infty$ and $1$, and satisfy
	\begin{equation}
	     (f|_k\gamma)(\tau) = \rho(\gamma)^{-1} (c\tau+d)^{-k}f(\gamma \tau) = f(\tau),  \quad  \forall \gamma\in \Gamma_\theta\,   .
	\end{equation}
where $\rho: \Gamma_\theta \rightarrow U(1)$ is a possible phase. To derive a valence formula for $\Gamma_\theta$, we   are interested in the  integration of $df/f $ along the contour in the fundamental domain of $\Gamma_\theta$ as shown in Figure \ref{fig:contour}. We follow the  argument in the derivation of the valence formula for ${\rm SL}_2(\mathbb{Z})$ in Section \ref{sec:SL2Z}.  The contour integration leads to the equalities,
\begin{align}
\label{eq:g1balance2}
-\frac{1}{2\pi i} \oint d\tau \frac{\frac{\text{d}}{\text{d}\tau} f(\tau)}{f(\tau)} & = - \sum_{p\in{\rm SL}_2(\mathbb{Z})\backslash \mathbb{H}
\atop p\neq i, 1,  i\infty} 
 \nu_{ p}(f)  \nonumber \\ 
&= 
   2\nu_{\infty}(f)+ \nu_1(f)+ \frac12 \nu_i(f)     - \frac{k}{4}  . 
\end{align}
  
When the contour gets shrunk, it receives contributions only from the zeros of $f$ lying on the interior region, that is, not a orbifold point or cusps.   This is the first equality. The second one comes from the contour contributions. One can read off the origin of each terms in the RHS of the above equation from the dashed line in Figure~\ref{fig:contour}. The contributions from  contour segments are   as follows:
\begin{align}
   & C_{EA}: \ 2\nu_\infty(f), \ \ \  C_{BB'}+C_{DD'} = \nu_1(f) , \ \ C_{CC'}:\ \frac12 \nu_i(f), \nonumber \\
   &   C_{AB}+C_{D'E}: \ 0,\ \ \   C_{BC}+C_{C'D}:\   -\frac{k}{4}, \ \ \  
\end{align}
For $f\sim q^n$ near $i\infty$, $\log(f) \sim 2\pi i n \log \tau$. The integration length of $C_{EA}$ is two  and so its contribution is  $2\nu_\infty(f)=2n$. At another cusp $1$, the contribution becomes just $\nu_1$. The structure of the zero and the contour at $1$ just comes from the $ST$ transformation of the $i\infty$ contour integration in ${\rm SL}_2(\mathbb{Z})$. On orbifold points $i$  the contribution gets only $\frac12$    fraction of $\nu_i(f)$ as one integrate over one half of $2\pi$ integration. On the arc $C_{BC}$ and $C_{C'C}$, a point $\tau$ and its S-dual points  $-\tau^{-1}$ are matched. One change the variable $\tau\rightarrow -1/\tau$ on $C_{B'C}$ and use the fact $f(-1/\tau)=\rho(S)\,\tau^k f(\tau)$ to get the contribution from the contour $C_{B'C}$ is minus of that from the contour $C_{C'D}$ and $-k/4$. The factor $1/4$ arises because the angle between $1$ and $i$ is one fourth of $2\pi$. 

\begin{table}[]
    \centering
     \begin{tabular}{c|c|cccl}
    $f(\tau)$ & $\frac{k}{4}$ & $2\nu_\infty$ & $\nu_1$ & $\frac12\nu_i$  & $\nu_{\rm interior}$  \\
    \hline
   $\vartheta_4^4-\vartheta_2^4$  &  $\frac12$ & 0 & 0 &   $\frac12$  & 0  \\
   $\vartheta_3^8$ &  1 & 0 &  1 & 0  & 0  \\
   $\vartheta_2^4\vartheta_4^4$ & 1 & 1 & 0 & 0 & 0 \\
   $E_4$ & 1 & 0 & 0 & 0 & 1   (at     $\omega$) \\
   $E_6$ & $\frac32$ & 0 & 0 & $\frac12$ & 1 (at $1+i$) \\
   $\vartheta_2^4\vartheta_3^8\vartheta_4^4$ & 2 & 1 & 1 & 0 & 0 \\
   $K(\tau)+24$ & 0 & $-1$   & 1 & 0 & 0  \\
    $K(\tau)-40$ & 0 & $-1$   & 0 & 1 & 0  \\
   $\Delta(\tau)$ & 3 & $2$ & 1 & 0 &0  
\end{tabular}
    \caption{Weight $k$ (weak) modular forms of $
    \Gamma_\theta$}
    \label{tab:valence2}
\end{table}

In short, we obtain the valence formula for $\Gamma_\theta$ as follows: 
 	\begin{equation}
 	\label{eq:valence_formula_gammatheta}
 	2\nu_{i\infty}(f)+\nu_1(f) + 	\frac{\nu_{i}(f)}{2}+\sum_{\substack{\tau\in\Gamma_\theta\backslash\mathbb H\\ \tau\neq i}}\nu_\tau(f)=\frac{k}{4}\,.
	\end{equation}
Table~\ref{tab:valence2} is a check for the valence formula for  modular forms of $\Gamma_\theta$. As studied in the previous subsection \ref{subsec:3.1}, the space $M_k(\Gamma_\theta)$ of modular forms is generated by $ (-\vartheta_2^4)^r\vartheta_4^{4s}+ (-\vartheta_2^4)^s\vartheta_4^{4r}, 0\le r\le s, r+s=k$. 
The space $S_k(\Gamma_\theta)$  of cusp forms for $\Gamma_\theta$ is made of the modular forms of $\Gamma_\theta$ which vanish at  cusps, $i\infty$ and $1$, given by
\begin{equation}
    S_{2k}(\Gamma_\theta)=\vartheta_2^4\vartheta_3^8\vartheta_4^4 M_{2k-8}(\Gamma_\theta)\, .
\end{equation}
The dimensions of $M_{2k}$ and $S_{2k}$ are given as follows:
\begin{align}
\label{dimension formula}
      {\rm dim } M_{2k}(\Gamma_\theta) & = \left\{ \begin{array}{cc}
           1+ \lfloor \frac{k}{2}\rfloor, & k\ge 0 \\
            0  , & k<0 ,
      \end{array} \right.
 \nonumber \\
 {\rm dim }S_{2k}(\Gamma_\theta) & = \left\{ \begin{array}{cc}
           -1+ \lfloor \frac{k}{2}\rfloor, & k\ge 4 \\
        0    ,  & k<4 .
      \end{array} \right. 
\end{align}
 
\subsection{Modular Linear Differential Equation for $\Gamma_\theta$}

Let us now see  how  to  use the  above valence formula  in the  context of fermionic  RCFTs. For that purpose, we consider a fermionic RCFT which has finite number of characters $\chi^{\rm NS}_i,i=0,1,...N-1$ with conformal weight $h_i$. We repeat the argument for MLDE of ${\rm SL}_2(\mathbb{Z})$ in Section \ref{sec:SL2ZMLDE} for these $N$ characters in NS-sector to get the MLDE for $\Gamma_\theta$: 
\begin{align}
    \Big[{\mathcal D}^N  + \sum_{k=0}^{N-1}\phi_k {\mathcal D}^k \Big]f^{\rm NS}=0\,.
\end{align}
The coefficient functions  $\phi_k=(-1)^{N-k} W_k/W_N$ are the weight $2N-2k$ rational functions of  modular forms of $\Gamma_\theta$. The Wronskian $W_k$ here is given in terms of the NS-sector characters $\chi^{\rm NS}_i$. We adopt the same strategy as in  i.e., studying or classifying MLDE's based on the order of poles of $\phi_k = (-1)^{N-k} W_k/W_N$.  Since the characters $\chi_i^\textsc{ns}$ hence $W_k$ are always holomorphic inside the fundamental domain, the only possibility of introducing a pole is through the zeros of the denominator $W_N$. It's easy to check that $W_N$ transforms as a (meromorphic) weight $N(N-1)$ modular form of $\Gamma_\theta$ up to a possible overall phase.  Therefore, we can safely apply the valence formula \eqref{eq:valence_formula_gammatheta} to $W_N$.

Due to the very nature of characters, we are able to further simplify the formula. Notice that at $\tau = i\infty$, by definition $\chi_{j}^{\textsc{ns}}$ have the following asymptotic expansion,
\begin{equation}
    \chi_{j}^{\textsc{ns}}|_{i\infty} = q^{-c/24 + h_j^{\textsc{ns}}}(\text{const.}+\mathcal O(q))\,.
\end{equation}
Taking Serre derivative on $\chi_{j}^{\textsc{ns}}$ doesn't change their order at $i\infty$, so we learn that 
	\begin{equation}
	    W_N|_{i\infty} = q^{-Nc/24 + \sum_{j}h_j^{\textsc{ns}}}(\text{const.}+\mathcal O(q))\,.
	\end{equation}
In other words, we have:
	\begin{equation}
		\nu_{i\infty}(W_N)=-\frac{Nc}{24} + \sum_{j}h_j^{\textsc{ns}}\,.
	\end{equation}
Concerning the cusp at $1$, a priori we need to find a suitable local coordinate to extract the leading order. However, there is a way to bypass this problem by noting the fact that $1$ is mapped to $i\infty$ under the $(S   T^{-1})$ transform. Then the local behaviour of $\chi_{j}^{\rm NS}$ around $1$ is equivalent to that of $S  T^{-1}(\chi_{j}^{\rm NS})\sim \chi_j^{\rm R} $ around $i\infty$.  Thus the Wronskian of NS characters near $\tau=1$ becomes that of R characters near $\tau=i\infty$:  
	\begin{equation}
	W_N|_1 \sim q^{-Nc/24 + \sum_{j}h_j^{\rm R}}(1+\mathcal O(q))\, .
	\end{equation}
Immediately, we see that 	
	\begin{equation}
		\nu_{1}(W_N)=-\frac{Nc}{24} + \sum_{j}h_j^{\textsc{r}}\,.
	\end{equation}
Mimicking what was done in \cite{Mathur:1988na} and also in \eqref{eq:ellsl2z}, if we define $l/2$ to be the``number" of zeros inside the fundamental domain, the valence formula (\ref{eq:valence_formula_gammatheta}) gives us then the following relation: 
\begin{equation}
\label{eq:valencetheta}
    -\frac{Nc}{8}+2\sum_j h_j^{\rm NS} +\sum_j h_j^{\rm R} + \frac{\ell}{2}= \frac{N(N-1)}{4},
\end{equation}
where the contribution from zeros  is given as
\begin{equation}
\label{eq:baltheta}
    \frac{\ell}{2}\equiv \frac12 \nu_i(W)+ \sum_{  \tau\in\Gamma_\theta\backslash \mathbb{H} 
    \atop \tau\neq i}\nu_\tau(W)\, .
\end{equation}

For a given fermionic RCFT with known $c, h_j^{\rm NS}$ and $ h_j^{\rm R}$, the value of the zero determines the structure of MLDE satisfied by the characters in NS sector. In this work, we will focus on the case where $\ell=0$. In this case the coefficient functions $\phi_k$ of the MLDE for $\Gamma_\theta$ become the modular functions of $M_{2N-2K}(\Gamma_\theta)$. We explore the possible values of $c,h^{\rm NS}_j, h^{\rm R}_j$ whose characters have the non-negative integer $q^\frac12$-expansion coefficient. For the conformal characters in $\rm NS$-sector, there  always exists the unique {\it vacuum} character with zero conformal weight $h=0$ and $q^\frac12$ expansion:
\begin{align}  \chi^{\rm NS}_0 &= q^{-\frac{c}{24}}(1+ a_1 q^\frac12+a_2 q+a_3 q^\frac32+a_4q^4+\cdots) \nonumber\\
&=q^{-\frac{c}{24}}\Big[ 1+  \sum_{n=1}^\infty a_n q^\frac{n}{2}\Big].
\end{align}
The MLDE for $\widetilde{\rm NS}, {\rm R}$ sector can be obtained by just making $T$ and $S$ transformation of the MLDE for the NS sector. The modular forms $\phi_k$ would transform accordingly.  
 
Before exploring the solutions of MLDE, let us compare bosonic and fermionic valence formulas \eqref{sl2zvalence} and \eqref{eq:valencetheta}. If we start with $N$ characters in the NS sector, we can bosonize and get $3N$ characters by combining all characters in $\rm NS$, $\widetilde{\rm NS}$, and $\rm R$ characters. By a linear combination of $\chi_a^{\rm NS}$ and $\chi_a^{\widetilde{\rm NS}}$ we get two characters with conformal weight $h_a$ for the sum and $h_a+s_a$ for the difference, where $s_a$ is a positive half integer, say, $\frac12$ or $\frac32$, in our case. So we could use the bosonic MLDE for the $3N$ characters with $h^{\rm NS}_a$ , $h^{\rm NS}_a+s_a$ and $h^{\rm R}_a$ with the number of zeros $\ell_b$. For simplicity, we are assuming that all $3N$ characters are linearly independent.  Then two valence formulas imply a consistency condition,
\begin{align}
\label{eq:consist}
   \frac{\ell_b}{6}= \frac{\ell_f}{2} +\frac{N^2}{2} - \sum_a s_a\,.  
\end{align}
As we consider the solutions of  MLDE for the fermionic RCFT in the $q$-expansion, we can read $s_a$ trivially. 
The above relation then provides the information on the $\ell_b$, that is, the pole structure of the coefficient functions for  MLDE of bosonic $3N$ characters.


\section{Fermionic First Order MLDE} \label{sec:1stMLDE}



As a warm-up exercise, we consider the fermionic RCFT whose torus partition function for each spin structure can be holomorphically factorized. That is to say, the theory has a single character in each sector. For convenience, let us focus on the NS sector. The vacuum character $f_0^\text{NS}(\tau)\sim q^{-\frac{c}{24}}(1+\cdots)$ then becomes a solution to the first order MLDE below 
\begin{align}
\label{eq:1storder}
    \Big[ {\cal D}   + \phi_0(\tau) \Big] f^\text{NS}(\tau) = 0, 
\end{align}
where $\phi_0(\tau)$ is the $\Gamma_\theta$ modular form of weight two. Restricting our attention to the case $l=0$, $\phi_0(\tau)$ can be generated by $\big( -\vartheta_2^4(\tau) + \vartheta_4^4(\tau) \big)$ since the space $M_2(\Gamma_\theta)$ is one-dimensional. Therefore the first order MLDE of our interest takes a form of 
\begin{align}
\label{eq:firstOrderMLDE}
  \Big[D_\tau + \mu  \left(-\vartheta_2^4(\tau)+\vartheta_4^4(\tau)\right) \Big]f^\textsc{NS}=0.
\end{align}
Note that its $T$ and $S$ transformations lead to the first order MLDE for $\Gamma^0(2)$ and $\Gamma_0(2)$:
\begin{align}
\label{eq:1stforG0s}
 & \Big[D_\tau - \mu  \left(\vartheta_3^4(\tau)+\vartheta_4^4(\tau)\right) \Big]f^{\widetilde{\rm NS}}=0, \nonumber \\
 & \Big[D_\tau + \mu  \left(\vartheta_2^4(\tau)+\vartheta_3^4(\tau)\right) \Big]f^{ \rm R}=0, 
\end{align}

To have a conformal character $f_0^{\rm NS} \sim q^{-\frac{c}{24}}(1+\cdots)$,one can determine the coefficient $\mu$ to be $\mu=\frac{c}{12}$. With the help of (\ref{eq:dlambda}), \eqref{eq:firstOrderMLDE} can be rewritten in terms of the $\lambda(\tau)$ variable as follows,
\begin{align}
    \bigg[\frac{\text{d}}{\text{d}\lambda} + \frac{c}{12} \frac{1-2\lambda}{\lambda(1-\lambda)}\bigg]f^{\rm NS} & =0\, .
\end{align}
The solution is simply given by
\begin{equation}
    f_0^\textsc{NS}(\lambda)= \Big[ \frac{16}{ \lambda (1 - \lambda )} \Big]^{\frac{c}{12}} 
    =(K+24)^{\frac{c}{12}} . 
\end{equation}
Demanding the solution to have non-negative integer coefficients for the $q$-expansion leads to $c=\frac{N}{2}$ for a positive integer $N$. It is consistent with the fact that the chiral central charge of a fermionic CFT has to be half-integral. This is because the $(2+1)$-dimensional fermionic gravitational Chern-Simons coupling can cancel certain (anomalous) $U(1)$ phases that arise from the modular transformation of partition functions of fermionic CFTs with $c_L-c_R\in \mathbb{Z}/2$. 

When $N=1$, one can see that $f_0^\text{NS}$ is nothing but the NS partition function of a single free Majorana-Weyl fermion, 
\begin{align}\label{aaa}
    \psi^\text{NS} = \Big[ \frac{16}{ \lambda (1 - \lambda )} \Big]^{\frac{1}{24}} = \sqrt{\frac{\vartheta_3}{\eta}} .
\end{align}
The other partition function for different spin structures can be obtained by acting $T$ and $S$ on $\psi^\text{NS}$,
\begin{align}
    & T:\psi^{\rm NS}\longleftrightarrow e^{-\pi i /24} \psi^{\widetilde{\rm NS}} = e^{-\pi i /24}  \sqrt{\frac{\vartheta_4}{\eta}},
    \nonumber \\ 
    & S:\psi^{\widetilde{\rm NS}} \longleftrightarrow \psi^{\rm R} = \sqrt{\frac{\vartheta_2}{\eta}}.
\end{align}
Note also that the characters of a free Majorana-Weyl fermion are related to the characters of the Ising model as follows
\begin{align}
\label{theta identity1}
   \psi^{\rm NS} &= \chi^{\rm Ising}_0+\chi^{\rm Ising}_\frac12,   \nonumber 
   \\
   \psi^{\widetilde{\rm NS}} & = \chi^{\rm Ising}_0-\chi^{\rm Ising}_\frac12  , 
   \nonumber\\
   \psi^{\rm R} &=  \sqrt{2}  \chi^{\rm Ising}_\frac{1}{16}.   
\end{align}
Therefore, the solution \eqref{aaa} can be identified with the NS partition function of $N$-copies of free Majorana-Weyl fermions,
\begin{align}
\label{free fermion}
    f^\textsc{NS}_0(\tau) = \Big(\psi^{\rm NS}(\tau)\Big)^{N} ,  
\end{align}
with the identities 
\begin{align}
\label{theta identity}
    (\psi^{\rm NS})^{24}= K(\tau) +24,\ (\psi^{\widetilde{\rm NS}})^{24}= -K(\tau+1) +24, \  (\psi^{\rm R})^{24}= -K(-\frac1\tau+ 1) +24.
\end{align}
%

    
\section{Fermionic Second Order MLDE}  \label{sec:2ndMLDE}


\subsection{The Second Order MLDE}

In this section, we consider the fermionic RCFT 
with two characters in each sector. Let us first focus on the NS sector. The characters for the  vacuum and primary state of conformal weight $h^{\rm{NS}}$ have a series expansion  
\begin{align}
\begin{split}
\label{eq:twochar}
    f^{\rm NS}_0(\tau) &=  q^{-\frac{c}{24}}\left( 1+ a_1 q^\frac12+ a_2q+a_3 q^\frac32+a_4q^2+\cdots \right),  \\
    f^{\rm NS}_1(\tau) &=  q^{h^{\rm NS}-\frac{c}{24}}\left( b_0 + b_1 q^\frac12+ b_2q+b_3 q^\frac32+b_4q^2+\cdots \right),
\end{split}
\end{align}
and they are independent solutions of the second order MLDE of the form
\begin{align}
   \Big[\mathcal{D}^2 +\phi_1(\tau)   \mathcal{D}  +\phi_0(\tau)\Big] f_j^{\rm NS}(\tau) = 0, \qquad j = 0, 1.
\end{align}
We restrict our attention to the zero-pole case $\ell=0$ where  the coefficients $\phi_0$ and $\phi_1$ are spanned by $\Gamma_\theta$ modular forms of weight four and two, respectively. The weight two modular form $M_{2}(\Gamma_\theta)$ is one-dimensional and generated by $-\vartheta_2^4+\vartheta_4^4$. The space of weight four modular form $M_4(\Gamma_\theta)$ is two-dimensional whose basis is chosen to be  $\vartheta_3^8$ and $E_4 $. Therefore, the second order MLDE for $\rm NS$ sector takes the expression: 
\begin{align}
\label{eq:2MLDEThe}
\begin{split}
& \Big[\mathcal{D}^2+ \mu_1  \left(-\vartheta_2^4(\tau)+\vartheta_4^4(\tau)\right) \mathcal{D} + \mu_2 \theta_3^8 + \mu_3 E_4 \Big]f^{\text{NS}}(\tau)=0, 
\end{split}
\end{align}
where $\mu_1,\mu_2$ and $\mu_3$ are three independent coefficients. We will mainly focus on the solutions of \eqref{eq:2MLDEThe} which have the non-negative integer coefficients in the $q$-expansion. Note that the first term of the vacuum solution should be given by $q^{-c/24}$ due to the uniqueness of the vacuum state. On the other hand, there is an ambiguity for the normalization of the primary solution denoted by $b_0$ in \eqref{eq:twochar}. We will show that the normalization $b_0$ for the NS sector solution can be fixed with the help of analytic expression of the $S$-matrix.

The MLDE for $\widetilde{\rm{NS}}$ and R sector can be obtained from \eqref{eq:2MLDEThe}  by applying $T$ and $ST$ transformation. Explicitly, the differential equations are given by
\begin{align}
\label{eq:2mldG0}
\begin{split}
& \Big[\mathcal{D}^2+ \mu_1 \left(\vartheta_3^4(\tau)+\vartheta_2^4(\tau)\right)\mathcal{D} + \mu_2 \vartheta_4^8(\tau)+ \mu_3 E_4(\tau)\Big]f^{\widetilde{\rm{NS}}}(\tau)=0 \qquad  \text{for} \ \widetilde{\text{NS}} \ \text{sector}, \\
& \Big[\mathcal{D}^2- \mu_1\left(\vartheta_3^4(\tau)+\vartheta_4^4(\tau)\right)\mathcal{D} + \mu_2 \vartheta_2^8(\tau)+ \mu_3 E_4(\tau)\Big]f^{\rm{R}}(\tau)=0 \qquad \text{for} \ {\text{R}} \ \text{sector},  
\end{split}
\end{align} 
which are consistent with the valence formula. In the case of  $\mu_1=\mu_2=0$, the above three MLDE reduce to the second-order MLDE for ${\rm SL}_2({\mathbb Z})$ whose solutions were discussed in Section \ref{sec:SL2ZMLDE}. Thus we restrict to the cases where at least one of $\mu_1,\mu_2$ is non-zero. 

Our strategy of solving the NS sector MLDE is to ask if the characters of the series form  \eqref{eq:twochar} are solutions of \eqref{eq:2MLDEThe} in every order of $q$. In addition, we demand the coefficients of the series expansion to be non-negative integers. Let us choose the input to be central charge $c$, weight of the primary $h^{\rm NS}$, and $a_1\ge 0$,
where $a_1$ is the coefficient of  $q^\frac12$ term in the vacuum character. For non-zero $a_1$, the corresponding theory contains primaries of dimension $1/2$ and spin $1/2$, i.e., free fermions. See, for instance, \cite{Lee:2019uen}. Once we require the NS-sector characters to satisfy the MLDE for $\Gamma_\theta$, the  coefficients $\mu_1,\mu_2,\mu_3$ are determined by inputting parameters $(c,h,a_1)$ as follows:
\begin{align}
\begin{split}
\label{mu for 2nd MLDE}
    \mu_1 &=\frac{c-12h^{\rm NS}+2}{12},  \\   
    \mu_2 &=-\frac{c(c-12h^{\rm NS}+2)+3a_1(1-2h^{\rm NS})}{192}, \\
   \mu_3 &=\frac{2c(2c-30h^{\rm NS}+3)+9a_1(1-2h^{\rm NS})}{576}.
\end{split}
\end{align}

Now let us move our attention to the R sector. The two characters of the R sector have series expansions of the form
\begin{align}
\label{eq:twocharR}
    f^{\rm R}_0(\tau) &=  q^{h^{\rm R}_{-}-\frac{c}{24}}\left( a'_0+  a'_1 q +a'_2 q^2 +a'_3 q^3 + a'_4 q^4 +\cdots \right), \nonumber \\
    f^{\rm R}_1(\tau) &=  q^{h^{\rm R}_{+}-\frac{c}{24}}\left( b'_0+ b'_1 q +b'_2q^2 +b'_3q^3 +b'_4q^4+\cdots \right),
\end{align}
where $h^{\rm R}_\pm$ denote the weights of two primaries in the R sector. Without loss of generality, we assume $h^{\rm R}_{+} \geq h^{\rm R}_{-}$, i.e., $h_-^\text{R}$ is the Ramond vacuum weight. With the help of \eqref{mu for 2nd MLDE}, one can show that the weights $h^{\rm R}_\pm$ of the R sector are determined from the MLDE \eqref{eq:2mldG0} as below,
\begin{equation}
\label{eq:rpmweight}
    h^{\rm R}_\pm = \frac18 \left( 2 + c - 8 h^{\rm{NS}} \pm   \sqrt{4(1-4h^{\rm NS})^2 +(2c -a_1)(1-2h^{\rm NS})} \right),
\end{equation}
and the weights $h^{\rm R}_\pm$ should be a rational number as far as we consider the RCFT. Note that the valence formula  \eqref{eq:valencetheta} is satisfied trivially with \eqref{eq:rpmweight}. When we assume the resulting theory has supersymmetry, there is a unitarity bound 
\begin{align}\label{SUSYbound}
    h^{\rm R} \ge \frac{c}{24}
\end{align} 
for the R sector weights which leads to the constraint on the initial data $(c, h^{\rm NS}, a_1)$,
\begin{align}
  \label{eq:bpsonchk}
    \mu_3 \sim 2c(2c-30 h^{\rm NS}+3) +9a_1(1-2h^{\rm NS}) \ge 0  .
\end{align}
For a specific case where $a_1=0$, or $h^{\rm NS}=\frac{1}{2}$, this bound takes a simpler form:   
\begin{align}
\left\{ \begin{array}{lcl}
 c\ge 15 h^{\rm NS} -\frac32    & \  {\rm for} & a_1=0, \\
 c\ge 6 &\ {\rm for} & h^{\rm NS}=\frac12.
\end{array} \right.
 \end{align}

Let us further comment on the RCFT with supersymmetry. A necessary condition for a fermionic RCFT to have supersymmetry is that there is at least one weight-$3/2$ supersymmetry current as a vacuum descendant. Such supersymmetry currents contribute to $a_3 q^\frac32 $ in the NS  vacuum character. If a supersymmetric RCFT further saturates a unitarity bound \eqref{SUSYbound}, the RCFT has the supersymmetric ground states. Otherwise one can say that the supersymmetry is spontaneously broken. For this reason,  we call $h^{\rm R}_{-} = \frac{c}{24}$ as the BPS condition. When the BPS condition is satisfied and $h^{\rm NS}\neq \frac{1}{2}$, one can fix $h^{\rm R}_+$ and $a_1$ in terms of $c$ and $h^{\rm NS}$ as follows.
\begin{align}
\label{BPS constraints}
     h_-^{\rm R} &= \frac{c}{24}, \ \ \ 
   h_+^{\rm R}  =\frac{5c}{24}+\frac12 -2h^{\rm NS},  \ \  \  
    2c(2c-30 h^{\rm NS}+3) +9a_1(1-2h^{\rm NS}) =0.
\end{align}
We will discuss later some examples of solutions that violate the unitarity bound, thus cannot be considered as the characters for a supersymmetric RCFT. 

\subsection{Solutions of the Second Order MLDE}
\label{sec:Classification 2ndOrder}

Let us reformulate the second order MLDEs in terms of $\lambda(\tau)$ variable to find a closed-form solution, 
\begin{align}
\label{MDEs by lambda}
\begin{split}
    \Big[ \frac{\text{d}^2}{\text{d}\lambda^2} +   \frac{2(1+3\mu_1)(1 -2\lambda)}{3\lambda(1-\lambda)} \frac{\text{d}}{\text{d}\lambda} +  \frac{4(\mu_2 + \mu_3)  -4\mu_3\lambda(1-\lambda)   }{\lambda^2(1-\lambda)^2}  \Big] f^{\rm{NS}}(\lambda) &=0,   \\
\Big[ \frac{\text{d}^2}{\text{d}\lambda^2} +   \frac{2(1-2\lambda) +6\mu_1 (1 + \lambda)}{3\lambda(1-\lambda)} \frac{\text{d}}{\text{d}\lambda} +  \frac{4(\mu_2 + \mu_3)  -4\mu_3\lambda(1-\lambda)   }{\lambda^2(1-\lambda)^2}  \Big] f^{\widetilde{\rm{NS}} }(\lambda) &=0,   \\
 \Big[ \frac{\text{d}^2}{\text{d}\lambda^2} +   \frac{2(1-2\lambda) +6\mu_1 ( \lambda-2)}{3\lambda(1-\lambda)} \frac{\text{d}}{\text{d}\lambda} +  \frac{4(\mu_2 + \mu_3)  -4\mu_3\lambda(1-\lambda)   }{\lambda^2(1-\lambda)^2}  \Big] f^{ \rm R }(\lambda) &= 0.
\end{split}
\end{align}
We obtain the MLDE for $f^{\widetilde{\rm{NS}} }(\lambda)$ and $f^{{\rm{R}} }(\lambda)$ by acting $T$ and $ST$ transformations on the MLDE for $f^{{\rm{NS}} }(\lambda)$. When $h^{\rm{NS}}\neq \frac{1}{2}$, the solutions can be expressed in terms of the hypergeometric function as follows,
\begin{align}
\label{NS characters 2nd}
f^{\rm{NS}}_0(\lambda) &= 2^{\frac{c}{3}}\lambda^{-\frac{c}{12}} (1-\lambda)^{-\frac{c}{12}} {_2F_1}\left(\beta^+,\beta^-;1-2h^{\rm{NS}} ; \lambda \right), \\
f^{\rm{NS}}_1(\lambda) &= b_0 2^{\frac{c}{3}-8h^{\rm{NS}}} \lambda^{2h^{\rm{NS}}-\frac{c}{12}} (1-\lambda)^{-\frac{c}{12}} {_2F_1}\left(\beta^+ + 2h^{\rm{NS}},\beta^- + 2h^{\rm{NS}};1+2h^{\rm{NS}} ; \lambda \right),
\nonumber
\end{align}
where $b_0$ denotes a normalization constant of the non-vacuum character and
\begin{align}
\beta^{\pm} = \frac{1}{4} \left( 2 - 8 h^{\rm{NS}} \pm \sqrt{4-a_1 + 2c-32 h^{\rm{NS}} + 2 a_1 h^{\rm{NS}} - 4 c h^{\rm{NS}} + 64 (h^{\rm{NS}})^2}\right).
\end{align}
From the known identities of the hypergeometric functions, one can see that \eqref{NS characters 2nd} transform under the $S$ transformation as follows
\begin{align}
\label{S-matrix of 2nd SMDE main}
\left(
\begin{array}{c}
f^{\rm{NS}}_0(1-\lambda)    \\[0.5em]
f^{\rm{NS}}_1(1-\lambda)
\end{array}
\right)
=
\left(
\begin{array}{cc}
\frac{\Gamma(1-2h^{\rm{NS}}) \Gamma(2h^{\rm{NS}})}{\Gamma(1-2h^{\rm{NS}}-\beta^+)\Gamma(1-2h^{\rm{NS}}-\beta^-)} & \frac{2^{8h^{\rm{NS}}}}{b_0} \frac{\Gamma(1-2h^{\rm{NS}})\Gamma(-2h^{\rm{NS}})}{\Gamma(\beta^+) \Gamma(\beta^-)}    \\[0.5em]
\frac{b_0}{2^{8h^{\rm{NS}}}} \frac{\Gamma(1+2h^{\rm{NS}})\Gamma(2h^{\rm{NS}})}{\Gamma(1-\beta^{+})\Gamma(1-\beta^{-})} & \frac{\Gamma(1+2h^{\rm{NS}})\Gamma(-2h^{\rm{NS}})}{\Gamma(\beta^+ +2h^{\rm{NS}})\Gamma(\beta^- + 2h^{\rm{NS}})}
\end{array}
\right)
\left(
\begin{array}{c}
f^{\rm{NS}}_0(\lambda)    \\[0.5em]
f^{\rm{NS}}_1(\lambda)
\end{array}
\right).
\end{align}
The derivation of \eqref{NS characters 2nd} and \eqref{S-matrix of 2nd SMDE main} was inspired by \cite{Naculich:1988xv}, and we briefly review it in the Appendix \ref{App:A} for completeness.  Whenever we encounter zeros or poles in the denominator or numerator, a careful limit has to be taken.

For the case of $h^{\rm NS}=\frac{1}{2}$, the MLDE takes a form of
\begin{align}
\left[\lambda^2(1-\lambda^2)  \frac{\text{d}^2}{\text{d}\lambda^2} +  \frac{c \lambda(1-\lambda)(1-2\lambda)}{6} \frac{\text{d}}{\text{d}\lambda} + \left(\frac{c(c-12)}{144} -\frac{1}{36} c(c-6) \lambda(1-\lambda)\right) \right] f^{\rm{NS}}(\lambda)=0 \nonumber
\end{align}
and there are two independent solutions for the above differential equation,
\begin{align}
g_1(\lambda) = \left(\lambda(1-\lambda)\right)^{-\frac{c}{12}}, \qquad g_2(\lambda) = \lambda\, \left(\lambda(1-\lambda)\right)^{-\frac{c}{12}}
\end{align}
Thus any NS sector character of two character RCFT with $h^{\rm{NS}}=\frac{1}{2}$ should be written as a combination of $g_1(\lambda)$ and $g_2(\lambda)$. 

The structure of NS sector partition function can be fixed by the $S$-matrix presented in \eqref{S-matrix of 2nd SMDE main}. More precisely, the NS sector partition function takes a form of
\begin{align}
\label{NS sector partition function analytic}
Z^{\rm{NS}}(\tau, \bar{\tau}) = |f^{\rm{NS}}_0(\lambda)|^2 + \mathcal{M} |f^{\rm{NS}}_1(\lambda)|^2,
\end{align}
where $\mathcal{M}$ denote the degeneracy of the NS primary states. Note that the multiplicity $\mathcal{M}$ can be determined from the $S$-matrix as below.
\begin{align}
S^T \cdot \text{diag}(1,\mathcal{M}) \cdot S = \text{diag}(1,\mathcal{M}) 
\end{align}

With the help of the identity of hypergeometric function
\begin{align}
\begin{split}
_2F_1(\alpha_1, \alpha_2; \alpha_3;\lambda) = (1-\lambda)^{-\alpha_1} {_2F_1}\left(\alpha_1, \alpha_3-\alpha_2; \alpha_3;\frac{\lambda}{\lambda-1}\right) ,
\end{split}
\end{align}
we find the analytic form of the $\widetilde{\rm{NS}}$ sector solutions that work for $h^{\rm{NS}} \neq \frac{1}{2}$.
\begin{align}
\begin{split}
f^{\widetilde{\rm{NS}}}_0(\lambda) &= 2^{\frac{c}{3}}\lambda^{-\frac{c}{12}} (1-\lambda)^{\frac{c}{6} + \beta^{+}} {_2F_1}\left(\beta^+,1-2h^{\rm{NS}}-\beta^-;1-2h^{\rm{NS}} ; \lambda \right), \\
f^{\widetilde{\rm{NS}}}_1(\lambda) &= b_0 2^{\frac{c}{3}-8h^{\rm{NS}}} \lambda^{2h^{\rm{NS}}-\frac{c}{12}} (1-\lambda)^{\beta^+ +\frac{c}{6}} {_2F_1}\left(\beta^+ + 2h^{\rm{NS}},1-\beta^- ;1+2h^{\rm{NS}} ; \lambda \right)
\end{split}
\end{align}
One can then express the partition function of $\widetilde{\rm{NS}}$ and R sector in terms of the $\widetilde{\rm{NS}}$ sector characters $f^{\widetilde{\rm{NS}}}_0(\lambda)$ and $f^{\widetilde{\rm{NS}}}_1(\lambda)$. Explicitly, 
\begin{align}
\begin{split}
Z^{\widetilde{\rm{NS}}}(\tau, \bar{\tau}) &= |f^{\widetilde{\rm{NS}}}_0(\lambda)|^2 + \mathcal{M} |f^{\widetilde{\rm{NS}}}_1(\lambda)|^2, \\
Z^{\rm{R}}(\tau, \bar{\tau}) &= |f^{\widetilde{\rm{NS}}}_0(1-\lambda)|^2 + \mathcal{M} |f^{\widetilde{\rm{NS}}}_1(1-\lambda)|^2 = \mathcal{M}_1 |f^{{\rm{R}}}_0(\lambda)|^2 + \mathcal{M}_2 |f^{{\rm{R}}}_1(\lambda)|^2 ,
\end{split}
\end{align}
where $\mathcal{M}_1$ and $\mathcal{M}_2$ are related to the degeneracy of two characters in the R sector. The $q$-expansion of characters are presented in the Appendix \ref{App:B}.

Let us now describe how we find the initial values $(c,h^{\rm{NS}},a_1)$ which generate the solutions $f^{\rm NS}_0(\tau)$ and $f^{\rm NS}_1(\tau)$ as the $q$-series with the non-negative integer coefficients. Our approach is first to focus on the vacuum character $f^{\rm NS}_0(\tau)$ by taking the series expansion of \eqref{NS characters 2nd}.  One can survey the solutions of MLDE which have rational $0<c = \frac{p'}{p}\le24$ and non-negative integers for the coefficients $a_n$ in \eqref{eq:twochar}. For instance, when we impose the BPS condition in the R sector, coefficients $a_2$ of the vacuum character is given as follow : 
\begin{align}
    a_2 &=  \frac{4 c^2-68 c h^{\rm{NS}}-4 c a_1+4 c+2 h^{\rm{NS}} a_1^2+32 h^{\rm{NS}} a_1-a_1^2}{4 (h^{\rm{NS}}-1)} ,
\end{align}
and it should be non-negative integer for a lot of positive rational numbers $c=\frac{p'}{p}$ and $h^{\text{NS}}=\frac{r}{s}<1$. Assuming the upper bound on the denominators $p \le p^*$, $s \le s^*$ and the spin half current $a_1 \le a_1^*$, we search for values $(c,h^{\rm{NS}},a_1)$ that make $a_2$ a non-negative integer.

Alternatively, one can choose the initial data to be $(c,a_2,a_1)$ and check $a_3$ being integral and non-negative for rational $0<c=\frac{p'}{p}<24$ with upper bounds on the denominator i.e., $p \le 12$. Then we find the initial data that furnishes non-negative integer $a_3$ with $a_2$ in the range $0\le a_2 \le 10^6$. As an illustrative example, let us consider the case of $a_1=0$. Then the vacuum solution takes a form of
\begin{align}
f^{\rm{NS}}_{0}(\tau) = q^{-\frac{c}{24}} \left( 1 +  a_2 q + \left( \frac{8 (a_2+c) (a_2 (c-10)+c (5 c+22))}{3 (a_2+c (49-2 c))} \right) q^\frac{3}{2} + \cdots \right),
\end{align}
and the coefficients $a_n$ for $n \ge 4$ are in general the rational function of two parameters $a_2$ and $c$. Here we demand at least one $a_{2\ell+1}\neq 0$ for $\ell\in \mathbbm{Z}_{>0}$ to obtain the fermionic characters. Then we search the values of $c$ and $a_2$ which give us the series with non-negative integer coefficients.

Following the process described above, our next goal is to find the solutions of the second order MLDE and classify them. We first divide the possible solutions of MLDE into two types: the BPS solution and the non-BPS solution. For the type of BPS solution, we explore both the cases of $a_1 = 0$ and $a_1 \neq 0$. For the non-BPS type, we work out only the case with $a_1 = 0$. In total, there are six types of solutions and their characteristic profiles are presented in Table \ref{tab:2mldesol}.  The detailed descriptions for each type will be given below, while the list of solutions can be found in the Appendix \ref{App:B}. \footnote{So far we expressed MLDEs in terms of the covariant Serre derivative $\mathcal{D}_k=\frac{1}{2i\pi}\frac{\text{d}}{\text{d}\tau}-\frac{k}{12}\,E_2$. However, for $\Gamma_\theta$ the most generic modular covariant derivative is given by: $\tilde{\mathcal{D}}_k=\frac{1}{2i\pi}\frac{\text{d}}{\text{d}\tau}-\frac{k}{12}\left(E_2+\alpha \left(\vartheta_2^4(\tau)-\vartheta_4^4(\tau)\right)\right)$ with $\alpha\in\mathbb C$. With the choice of connection $\alpha=\frac{c}{2}-6h^{\rm{NS}}+1$, one can show that \eqref{eq:2MLDEThe} is equivalent to:
\begin{equation}
	\left\{\tilde{\mathcal{D}}_0^2-\frac{c(1+6h^{\rm{NS}})}{288}\,E_4+\frac{c(\frac{c}{2}-6h^{\rm{NS}}+1)}{288}\,\left(\vartheta_2^4(\tau)-\vartheta_4^4(\tau)\right)^{\ 2}\right\}\chi=0
\end{equation}
The deformation $\alpha$ in the covariant derivative also precisely appears in front of the last term. Therefore this second order MLDE can be viewed in some sense as a direct generalization of the Kaneko-Zagier equation. Indeed, the standard Deligne-Cvitanovic exceptional series is reproduced as a particular case of our equation, for which $\alpha$ is simply set to zero, namely when $c=12h^{\rm{NS}}-2$.}

\begin{table}[t!]
    \centering
    {\setlength{\extrarowheight}{5pt}
    \begin{tabular}{c|c|c}
        type & property & central charge $c$ \\ \hline
      BPS, {\bf I}   &      $h^{\text{R}}_-=\frac{c}{24},a_1=0$ & $1,\frac94,6,\frac{39}{4},11,12  $ \\ [2mm]
      BPS, {\bf II} & $h^{\rm R}_{-} =\frac{c}{24}, a_1\neq 0 $ & $\frac34, \frac32, 3, 6, 9, \frac{21}{2}, \frac{45}{4}  $ ,12\\ [2mm] \hline 
      non-BPS, {\bf I} & $h^{\rm R}_{-} >\frac{c}{24}, h^{\rm NS}\neq \frac12 $ & $ \frac{7}{10}, \frac{133}{10}, \frac{91}{5} , \frac{102}{5} , 21, \frac{85}{4}, 22 ,  \frac{114}{5} $   \\ [2mm]
      non-BPS, {\bf II} & $h^{\rm R}_{-} >\frac{c}{24}, h^{\rm NS}=\frac12$ & $\frac92,5,\frac{11}{2},6, \frac{13}{2}, 7, \frac{15}{2}$ \\ [2mm]
      non-BPS, {\bf III} & one-parameter family & 16 \\ [2mm]
      non-BPS, {\bf IV} & single-character & $\frac{17}{2}, 9, \frac{19}{2}, 10,  \cdots, \frac{47}{2}  $ 
    \end{tabular}}
    \caption{Classification for the solutions of the second order MLDE. We impose the BPS condition $h^{\text{R}}_-=\frac{c}{24}$ for the first two classes and search the solutions up to $a_1 \le 15$. The solutions in BPS type {\bf{I}} are obtained by imposing both BPS condition and $a_1 = 0$. On the other hand, we impose the BPS condition and $a_1 \neq 0$ for the BPS type {\bf{II}}. We further classify four classes by relaxing the BPS condition thus we refer to them as non-BPS solutions. We search the non-BPS solutions under the constraint $a_1 = 0$. The non-BPS type {\bf{I}} involves the non-BPS solutions with $h^{\rm{NS}} \neq \frac{1}{2}$ and some of them are identified to the WZW models. The class with constraint $h^{\rm{NS}} = \frac{1}{2}$ is labeled as non-BPS type {\bf{II}}. The $c=6$ solution in BPS type {\bf{I}} and non-BPS {\bf{II}} are same, as they satisfy both BPS condition and $h^{\rm{NS}} = \frac{1}{2}$. The one-parameter family appear for $c=16$ and we label them as non-BPS type {\bf{III}}. We also find that only the vacuum solutions of the non-BPS type {\bf{IV}} can have the positive integer coefficients and describe the partition function of a single-character RCFT.}
    \label{tab:2mldesol}
\end{table}

\subsection{BPS solutions}
\subsubsection*{BPS Type \bf{I} : Supersymmetric BPS pairs}
Let us discuss the class of solutions whose R sector ground state saturates the unitarity bound $h^{\rm R}_{-}=c/24$, i.e., the ground state of the supersymmetric theory preserves the supersymmetry. In this subsection, we first focus on the case of $a_1 = 0$. Sensible solutions appear for the five central charge values of $c=1,\frac{9}{4},6,\frac{39}{4},11$ and they are listed in Table~\ref{type I}. For the solutions in this type, the NS sector weight is given by $h^{\rm{NS}} =(2c+3)/30$. On the other hand, the R sector weights read $h^{\rm R}_{-} = c/24$ and $h^{\rm R}_{+} = 3(c+4)/40$. The explicit $q$-series of the solutions can be found in Table \ref{NS type I} and Table \ref{R type I} in Appendix \ref{App:B}.
\begin{table}[t!]
\begin{center}
\begin{tabular}{c|cccccc}
$c$ &  1 & $\frac{9}{4}$ & 6 & $\frac{39}{4}$ & 11  & 12\\ [1mm]
\hline
$h^{\textsc{ns}}  $& $\frac16$ & $\frac14$  & $\frac12$ & $\frac34$ & $\frac56$ & $[1]$\\
$h^{\textsc{r}} $ & $\frac{1}{24},\frac38 $ & $\frac{3}{32}, \frac{15}{32}$  & $\frac{1}{4},\frac34$ & $\frac{13}{32},\frac{33}{32} $ & $\frac{11}{24},\frac98 $& $\frac{1}{2},[1]$\\
$\mathcal{M}$ & 2 & 1 & 15 & 1 &  2 & $[0]$
\end{tabular}
\caption{\label{type I} The five solutions of BPS type {\bf{I}}. We obtain these solutions by imposing BPS condition $h^{\rm{R}}=\frac{c}{24}$ and no free fermion currents in the NS sector vacuum solution. Here $\mathcal{M}$ denotes the degeneracy of the non-vacuum primary. When $\mathcal{M}=1$, the $S$-matrix is symmetric otherwise one should find the extended $S$-matrix for the consistent fusion rule algebra. The bracket means that the corresponding theory is effectively a single character theory, appearing as a solution to the second order MLDE, with artificial second solution.}
\end{center}
\end{table}%

\paragraph{Fermionic RCFT with $c=1$}
The solution with $c=1$ can be expressed in terms of the characters of the $\mathcal{N}=1$ supersymmetric minimal models. We denote the unitary supersymmetric minimal model by $\mathcal{SM}(\ell+2,\ell)$ where the central charge is given by
\begin{align}
    c = \frac{3}{2} \left(1- \frac{8}{\ell(\ell+2)}\right), \qquad \ell=2,3,\cdots.
\end{align}
The partition function of a unitary supersymmetric minimal model can be expressed by the characters $\chi_{m,n}^{\text{NS}}$ and $\chi_{m,n}^{\text{R}}$ \cite{Goddard:1986ee},
\begin{align}
\label{character}
\begin{split}
 \chi^{\text{NS}}_{m,n}(\tau;\ell) &= \zeta^{k}_{m,n}(q) \prod_{\ell=1}^{\infty}\left( \frac{1+q^{\ell-\frac{1}{2}}}{1-q^\ell} \right), \quad  \chi^{\text{R}}_{m,n}(\tau;\ell) = \zeta^{k}_{m,n}(q) q^{\frac{1}{16}} \prod_{\ell=1}^{\infty}\left( \frac{1+q^{\ell-\frac{1}{2}}}{1-q^\ell} \right),
\end{split}
\end{align}
where
\begin{align}
\begin{split}
\zeta^k_{m,n}(q) &= \sum_{\alpha \in \mathbb{Z}} \left( q^{\gamma^k_{m,n}(\alpha)} - q^{\delta^k_{m,n}(\alpha)}\right), \\
\gamma^k_{m,n}(\alpha)&= \frac{\left(2k(k+1)\alpha - m(k+2) + pk\right)^2-4}{8k(k+2)}, \\
\delta^k_{m,n}(\alpha) &=  \frac{\left(2k(k+1)\alpha + m(k+2) + pk\right)^2-4}{8k(k+2)} .
\end{split}
\end{align}
For the $\mathcal{N}=1$ supersymmetric minimal model, the conformal weights are allowed to have the values 
\begin{equation}
\label{superminimal h}
    h_{m,n} = \frac{\big[(\ell+2)m-\ell n\big]^2-4}{8 \ell(\ell+2)}
  +\frac{1}{32}\big[1-(-1)^{m-n}\big],  
\end{equation}
where $m,n$ are integers subject to $1\le m <\ell, 1\le n<\ell+2 $. Here $(m-n)$ is even for the NS sector while $(m-n)$ is odd for the R sector.

The ${\cal N}=1$ supersymmetric minimal model $\mathcal{SM}(6,4)$ has central charge $c=1$ and the NS sector involves four primaries of weight $h^{\rm{NS}} = 0, 1/16, 1/6, 1$. One can show that the $q$-series solutions to the MLDE with $c=1$ in Table \ref{NS type I} can be expressed in terms of $\chi^{\text{NS}}_{m,n}$ as follows. 
\begin{align}
    \label{characters of SM(6,4)}
    \begin{split}
    f_0(\tau) &= \chi^{\text{NS}}_{1,1}(\tau;\ell=4) + \chi^{\text{NS}}_{1,5}(\tau;\ell=4) = q^{-\frac{1}{24}} \Big( 1 + q + 2 q^\frac{3}{2} + 2 q^2 + 2 q^\frac{5}{2} + \cdots \Big) , \\
    f_1(\tau) &= \chi^{\text{NS}}_{1,3}(\tau;\ell=4) = q^{\frac{1}{6} - \frac{1}{24}} \Big( 1 + q^{\frac{1}{2}} + q + q^{\frac{3}{2}} + 2 q^2 + 3 q^{\frac{5}{2}} + \cdots \Big)
    \end{split}
\end{align}
Moreover, using the known $S$-matrix of  $\mathcal{SM}(6,4)$, one can verify that these two solutions transform into themselves under $S$ transformation, 
\begin{align}
    \left(
    \begin{matrix}
        f_0(-\frac{1}{\tau}) \\ f_1(-\frac{1}{\tau})
    \end{matrix}
    \right)
    = 
    \frac{1}{\sqrt3} \left( 
    \begin{matrix}
        1 & 2 \\ 1 & -1 
    \end{matrix}
    \right)
    \left(
    \begin{matrix}
        f_0(\tau) \\ f_1(\tau)
    \end{matrix}
    \right). 
\end{align}  
The above modular S-transformation matrix is neither unitary nor symmetric, but leaves the matrix $\text{diag}(1,2)$ invariant. Thus $\Gamma_\theta$ invariant NS partition function is given by
\begin{align}\label{zz}
    Z^\text{NS} = \Big| f_0(\tau) \Big|^2 + 2  \Big| f_1(\tau) \Big|^2. 
\end{align}
The multiplicity ${\cal M}= 2$ in \eqref{zz} suggests that there are two different primaries associated with the same character $f_1(\tau)$. In other words, the theory of our interest has three NS characters, denoted by $\hat{f}_0, \hat{f}_1$ and $\hat{f}_2$ with $\hat{f}_0 = f_0^{(1)}$ and $\hat{f}_{1,2} = f_1^{(1)}$, that transform under the $S$ transformation as follows 

\begin{equation}
\left(
\begin{matrix}
\hat{f}_0(-\frac{1}{\tau})\\
\hat{f}_1(-\frac{1}{\tau})\\
\hat{f}_2(-\frac{1}{\tau})
\end{matrix}\right) = S \left(\begin{matrix}
\hat{f}_0(\tau)\\
\hat{f}_1(\tau)\\
\hat{f}_2(\tau)
\end{matrix}\right),
\end{equation}
with
\begin{align}\label{zxzx}
    S = \renewcommand{\arraystretch}{2}
\left(\begin{matrix}
\frac{1}{\sqrt{3}} & \frac{1}{\sqrt{3}} & \frac{1}{\sqrt{3}}\\
\frac{1}{\sqrt{3}} & \frac{1}{\sqrt{3}}\,e^{\pm 2 \pi i/3 } & \frac{1}{\sqrt{3}}\,e^{\mp 2 \pi i/3 } \\
\frac{1}{\sqrt{3}} & \frac{1}{\sqrt{3}}\,e^{\mp 2 \pi i/3 } & \frac{1}{\sqrt{3}}\,e^{\pm 2 \pi i/3 }
\end{matrix}\right).   
\end{align}
Here the sign of exponents can be flipped as the role of $\hat f_1$ and $\hat f_2$ exchanges. Note that the modular $S$-matrix \eqref{zxzx} is symmetric and  satisfies $S^2=1$. Once we have the symmetric $S$-matrix, the fusion rule algebra coefficients are followed by the Verlinde formula
\begin{align}
    N_{ij}^{k} = \sum_{\ell} \frac{S_{i\ell}S_{j\ell}(S^{-1})_{\ell k}}{S_{0 \ell}},
\end{align}
where the index 0 denotes the vacuum. Using the Verlinde formula, we can obtain the well-defined  fusion algebra given below.
\begin{gather}\label{fusion rule 1}
    N_{00}^{0} = N_{01}^{1}  = N_{10}^{1} = N_{02}^{2} = N_{20}^{2} = 1, \nonumber \\
    N_{11}^{2} = N_{22}^{1} = N_{12}^{0} = N_{21}^{0} = 1,\quad N_{ij}^{k} = 0 \quad \text{otherwise}\,.
\end{gather}
Let us comment on the $\widetilde{\text{NS}}$ sector and R sector partition function. The transformation rule of the $\widetilde{\text{NS}}$ sector and R sector characters are given by
\begin{align}
\label{S transform of c=1}
    \left(
    \begin{matrix}
        f^{\widetilde{\rm{NS}}}_0(-\frac{1}{\tau}) \\  f^{\widetilde{\rm{NS}}}_1(-\frac{1}{\tau})
    \end{matrix}
    \right)
    = 
    \frac{1}{\sqrt3} \left( 
    \begin{matrix}
        \sqrt{2} & 1 \\ \frac{1}{\sqrt{2}} & -1 
    \end{matrix}
    \right)
    \left(
    \begin{matrix}
        f^R_0(\tau) \\ f^R_1(\tau)
    \end{matrix}
    \right). 
\end{align}  
One can read the $\widetilde{\rm{NS}}$ sector partition function by taking $T$-transformation to \eqref{zz}.
\begin{align}
    Z^{\widetilde{\text{NS}}} = \Big| f^{\widetilde{\rm{NS}}}_0(\tau) \Big|^2 + 2  \Big| f^{\widetilde{\rm{NS}}}_1(\tau) \Big|^2. 
\end{align}
The Ramond sector partition function can be obtained from the $S$-matrix \eqref{S transform of c=1}. Because the $S$-matrix leaves diag$(1,1)$ invariant, the R sector partition function takes a form of
\begin{align}
    Z^{\text{R}} =  \Big|  f^R_0(\tau) \Big|^2 +   \Big|  f^{R}_1(\tau) \Big|^2\,.
\end{align}

In fact, the above fermionic model with $c=1$ can be understood with the $\mathcal N=2$ supersymmetric $A_1$ minimal model, {\it a.k.a.} Kazama-Suzuki model with level one. The $\mathcal N=2$ supersymmetric $A_k$ minimal model has the central charge
\begin{equation}
\label{N2 minimal model cc}
    c(k)= \frac{3k}{k+2} \qquad (k=1,2,3,...), 
\end{equation}
and super-Virasoro characters of weight $h^{{\rm NS}}(a,b)$ and $U(1)$ R-charge  $Q^{{\rm NS}}(a,b)$ in the NS sector read  
\begin{align}
    h^{{\rm NS}}(a,b)  =  \frac{1}{k+2} \Big(ab-\frac14\Big), \qquad
    Q^{{\rm NS}}(a,b)  = \frac{a-b}{k+2} ,
\end{align}
with $a,b\in {\mathbb Z}+\frac12$ and $0<a,b,(a+b)\le k+1$. The explicit form of the ${\cal N}=2$ super-Virasoro character in the NS sector is given by 
\begin{align}
    {\rm ch}^{{\rm NS}}_{h ,Q }(\tau,z) & 
    =  q^{-\frac{c(k)}{24}+h(a,b)} y^{Q(a,b) } \cdot \varphi^{\rm NS}(\tau,z)\cdot\Gamma^{(k)}_{a,b}(\tau,z),
\end{align}
with
\begin{align}
    \varphi^{\rm NS}  &  = \prod_{n=1}^\infty \frac{(1+yq^{n-\frac12})(1+y^{-1} q^{n-\frac12})}{(1-q^n)^2},   \nonumber \\
    \Gamma^{(k=\ell-2)}_{a,b} & = \prod_{n=1}^\infty \frac{ (1-q^{ \ell (n-1)+a+b})(1-q^{\ell n-a-b})(1-q^{\ell n})^2}{(1+yq^{ \ell n-a})(1+y^{-1}q^{\ell(n-1)+a})(1+y^{-1}q^{\ell n -b})(1+yq^{\ell(n-1)+b})} , 
\end{align}
Here $y=e^{2\pi i z}$ with  the chemical potential $z$ for the $U(1)_R$ charge. The Kazama-Suzuki model with $k=1$ has $c=1$, and three NS characters of $(h^{\rm NS},Q^{\rm NS})=(0,0),(\frac16,\pm\frac13)$. One can see that, after turning off the chemical potential, $f^{\rm NS}_0(\tau)$ and $f^{\rm NS}_1(\tau)$ become the ${\mathcal N}=2$ NS characters  
\begin{align}
    \chi^{\rm NS}_{0,0}(\tau,z=0) = f^{\rm NS}_0(\tau), \ \  
    \chi^{\rm NS}_{\frac16,\frac13}(\tau,0)=  
    \chi^{\rm NS}_{\frac16,-\frac13}(\tau,0)= f^{\rm NS}_1(\tau) .
\end{align}

\paragraph{Fermionic RCFT with $c=9/4$}
The $c=9/4$  fermionic RCFT with two NS characters can be understood as  the fermionization of the $\mathfrak{a}_1$ WZW model with level six. 

The $\mathfrak{a}_1$ WZW model with level six has central charge $c=9/4$ and its torus partition function can be expressed in terms of seven characters $\chi^{\widehat{su}(2)_6}_h(\tau)$ of  conformal weights $h=0$, $3/32$, $1/4$, $15/32$, $3/4$, $35/32$, $3/2$. In terms of characters of the WZW model, $f_0^\text{NS}(\tau)$ and $f_1^\text{NS}(\tau)$ with $c=9/4$ can be written as 
\begin{align}\label{adad}                           
  f^\text{NS}_0 & = \chi^{\widehat{su}(2)_6}_0 + \chi^{\widehat{su}(2)_6}_{\frac32} =   
  q^{-\frac{3}{32}} \Big( 1 + 3 q + 7 q^{\frac{3}{2}} + 9 q^2 + 12 q^{\frac{5}{2}} + \cdots \Big), 
  \nonumber \\ 
  f^\text{NS}_1 & = \chi^{\widehat{su}(2)_6}_{\frac14} + \chi^{\widehat{su}(2)_6}_{\frac34} 
  =  q^{\frac{1}{4}-\frac{3}{32}} \Big( 3 + 5 q^{\frac{1}{2}} + 9 q + 15 q^{\frac{3}{2}} + 27 q^2 + 45 q^{\frac{5}{2}} + \cdots \Big). 
\end{align}  
The modular $S$-matrix now becomes 
\begin{equation}\label{smatrix9/4}
    S= \left(
    \begin{matrix}
    \sin\frac{\pi}{8} & \cos\frac{\pi}{8}\\
    \cos\frac{\pi}{8} & -\sin\frac{\pi}{8}\\
    \end{matrix}\right)\,,
\end{equation}
which is already symmetric, meaning that there are only two primaries in the NS sector and no need to find an extended $S$-matrix. Then the fusion coefficients can be read off from the Verlinde formula,
\begin{equation}
\label{fusion rule 2}
    N_{00}^{0} = N_{10}^{1} = N_{01}^{1} = N_{11}^{0} = 1,\quad N_{11}^{1} = 2\,,\quad N_{ij}^{k} = 0 \ \text{otherwise}\,.
\end{equation}
From \eqref{smatrix9/4}, the $\Gamma_\theta$ invariant NS partition function is given by  
\begin{align}
    Z_\text{NS}  =  \Big| f_0^\text{NS} (\tau)\Big|^2 +  \Big| f_1^\text{NS} (\tau)\Big|^2 =
    \Big|\chi^{\widehat{su}(2)_6}_0 + \chi^{\widehat{su}(2)_6}_{\frac32} \Big|^2 + \Big| \chi^{\widehat{su}(2)_6}_{\frac14} + \chi^{\widehat{su}(2)_6}_{\frac34}  \Big|^2. 
\end{align}
Applying suitable modular transformations on $f^\text{NS}_{0,1}(\tau)$ gives other partition functions for different spin structures, 
\begin{align}
    Z_{\widetilde{\text{NS}}} & = 
    \Big|\chi^{\widehat{su}(2)_6}_0 - \chi^{\widehat{su}(2)_6}_{\frac32} \Big|^2 + \Big| \chi^{\widehat{su}(2)_6}_{\frac14} - \chi^{\widehat{su}(2)_6}_{\frac34}  \Big|^2 ,
    \nonumber \\ 
    Z_\text{R} & = \Big| \chi^{\widehat{su}(2)_6}_{\frac{3}{32}} + \chi^{\widehat{su}(2)_6}_{\frac{35}{32}}\Big|^2 + \Big|  \sqrt2 \chi^{\widehat{su}(2)_6}_{\frac{15}{32}} \Big|^2 . 
\end{align}
Finally, the SL$_2(\mathbb{Z})$ invariant $\tilde{\text{R}}$ partition function is
\begin{align}
    Z_{\tilde{\text{R}}} = \Big| \chi^{\widehat{su}(2)_6}_{\frac{3}{32}} - \chi^{\widehat{su}(2)_6}_{\frac{35}{32}}\Big|^2 = 4, 
\end{align}
which suggests that the theory of our interest could be supersymmetric. The GSO projection (or equivalently the bosonization) leads to the torus partition function of the $su(2)$ WZW model with level six, 
\begin{align}
    Z(\tau,\bar \tau) &= \frac12 \Big[ Z_\text{NS}(\tau,\bar \tau) +  Z_{\widetilde{\text{NS}}}(\tau,\bar \tau) + Z_\text{R}(\tau,\bar \tau) + Z_{\tilde{\text{R}}}(\tau,\bar \tau) \Big], 
    \\ & = 
    \Big| \chi^{\widehat{su}(2)_6}_{0} \Big|^2+ \Big| \chi^{\widehat{su}(2)_6}_{\frac{3}{32}} \Big|^2 +
    \Big| \chi^{\widehat{su}(2)_6}_{\frac{1}{4}} \Big|^2+ \Big| \chi^{\widehat{su}(2)_6}_{\frac{15}{32}} \Big|^2 + 
    \Big| \chi^{\widehat{su}(2)_6}_{\frac{3}{4}} \Big|^2+ \Big| \chi^{\widehat{su}(2)_6}_{\frac{35}{32}} \Big|^2 +
    \Big| \chi^{\widehat{su}(2)_6}_{\frac{3}{2}} \Big|^2, \nonumber 
\end{align}
as expected. It was shown recently in \cite{Johnson-Freyd:2019wgb} that this model indeed has 
${\cal N}=1$ supersymmeteric vertex operator algebra. 

We find that two solutions \eqref{adad} can be expressed in terms of the characters of the $\mathcal{N}=1$ supersymmetric minimal models $\mathcal{SM}(6,4)$ and $\mathcal{SM}(8,6)$ as follows.
\begin{align}
\begin{split}
f_0(\tau) &= (\chi^{\ell = 4}_{1,1} + \chi^{\ell = 4}_{1,5}) (\chi^{\ell = 6}_{1,1} + \chi^{\ell = 6}_{1,7}) + 2 \chi^{\ell = 4}_{1,3} \chi^{\ell = 6}_{3,1}, \\
f_1(\tau) &= (\chi^{\ell = 4}_{1,1} + \chi^{\ell = 4}_{1,5}) (\chi^{\ell = 6}_{1,3} + \chi^{\ell = 6}_{1,5}) + 2 \chi^{\ell = 4}_{1,3} \chi^{\ell = 6}_{3,3}
\end{split}
\end{align}
We also note that $c=\frac94$ is one of the central charge for the minimal ${\mathcal N}=2$ model \eqref{N2 minimal model cc}. Indeed we can show that  two NS characters with the $\text{SU}(2)$ chemical potential turned on can be expressed in terms of the ${\mathcal N}=2$ NS characters with $U(1)_R$ chemical potential tuned on:
\begin{align}
    \Big[\chi^{\widehat{su}(2)_6}_0  + \chi^{\widehat{su}(2)_6}_\frac32\Big](\tau,\frac{z}{2} ) &= \Big[ {\rm ch}^{(\frac49){\rm NS}}_{0,0}  + {\rm ch}^{(\frac49){\rm NS}}_{1,\frac12} + {\rm ch}^{(\frac49){\rm NS}}_{1,-\frac12}  + {\rm ch}^{(\frac49){\rm NS}}_{\frac32,0}\Big](\tau,z)\,, \nonumber \\
    \Big[\chi^{\widehat{su}(2)_6}_\frac14  + \chi^{\widehat{su}(2)_6}_\frac34\Big] (\tau,\frac{z}{2} ) &= \Big[{\rm ch}^{(\frac49){\rm NS}}_{\frac14,0}  + {\rm ch}^{(\frac49){\rm NS}}_{\frac14,\frac12} + {\rm ch}^{(\frac49){\rm NS}}_{\frac14,-\frac12}  + {\rm ch}^{(\frac49){\rm NS}}_{\frac34,0}\Big](\tau,z)\,.
\end{align}

\paragraph{Fermionic RCFT with $c=6$}

One can identify the solutions to the second order MLDE with $c=6$ as the fermionization of the six-fold product of the $\mathfrak{a}_1$ WZW model with level-one. This model has been recently studied in \cite{Harvey:2020jvu} toward understanding the origin of the Mathieu moonshine of the K3 CFT. In particular, the quantum hexacode was utilized to construct the $N=1$ supersymmetry current out of $2^6$ primaries of weight $3/2$. The level-one $\mathfrak{a}_1$ WZW model, denoted by $\widehat{su}(2)_1$ for later convenience, has central charge $c=1$ and contains two characters, one of which corresponds to the vacuum and the other to the primary of conformal weight $h=1/4$. The explicit form of two characters is  
\begin{align}
    \psi_0(\tau) &= \sqrt{\frac{\vartheta_3^2(\tau)+\vartheta_4^2(\tau)}{2\eta^2(\tau) }}= q^{-\frac{1}{24}} \left( 1 + 3 q + 4 q^2 + 7 q^3 
    + \cdots \right),\nonumber\\
    \psi_{\frac14}(\tau) &=  \sqrt{\frac{\vartheta_3^2(\tau)-\vartheta_4^2(\tau)}{2\eta^2(\tau) }}=  2 q^{\frac{5}{24}} \left( 1 + q + 3 q^2 + 4 q^3 + 
    \cdots \right). 
\end{align}
Under $S: \tau \to - 1/\tau$, $\psi_{0}$ and $\psi_1$ transform as 
\begin{align}\label{S transform of SU(2)}
    \begin{split} \left(
    \begin{array}{c}
    \psi_0(-\frac{1}{\tau})\\
    \psi_1(-\frac{1}{\tau}) 
    \end{array} \right)
    =  \left(
    \begin{array}{cc}
    \frac{1}{\sqrt{2}} & \frac{1}{\sqrt{2}}\\
    \frac{1}{\sqrt{2}} & -\frac{1}{\sqrt{2}}
    \end{array} \right)
    \left(
    \begin{array}{c}
    \psi_0(\tau)\\
    \psi_1(\tau) 
    \end{array} \right).
\end{split}
\end{align}

The six-fold product of $\widehat{su}(2)_1$ has $2^6$ primaries, many of which become degenerate. There are eventually seven different characters of conformal weights $h=0$, $1/4$, $1/2$, $3/4$, $1$, $5/4$, $3/2$, given by 
\begin{equation}
    \chi_{\frac{n}{4}}(\tau) = \psi_0^{6-n} (\tau) \cdot  \psi_{\frac14}^n(\tau) , \quad (n=0,1,...,6). 
\end{equation} 
We can express two solutions with $c=6$ in Table \ref{type I} in terms of $\chi_{\frac n4}(\tau)$,
\begin{align}
    f_0^\text{NS}(\tau) & = \chi_0(\tau) + \chi_{\frac32}(\tau) = 
    q^{-\frac{1}{4}} \left( 1 + 18 q + 64 q^{\frac{3}{2}} + 159 q^2 + 384 q^{\frac{5}{2}}  + \cdots \right),
    \nonumber \\
    f_1^\text{NS}(\tau) & = \chi_{\frac12}(\tau) + \chi_{1}(\tau) = 
    4q^{\frac{1}{4}} \left( 1 + 4 q^{\frac{1}{2}} + 14 q + 40 q^{\frac{3}{2}} + 101 q^2 + \cdots \right). 
\end{align}
The modular $S$-matrix can be read off from \eqref{S transform of SU(2)} as 
\begin{align}\label{fgh}
    \left(
    \begin{matrix}
        f_0^\text{NS}(-\frac{1}{\tau}) \\ f_1^\text{NS}(-\frac{1}{\tau})
    \end{matrix}
    \right)
    = \frac14 \left( 
    \begin{matrix}
        1 & 15 \\ 1 & -1 
    \end{matrix}
    \right)
    \left(
    \begin{matrix}
        f_0^\text{NS}(\tau) \\ f_1^\text{NS}(\tau)
    \end{matrix}
    \right), 
\end{align}  
which implies that the NS partition function has to be
\begin{align}\label{qsqs}
    Z_\text{NS} =  \Big|f_0^\text{NS}(\tau)\Big|^2 +15 \Big| f_1^\text{NS}(\tau) \Big|^2. 
\end{align}
The multiplicity $15$ in \eqref{qsqs} which says there are $15$  primaries associated with $f_1^\text{NS}(\tau)$ is reflected in the non-symmetric and non-unitary modular matrix in \eqref{fgh}. It implies that we have to find a $16 \times 16$ extended $S$-matrix $\hat S$. To this end, let us first denote the extended characters as
\begin{align}
\begin{split}
\widehat{f}_{0}^\text{NS}(\tau) &= {f}_{0}^\text{NS}(\tau), \\
\widehat{f}_{1,\alpha}^{(3)}(\tau) &= \psi_0 \otimes \cdots \overset{i}{\psi_{1}} \cdots \overset{j}{\psi_{1}}\cdots \otimes \psi_0 + (0 \leftrightarrow 1),
\end{split}
\end{align}
where the index $\alpha$ means $\alpha = \{i,j\}$ for $1 \leq i < j \leq 6$. For the 16 component vector-valued modular form $\left(\widehat{f}_0^{(3)}, \widehat{f}_{1,\alpha}^{(3)}(\tau)\right)$, the below $16 \times 16$ matrix whose components are
\begin{equation}
\begin{aligned}
&S_{00} = S_{\alpha\alpha} = +\tfrac{1}{4}\,, \quad S_{0\alpha} = S_{\alpha0} = +\tfrac{1}{4}\,,\\
&S_{\alpha\beta} = \begin{cases}+\frac{1}{4} &\text{if}\ \alpha \cap \beta = \varnothing,\\
-\frac{1}{4}\ &\text{otherwise}\,, \end{cases}
\end{aligned}
\end{equation}
acts as the extended $\hat{S}$-matrix. Now the fusion rules follow from the Verlinde formula,
\begin{equation}
\begin{aligned}
&N_{00}^{0} = N_{0\alpha}^{\alpha} = N_{\alpha0}^{\alpha} = N_{\alpha\alpha}^{0} = 1,\\
&N_{\alpha\beta}^{\gamma} = \begin{cases}1 \ \text{if}\ \alpha \cup \beta \cup \gamma = \{1,\cdots,6\}\ \text{or} \ \alpha \cup \beta = \gamma \cup (\alpha \cap \beta)\,,\\ 0\  \text{otherwise}.\, \end{cases}
\end{aligned}
\end{equation}

Performing $T$ and $S$ transformation on the NS characters, one can obtain the $\widetilde{\text{NS}}$ and R partition functions
\begin{align}  
    Z_{\widetilde{\text{NS}}} & = \big|f^{\widetilde{\text{NS}}}_0(\tau)\big|^2 + 15 \big|f^{\widetilde{\text{NS}}}_1(\tau)\big|^2 = \Big| \chi_0 - \chi_{\frac32} \Big|^2 + 15\Big| \chi_{\frac12} - \chi_{1} \Big|^2,
    \nonumber \\
    Z_\text{R} & = 6 \big|f^{{\text{R}}}_0(\tau)\big|^2 + 40 \big|f^{{\text{R}}}_1(\tau)\big|^2 = 6 \Big| \chi_{\frac14} + \chi_{\frac54} \Big|^2 + 20 \Big| \sqrt 2 \chi_{\frac34} \Big|^2. 
\end{align}  
Analogous to the previous example of $c=9/4$, the $\tilde{\text{R}}$ partition function becomes constant,
\begin{align}
    Z_{\tilde{\text{R}}} = 6 \Big|\chi_{\frac14} - \chi_{\frac54}  \Big|^2 = 24, 
\end{align}
which is consistent to the fact that there is a unique  $N=1$ supersymmetry current in the $\big(\widehat{su}(2)_1\big)^6$ \cite{Harvey:2020jvu}. 
One can also see that the GSO projection of the theory with $c=6$ leads to 
the $\big(\widehat{su}(2)_1\big)^6$.

\paragraph{Dual fermionic RCFTs with $c=11$ and $c=39/4$}

Let us make comments on the duality relation among the solutions. We take the three pairs for the BPS type $\bf{I}$ solutions
\begin{align}
    \left(c, \tilde{c} = 12 - c\right) = \left(1,11\right), \quad \left(\frac{9}{4}, \frac{39}{4}\right), \quad \left(6,6\right),
\end{align}
where the $c=6$ solution is self-dual. Note that the NS weights $h^{\rm NS}$ and $\tilde h^{\rm NS}$ of the dual pairs satisfy $h^{\rm NS}+\tilde h^{\rm NS} =1$. We find that the  NS sector solutions of BPS type $\bf{I}$ ($f_0^{(c)}(\tau),f_1^{(c)}(\tau)$) and their dual pairs ($\widetilde f_0^{(12-c)}(\tau),\widetilde f_{1}^{(12-c)}(\tau)$) are combined to produce bilinear relation:
\begin{align}
\label{dual relation}
    f_0^{(c)} \tilde f_0^{(12-c)}+ \mathcal{M}^{(c)} f_1^{(c)}\tilde f_{1}^{(12-c)}=K(\tau), 
\end{align}
where $\mathcal{M}^{(c=1)} =2$, $\mathcal{M}^{(c=9/4)} = 2$, $\mathcal{M}^{(c=6)} = 15$. The coefficient $\mathcal{M}^{(c)}$ agrees with the degeneracy of the primaries in each case, and it is thus guaranteed that the bilinear relation holds even after introducing the extended $S$-matrix. Since $K(\tau)$ is invariant under the $S$ transformation,  the dual pairs $\left(\widetilde f_0^{(12-c)}(\tau),\widetilde f_{1}^{(12-c)}(\tau)\right)$ share the same $S$-matrix of $\left(f_0^{(c)}(\tau),f_1^{(c)}(\tau) \right)$. Therefore, the bilinear relation suggests that the two theories in a given pair have identical fusion rule algebra.  Explicitly, the fusion rules of the $c=11$ putative theory are given by \eqref{fusion rule 1} while the hypothetical theory of $c=\frac{39}{4}$ has fusion rules \eqref{fusion rule 2}.

The $\mathcal{N}=1$ superconformal field theory with $c=12$  has been constructed by \cite{Frenkel:1988xz} and its partition function can be factorized into $K(\tau)K(\bar{\tau})$. Analogous to the deconstruction of the Monster CFT by two bosonic RCFTs discussed in \cite{Bae:2018qfh,Bae:2020pvv}, \eqref{dual relation} suggests that one can deconstruct the $c=12$ superconformal field theory by two fermionic RCFTs. On the one hand, the $c=12$ superconformal theory is known to has symmetry group $\rm Co_0 = 2. \rm Co_1$ \cite{2005math2267D}. Therefore, it is natural to expect that the dual pairs exhibit moonshine phenomena for some subgroup of $\rm Co_0$. For instance let us consider the putative $c=11$ theory. Our solution for $c=11$ can be identified to the characters of $\mathcal{N}=1$ SVOA with $c=11$ given in \cite{Johnson-Freyd:2019wgb}. The automorphism group of $\mathcal{N}=1$ SVOA with $c=11$ is known as the Suzuki group, which is one of the maximal subgroup of $\rm Co_0$. 
It would be interesting to investigate further deconstruction examples of $c=12$ superconformal theory, and the details will be discussed in an upcoming paper \cite{ONP20}.

\subsubsection*{BPS Type {\bf{II}} : BPS solutions with free fermion currents}

Let us then consider a mild generalization to explore the space of BPS solutions that contain free fermions. To do so, we consider the case that each NS sector vacuum character accommodates for the $q^{1/2-c/24}$ term, namely, 
\begin{equation}
	    f^{\rm{NS}}_0(q)=q^{-\frac{c}{24}}(1+a_1q^{\frac{1}{2}}+\mathcal O(q))\,.
\end{equation}
As shown in \eqref{BPS constraints}, the weights of the primaries in each sector are given by
\begin{equation}
h^\textsc{r}_- = \frac{c}{24}, \quad h_+^\textsc{r}=\frac{5 c}{24}-2 h^\textsc{ns}+\frac{1}{2}, \quad h^\textsc{ns} = \frac{4 c^2+6 c+9 a_1}{6 (10 c+3 a_1)},
\end{equation}
when the unitarity bound is saturated. As a consequence, the solutions of the second order MLDE are in general expressed as the rational function of $c$ and $a_1$. We search for the values of $c$ and $a_1$ leading to non-negative integer Fourier coefficients of each solution. In what follows, we discuss BPS solutions by imposing $c>0$ and $a_1 \neq 0$ which are summarized in Table \ref{type VI list}. The explicit solutions in $q$-series can be found in Table \ref{NS type VI} and \ref{R type VI}.
\begin{table}[t!]
\begin{center}
\begin{tabular}{c|cccccccc}
	$c$ & $\frac{3}{4}$ & $\frac{3}{2}$&$3$ & $6$ & $9$ &$\frac{21}{2}$ & $\frac{45}{4}$& 12\\[1mm] \hline
	$h^\textsc{ns}$ &$\frac{1}{4}$&$\frac{1}{4}$&$\frac{1}{3}$&$\frac{1}{2}$& $\frac{2}{3}$&$\frac{3}{4}$&$\frac{3}{4}$&$\left[\frac{5}{6}\right]$\\[1mm]
    $h^\textsc{r}_{\pm}$ &$\frac{1}{32},\frac{5}{32}$&$\frac{1}{16},\frac{5}{16}$&$\frac{1}{8},\frac{11}{24}$& $\frac{1}{4}, \frac{3}{4}$ & $\frac{3}{8},\frac{25}{24}$&$\frac{7}{16},\frac{19}{16}$&$\frac{15}{32},\frac{43}{32}$& $\frac{1}{2},\left[\frac{4}{3}\right]$\\[1mm]
	$a_1$ &1&1&2&4&6&7&15&8\\[1mm]
	$\mathcal{M}$ & 1 & 1 & 2 & 3 & 2& 1& 1& $[0]$
\end{tabular}
\caption{\label{type VI list} The list of eight BPS solutions with $a_1 \neq 0$.}
\end{center}
\end{table}

\paragraph{Supersymmetric Lee-Yang model}

For the solutions with $(c,h^\textsc{ns})=(3/4,1/4)$, one of the fusion rules turns out to be a negative number. It means that the solutions cannot be regarded as conformal characters, as they stand. To reconcile this, we exchange the role of vacuum and non-vacuum characters. After interchange of characters, the theory can be identified to the supersymmetric version of the Lee-Yang edge model, namely the ${\cal N}=1$ supersymmetric non-unitary minimal model $\mathcal{SM}(8,2)$, with negative central charge $c=-\frac{21}{4}$.

The supersymmetric Lee-Yang edge model has two NS characters of weight $h=0$ and $h=-1/4$,
\begin{align}
    \chi_0^\textsc{ns}(q)&=q^{\frac{7}{32}} \prod _{n=0}^{\infty} \frac{1}{\left(1-q^{\frac{1}{2} (8 n+3)}\right) \left(1-q^{\frac{1}{2} (8 n+4)}\right) \left(1-q^{\frac{1}{2} (8 n+5)}\right)}, 
    \nonumber \\ 
    \chi_{-\frac{1}{4}}^\textsc{ns}(q)&=q^{-\frac{1}{32}} \prod _{n=0}^{\infty} \frac{1}{\left(1-q^{\frac{1}{2} (8 n+1)}\right) \left(1-q^{\frac{1}{2} (8 n+4)}\right) \left(1-q^{\frac{1}{2} (8 n+7)}\right)},
\end{align}
and two R characters of weight $h=-3/32$ and $h=-7/32$, 
\begin{align}
    \chi_{-\frac{3}{32}}^\textsc{r}(q)&=q^{\frac{1}{8}} \prod _{n=1}^{\infty} \left(q^n+1\right) \left(q^{2 n}+1\right)
    , \quad \chi_{-\frac{7}{32}}^\textsc{r}(q)=\prod _{n=1}^{\infty} \left(q^{2 n-1}+1\right) \left(q^n+1\right). 
\end{align}
The solution of $c=3/4$ in Table \ref{type VI list} can then be identified as follows,
\begin{align}
    f_0^{\rm{NS}}(\tau) = \chi_{-\frac{1}{4}}^\textsc{ns}(q), \quad f_1^{\rm{NS}}(\tau) = \chi_{0}^\textsc{ns}(q), \quad 
    f_0^{\rm{R}}(\tau) = \chi_{-\frac{7}{32}}^\textsc{r}(q), \quad f_1^{\rm{R}}(\tau) = \chi_{-\frac{3}{32}}^\textsc{r}(q). 
\end{align}
From the modular $S$-matrix below, 
\begin{equation}
    \left( 
    \begin{array}{c}
        f_1^{\rm{NS}}(-\frac{1}{\tau})\\
        f_0^{\rm{NS}}(-\frac{1}{\tau}) 
    \end{array}
    \right)
    =
    \left(
    \begin{matrix}
    -\cos\frac{\pi}{8} & \sin\frac{\pi}{8}\\
    \sin\frac{\pi}{8} & \cos\frac{\pi}{8}\\
    \end{matrix}\right)
    \left(
    \begin{array}{c}
        f_1^{\rm{NS}}(\tau)\\
        f_2^{\rm{NS}}(\tau) 
    \end{array}
    \right),
\end{equation}
we can read off the fusion rules
\begin{equation}
    N_{00}^{0} = N_{10}^{1} = N_{01}^{1} = N_{11}^{0} = 1,\quad N_{11}^{1} = 2, \quad N_{ij}^{k} = 0 \ \text{otherwise}.
\end{equation}
Note that all fusion coefficients become non-negative integers.

\paragraph{Supersymmetric ADE WZW models with level one}

We claim that the solutions with  $c=3/2,3,6,9,21/2$ can be understood as the ${\cal N}=1$ supersymmetric ADE WZW model with level one.

Let us start with the tensor product of free Majorana fermions 
and the WZW model with level-one. The number of Majorana fermions is given by the rank of a group on which the WZW model is defined. The theory then has two NS characters which take the forms,  
\begin{align}\label{WZW x free MW}
    f_0^\text{NS}(\tau) & = \Big( \psi^\text{NS}(\tau) \Big)^{\text{rk}(\mathfrak{g})} \cdot 
    \chi_0^{\mathfrak{g}} (\tau), 
    \nonumber \\
    f_1^\text{NS}(\tau) & = \Big( \psi^\text{NS}(\tau) \Big)^{\text{rk}(\mathfrak{g})} \cdot 
    \chi_1^{\mathfrak{g}} (\tau),   
\end{align}  
where $\psi^\text{NS}(\tau)=\sqrt{\vartheta_{3}(\tau)/\eta(\tau)}$ is the NS character for a single Majorana fermion as given in \eqref{aaa}. $\chi_0^{\mathfrak{g}}(\tau)$ and $\chi_1^{\mathfrak{g}}(\tau)$ denote the vacuum and non-degenerate characters of level-one WZW model on $\mathfrak{g}$. When $\mathfrak{g}=\mathfrak{a}_1$, $\mathfrak{a}_2$, $\mathfrak{d}_4$, $\mathfrak{e}_6$, $\mathfrak{e}_7$, the above characters \eqref{WZW x free MW} reproduce the BPS solutions for $c=3/2$, $3$, $6$, $9$, $21/2$, respectively.

We notice that the level-one WZW model on a simply-laced $\mathfrak{g}$ can be realized as free bosons on the root-lattice of $\mathfrak{g}$, which is  referred to as the Frenkel-Kac construction \cite{Frenkel:1980rn, segal1981}. Thus, we can describe the fermionic RCFTs with $c=3/2$, $3$, $6$, $9$, $21/2$ as the ${\cal N}=1$ supersymmetric theories of rank rk$(\mathfrak{g})$  pairs of a free boson and a fermion.  

By construction, the modular matrices of the above BPS solutions are governed by those of the corresponding WZW model:
\begin{enumerate}

\item For instance, the modular $S$-matrix 
of the $c=3/2$ BPS solution becomes 
\begin{equation}
    S = \frac{1}{\sqrt{2}}\left(
\begin{array}{cc}
1 & 1\\
1 & -1 
\end{array}
\right).
\end{equation}
It is straightforward to read the fusion rule algebra coefficients using the Verlinde formula. The non-vanishing coefficients are given by
\begin{equation}
N_{00}^{0} = N_{10}^{1} = N_{01}^{1} = N_{11}^{0} = 1,
\end{equation}
thus we have the consistent fusion rule as expected. The $\Gamma_\theta$ invariant 
NS partition function is 
\begin{align}
    Z_\text{NS} = \Big| f_0^\text{NS}(\tau)\Big|^2 + \Big| f_1^\text{NS}(\tau) \Big|^2. 
\end{align}

\item The $c=3$ fermionic RCFTs has two primaries associated with the same characters $f_1^\text{NS}(\tau)$. They rotate 
as ${\bf 3}$ and $\bar{\bf 3}$ under $\text{SU}(3)$. In the basis of $({\hat f}_0^\text{NS}, {\hat f}_1^\text{NS}, 
{\hat f}_2^\text{NS})= (f_0^\text{NS}, f_1^\text{NS}, 
f_1^\text{NS})$, the modular $S$-matrix is given by
\begin{equation}
\hat{S} = \renewcommand{\arraystretch}{2}
\left(\begin{matrix}
\frac{1}{\sqrt{3}} & \frac{1}{\sqrt{3}} & \frac{1}{\sqrt{3}}\\
\frac{1}{\sqrt{3}} & \frac{1}{\sqrt{3}}\,e^{ 2 \pi i/3 } & \frac{1}{\sqrt{3}}\,e^{- 2 \pi i/3 } \\
\frac{1}{\sqrt{3}} & \frac{1}{\sqrt{3}}\,e^{- 2 \pi i/3 } & \frac{1}{\sqrt{3}}\,e^{ 2 \pi i/3 }
\end{matrix}\right).
\end{equation}
It gives consistent fusion rules below 
\begin{equation}
\begin{aligned}
&N_{00}^{0} = N_{01}^{1}  = N_{10}^{1} = N_{02}^{2} = N_{20}^{2} = 1\,,\\
&N_{11}^{2} = N_{22}^{1} = N_{12}^{0} = N_{21}^{0} = 1\,,\quad N_{ij}^{k} = 0 \ \text{otherwise}\,.
\end{aligned}
\end{equation}
The NS partition function is thus given by 
\begin{align}
    Z_\text{NS} = \Big| f_0^\text{NS}(\tau)\Big|^2 + 2\Big| f_1^\text{NS}(\tau) \Big|^2.
\end{align} 

\item For the ${\cal N}=1$ supersymmetric level-one WZW model on $\mathfrak{d}_4$, we have three primaries that are associated with the same character $f_1^\text{NS}(\tau)$ but transform differently under $SO(8)$ as the vector ${\bf 8}_v$, the chiral spinor ${\bf 8}_s$, and the anti-chiral spinor ${\bf 8}_c$. In the basis of $({\hat f}_0^\text{NS}, {\hat f}_1^\text{NS}, {\hat f}_2^\text{NS},{\hat f}_3^\text{NS})= (f_0^\text{NS}, f_1^\text{NS}, f_1^\text{NS},f_1^\text{NS})$, the modular $S$-matrix becomes 
\begin{equation}
\hat{S} = \renewcommand{\arraystretch}{2}
\left(\begin{matrix}
    \frac12 & \frac12 & \frac12 & \frac12 \\
    \frac12 & \frac12 & -\frac12 & -\frac12 \\
    \frac12 & -\frac12 & \frac12 & -\frac12 \\
    \frac12 & -\frac12 & -\frac12 & \frac12
\end{matrix}\right),
\end{equation}
which leads to the fusion rules below
\begin{align}
    N_{00}^0 = N_{01}^1 = N_{02}^2 = N_{03}^3 = N_{12}^3 =N_{23}^1 = N_{31}^2=1, 
\end{align}
with all other $N_{ij}^k=0$. The $\Gamma_\theta$ invariant NS partition function takes the form 
\begin{align}
    Z_\text{NS} =  \Big| f_0^\text{NS}(\tau)\Big|^2 + 3\Big| f_1^\text{NS}(\tau) \Big|^2.
\end{align}

\item One can easily show that the modular $S$-matrices of the $c=9$ and $c=21/2$ solutions are shared by the $c=3$ and $c=3/2$ solutions, respectively. In fact, the ${\cal N}=1$ supersymmetric level-one WZW model on $\mathfrak{e}_6$ ($\mathfrak{e}_7$) is dual to the supersymmetric $\mathfrak{a}_2$ ($\mathfrak{a}_1$) WZW model with level one in a sense that their characters obey a bilinear relation \eqref{dhs} below. 

\end{enumerate}
Amusing examples are given by a one-parameter family of $c=12$ solutions, each of which has the vacuum character below 
\begin{align}\label{ghj}
    f_0^\text{NS}(\tau) = K(\tau) + n \quad (n \in \mathbb{Z}).  
\end{align}
They do not  provide consistent two-character theories, because the other character $f_1^\text{NS}(\tau)$ does not admit positive integer coefficients for any $n$. Nonetheless, the character $f_0^\text{NS}(\tau)$ is $\Gamma_\theta$ invariant and serves as the NS partition function of a certain putative fermionic RCFT. 

We argue that the integer parameter $n$ can be further restricted to $n=0$, $2$, $4$, $8$, and $24$ in order to make the corresponding theories physically well-behaved. To see this, note that the parameter $n$ of \eqref{ghj} is the number of free fermions contained in a given model. \eqref{ghj} can thus split into two pieces as follows
\begin{align}
    K(\tau) + n = \Big( \psi^\text{NS}(\tau) \Big)^n \cdot {\tilde f}^\text{NS}(\tau)\ ,
\end{align}
where ${\tilde f}^\text{NS}(\tau)$ should be understood as the NS partition function of a $c=(12-n/2)$ CFT with no free fermion. We found that ${\tilde f}^\text{NS}(\tau)$ fails to have positive integer coefficients in the $q$-expansion for $n > 8$ except $n=24$. Another simple diagnostic to rule out an unphysical NS partition function is to analyse the degeneracy and weights in its R sector partition function \cite{Benjamin:2020zbs}. One can show that, for $n=1$, $3$, $5$, $6$, and $7$, the Ramond partition function has non-integer degeneracy and thus any of fermionic RCFTs without free fermions cannot have the ${\tilde f}^{NS}(\tau)$ as its NS partition function. 

As discussed in the previous section, we can identify $f^\text{NS}(\tau)$ with $n=24$ as the NS partition function of the $24$ free Majorana-Weyl fermions \eqref{theta identity}. We also note that a system of $24$ free Majorana fermions can be related to the $O(24)$ WZW model with level one via the non-abelian bosonization \cite{Witten:1983ar}. More explicitly, one can express $f_0^\text{NS}(\tau) = K(\tau)+ 24$ in terms of the WZW characters as follows 
\begin{align}
  f_0^\text{NS} (\tau) = \chi_0^{O(24)}(\tau) + \chi_{\frac12}^{O(24)}(\tau) , 
\end{align}
where $\chi_0^{O(24)}(\tau)$ and $\chi_{1/2}^{O(24)}(\tau)$ are the WZW characters of the vacuum and conformal weight $h=1/2$. The left-over character $\chi_{3/2}^{O(24)}(\tau)$ of the WZW model appears as the Ramond partition  function,
\begin{align}
  f_0^\text{R}(\tau) = \sqrt2 \chi_{\frac32}^{O(24)}(\tau). 
\end{align}
One can easily show that the GSO projection of the $24$ free Majorana fermions gives the partition function of the level-one WZW model for $O(24)$. 

It is worth to mention that \eqref{ghj} for $n=0$ is nothing but the NS partition function of the $c=12$ extremal SCFT, {\it a.k.a.} the Conway extremal CFT \cite{FLM85,2005math2267D}. There are two different but equivalent constructions of it. One of them is based on the ${\cal N}=1$ supersymmetric $\mathfrak{e}_8$ WZW model with level one, followed by a $\mathbb{Z}_2$ orbifold \cite{FLM85}. The other is a $\mathbb{Z}_2$ orbifold of $24$ free chiral fermions \cite{2005math2267D}. In both constructions, $\mathbb{Z}_2$ action is the charge conjugation. The second construction implies that the Conway extremal SCFT is the fermionization of the $SO(24)$ WZW model with level one \cite{Johnson-Freyd:2019wgb}. To be concrete, the NS and R partition function of the extremal SCFT with $c=12$ can be expressed in terms of the WZW characters,
\begin{align}
    f_0^\text{NS} (\tau) & = \chi_0^{SO(24)}(\tau) + \chi_{\frac32}^{SO(24)}(\tau) , 
    \nonumber \\
    f_0^\text{R}(\tau) & = \chi_{\frac12}^{SO(24)}(\tau) + \chi_{\frac32}^{SO(24)}(\tau) . 
\end{align}
From the fact that $f_0^{\widetilde{\text{R}}}(\tau) =\chi_{\frac12}^{SO(24)}(\tau) -\chi_{\frac32}^{SO(24)}(\tau)=24$, the bosonization of the Conway extremal CFT indeed gives the torus partition function of the level-one WZW model on $SO(24)$. In fact $O(24)$ and $SO(24)$ WZW models share the same characters, and it is noteworthy that the SL$_2(\mathbb{Z})$ invariant partition function of the $SO(24)_1$ ($O(24)_1$) WZW model allows the two different fermionizations. 

For $n=8$, one can see that the ${\cal N}=1$ supersymmetric  level-one WZW model for $\mathfrak{e}_8$ takes \eqref{ghj} as the NS partiton function, namely 
\begin{align}
    {\tilde f}^\text{NS}(\tau) = \Big( j(\tau) \Big)^{\frac13}.
\end{align}
However, it is not clear yet if there would exist fermionic RCFTs having ${\tilde f}^\text{NS}(\tau)$ with $n=2,4$ as the NS partition function, although they arise as solutions to second order MLDEs which do not saturate the supersymmetric unitarity bound $h^\text{R} \geq c/24$. See \eqref{qwe} for details.

One can also observe that the supersymmetric $\mathfrak{e}_8$ WZW model with level one exhibits various deconstructions into a pair of two supersymmetric WZW models: if $f^\text{NS}_i(\tau)$ and ${\widetilde f}^\text{NS}_i(\tau)$ ($i=0,1$) are the NS characters of the central charge $c$ and $(12-c)$, then 
\begin{align}\label{dhs}
    K(\tau) + 8 = f_0^\text{NS}{\widetilde f}^\text{NS}_0(\tau) 
    + {\cal M}^{(c)} f_1^\text{NS}{\widetilde f}^\text{NS}_1(\tau),  
\end{align}
where ${\cal M}^{(c=3/2)}=1$, ${\cal M}^{(c=3)}=2$, and ${\cal M}^{(c=6)}=3$. Indeed, \eqref{dhs} is a supersymmetric generalization of the bilinear relation \eqref{e8bilinear}, and thus potentially gives new insight into the supersymmetric modular tensor category. 

\begin{table}[t]
\begin{center}
\begin{tabular}{ccccc|c}
	$c$ & $h^\textsc{ns}$ & $h^\textsc{r}_{-}$ & $h^\textsc{r}_{+}$ & $a_1$ & Identification \\[1mm]
			\hline 
	$-3$ & $-\frac{1}{6}$ & $-\frac{1}{8}$ & $\frac{5}{24}$ & $1$ & $c = 1$ solution of BPS type {\bf{I}} \\[1mm]
	$-\frac{9}{2}$ & $-\frac{1}{4}$ & $-\frac{3}{16}$ & $-\frac{1}{16}$ & $1$ & $c = \frac{3}{2}$ solution of BPS type {\bf{II}} \\[1mm]
	$-5$ & $-\frac{1}{3}$ & $-\frac{5}{24}$ & $\frac{1}{8}$ & $2$ & $c = 3$ solution of BPS type {\bf{II}} \\[1mm]
	$-\frac{21}{4}$ & $-\frac{1}{4}$ & $-\frac{7}{32}$ & $-\frac{3}{32}$ & 0 & $c = \frac{3}{4}$ solution of BPS type {\bf{II}} \\[1mm]
	$-7$ & $-\frac{2}{3}$ & $-\frac{7}{24}$ & $\frac{3}{8}$ & $6$ &  $c = 9$ solution of BPS type {\bf{II}}
\end{tabular}
\caption{\label{type VI list c<0} The list of $c<0$ BPS solutions with free fermion currents. Eventually every five solutions in this list are mapped to the unitary solution upon exchanging the vacuum solution and non-vacuum solution.}
\end{center}
\end{table}
We finally remark on the BPS solutions for $c<0$. A complete list of BPS solutions with $c<0$ and $a_1 \neq 0$ is presented in Table \ref{type VI list c<0}. Here both NS sector and R sector solutions have non-negative integer coefficients in $q$-series. One can show that every solution in Table \ref{type VI list c<0} actually describes a unitary theory upon exchanging the vacuum and non-vacuum character. The precise mapping between BPS $c<0$ solutions and BPS unitary solutions is given in Table \ref{type VI list c<0}.

\subsection{Non-BPS solutions}

A typical type of fermionic RCFTs can be generated by tensoring  the WZW models for a Lie algebra $\mathfrak{g}$ with arbitrary number of free fermions. Those models do not saturate the supersymmetric unitarity bound $h^\text{R}\geq c/24$ in general except the cases discussed in the previous section. For instance, let us consider a tensor product of the (bosonic) $\mathfrak{a}_1$ WZW model with level $k$ and three free Majorana fermions. Although the model is known to preserve the ${\cal N}=1$ supersymmetry, the Ramond vacuum energy is greater than $c/24$, i.e., the bound is not saturated for any $k$. 

Although the space of solutions which contain free fermions but do not saturate the supersymmetric unitarity bound deserves further  investigation, we restrict our attention to a simpler problem of classifying the fermionic RCFTs with no free fermion where $h^\text{R}>c/24$ in the present work.

\subsubsection*{Non-BPS Type {\bf{I}} : Non-BPS solutions without free fermions}
\label{class6}

In this subsection, we explore the solutions of the second order MLDE with $a_1=0$ and $h^{\rm{NS}} \neq \frac{1}{2}$. We refer such solutions to as the non-BPS solutions because they do not saturate the supersymmetric unitarity bound \eqref{SUSYbound}. The central charge and weights for the solutions in this class are summarized in Table \ref{type III}. The $q$-series of the solutions are also given in Table \ref{NS type III} and \ref{R type III} for the convenience. However, it turns out that the solutions with $c=102/5$, $21$, $85/4$ have negative fusion coefficients and thus cannot be regarded as well-defined physical models. For the rests, the corresponding fermionic CFTs are identified in what follows.  
\begin{table}[t]
\begin{center}
\begin{tabular}{c|ccccccccc}
$c$ & $\frac{7}{10}$ & $\frac{133}{10}$  & $\frac{91}{5}$  & $\frac{39}{2}$ & $\frac{102}{5}$ & 21 & $\frac{85}{4}$ & 22 &  $\frac{114}{5}$ \\ [1mm]
\hline 
$h^{\rm {NS}} $&  $\frac{1}{10}$ & $\frac{9}{10}$  & $\frac{11}{10}$ & $\frac{7}{6}$ & $\frac{6}{5}$ & $\frac{5}{4}$ & $\frac{5}{4}$ & $\frac{4}{3}$ & $\frac{7}{5}$ \\[1mm]
$h^{\rm R}$ & $\frac{3}{80},\frac{7}{16}$ & $\frac{57}{80},\frac{21}{16}$ & $\frac{49}{40},\frac{13}{8}$ & 
$\frac{65}{48},\frac{27}{16} $ &   $\frac{3}{2},\frac{17}{10}$ & $\frac{3}{2},\frac{7}{4}$ & $\frac{51}{32},\frac{55}{32}$ & $\frac{3}{2},\frac{11}{6}$& $\frac{3}{2},\frac{19}{10}$ \\
$\mathcal{M}$ & 1 & 1 & 1 & 2 & 1 & 1 & 1 & 2 & 1
\end{tabular}
\caption{\label{type III} The non-BPS type ${\bf{I}}$ involves nine non-BPS solutions which obtained by imposing no free fermions and $h^{\rm{NS}} \neq \frac{1}{2}$. The GSO projection of the solutions with $c=\frac{133}{10}, \frac{91}{5}, \frac{39}{2}$ leads to the WZW models for $\widehat{E}_{7,2},\widehat{D}_{7,3},\widehat{E}_{6,4}$, respectively. On the other hand, the identification for the solutions with $c=\frac{102}{5}$, $\frac{85}{4}$, $22$ and $\frac{114}{5}$ remains unclear.}
\end{center}
\end{table}

\paragraph{Tricritical Ising model}
Let us first start with the solution with $c=\frac{7}{10}$. One can show that the solution describes the fermionization of the unitary minimal model $\mathcal{M}(5,4)$, namely the tricritical Ising model. Note that the model has six characters of weight
\begin{align}
h = \left\{ 0, \frac{3}{2},\frac{1}{10}, \frac{3}{5}, \frac{3}{80}, \frac{7}{16} \right\},
\end{align}
denoted by  $\chi_{h}(\tau)$. Then, the solution of $c=\frac{7}{10}$ can be written as follow,
\begin{align}
    f^{\rm{NS}}_{0}(\tau) &= \chi_{0}(\tau) + \chi_{\frac{3}{2}}(\tau) ,
    \nonumber \\
    f^{\rm{NS}}_{1}(\tau) & = \chi_{\frac{1}{10}}(\tau) + \chi_{\frac{3}{5}}(\tau),
\end{align}
and the characters of other sectors are given by
\begin{gather}
    f^{\widetilde{\rm{NS}}}_{0}(\tau) = \chi_{0}(\tau) - \chi_{\frac{3}{2}}(\tau), 
    \quad
    f^{\widetilde{\rm{NS}}}_{1}(\tau) = \chi_{\frac{1}{10}}(\tau) - \chi_{\frac{3}{5}}(\tau), 
    \nonumber \\
    f^{\rm{R}}_{0}(\tau) =  \sqrt{2}\chi_{\frac{3}{80}}(\tau), 
    \quad
    f^{\rm{R}}_{1}(\tau) =  \sqrt{2}\chi_{\frac{7}{16}}(\tau). 
\end{gather}
In terms of the above characters, the torus partition functions for various spin structures become
\begin{align}
    Z_{\text{NS}} (\tau,\bar \tau) & = \Big|f^{\rm{NS}}_{0}(\tau)\Big|^2 + \Big|f^{\rm{NS}}_{1}(\tau)\Big|^2,
    \nonumber \\ 
    Z_{\widetilde{\text{NS}}}(\tau,\bar \tau) & = \Big|f^{\widetilde{\text{NS}}}_{0}(\tau)\Big|^2 + \Big|f^{\widetilde{\text{NS}}}_{1}(\tau)\Big|^2, 
    \\ 
    Z_{\text{R}} (\tau,\bar \tau) & = \Big|f^{\rm{R}}_{0}(\tau)\Big|^2 + \Big|f^{\rm{R}}_{1}(\tau)\Big|^2.
    \nonumber 
\end{align}
Assuming $Z_{\widetilde{\text{R}}}(\tau,\bar \tau) = 0$, we can see that the GSO projection of the fermionic CFT of our interest leads to the tricritical Ising model, 
\begin{align}
\begin{split}
Z(\tau, \bar{\tau}) & = \frac{1}{2} \Big[ Z_{\text{NS}}(\tau,\bar \tau) +Z_{\widetilde{\text{NS}}}(\tau,\bar \tau) +  Z_{\text{R}} (\tau,\bar \tau) + Z_{\widetilde{\text{R}}} (\tau,\bar \tau) \Big] 
\nonumber \\ & = 
\Big|\chi_{0} (\tau)\Big|^2 + \Big|\chi_{\frac{3}{2}} (\tau)\Big|^2 + \Big|\chi_{\frac{1}{10}} (\tau)\Big|^2+ \Big|\chi_{\frac{3}{5}} (\tau)\Big|^2+ \Big|\chi_{\frac{3}{80}} (\tau)\Big|^2+ \Big|\chi_{\frac{7}{16}} (\tau)\Big|^2.
\end{split}
\end{align}
The tricritical Ising model is known to preserve the ${\cal N}=1$ supersymmetry, but has the Ramond ground state which breaks the supersymmetry spontaneously \cite{Friedan:1983xq}. It justifies our assumption $Z_{\widetilde{\text{R}}}(\tau,\bar \tau) = 0$. We also note that the solution with $c=\frac{7}{10}$ simply corresponds to the characters of the first unitary $\mathcal{N}=1$ supersymmetic minimal model $\mathcal{SM}(5,3)$.

\paragraph{$\mathfrak{e}_7$ WZW model with level two} 
We next move on to the solution of $c=133/10$. We claim that the bosonization of the fermionic CFT results in the level-two WZW model for $\mathfrak{e}_7$. To see this, let us first note that the NS, $\widetilde{\rm{NS}}$ and R sector 
characters can be expressed in terms of the WZW characters:  
\begin{align}
\begin{split}
    f^{\rm{NS}}_{0}(\tau) &= \chi^{{\mathfrak e}_{7,2}}_0(\tau) + \chi^{{\mathfrak e}_{7,2}}_{\frac{3}{2}}(\tau),
    \\ 
    f^{\rm{NS}}_{1}(\tau) & = \chi^{{\mathfrak e}_{7,2}}_{\frac{9}{10}}(\tau) + \chi^{{\mathfrak e}_{7,2}}_{\frac{7}{5}}(\tau),
\end{split}
\end{align}
and 
\begin{align}
\begin{split}
    f^{\widetilde{\rm{NS}}}_{0}(\tau) &= \chi^{{\mathfrak e}_{7,2}}_0(\tau) - \chi^{{\mathfrak e}_{7,2}}_{\frac{3}{2}}(\tau),
    \\
    f^{\widetilde{\rm{NS}}}_{1}(\tau) & = \chi^{{\mathfrak e}_{7,2}}_{\frac{9}{10}}(\tau) - \chi^{{\mathfrak e}_{7,2}}_{\frac{7}{5}}(\tau).   
\end{split}
\end{align}
The two solutions in the Ramond sector become 
\begin{align}
    \tilde{f}^{\rm{R}}_{0}(\tau)  =  \sqrt{2}\chi^{{\mathfrak e}_{7,2}}_{\frac{21}{16}}(\tau),
    \quad
    \tilde{f}^{\rm{R}}_{1}(\tau)  =  \sqrt{2}\chi^{{\mathfrak e}_{7,2}}_{\frac{57}{80}}(\tau),    
\end{align}
where we can see that the Ramond vacuum energy does not saturate the bound \eqref{SUSYbound}. The torus partition function for each spin structure is given by 
\begin{align}
    Z_{\text{NS}} (\tau,\bar \tau) & = \Big|f^{\rm{NS}}_{0}(\tau)\Big|^2 + \Big|f^{\rm{NS}}_{1}(\tau)\Big|^2,
    \nonumber \\ 
    Z_{\widetilde{\text{NS}}}(\tau,\bar \tau) & = \Big|f^{\widetilde{\text{NS}}}_{0}(\tau)\Big|^2 + \Big|f^{\widetilde{\text{NS}}}_{1}(\tau)\Big|^2, 
    \\ 
    Z_{\text{R}} (\tau,\bar \tau) & = \Big|f^{\rm{R}}_{0}(\tau)\Big|^2 + \Big|f^{\rm{R}}_{1}(\tau)\Big|^2.
    \nonumber 
\end{align}
Under the assumption that $Z_{\widetilde{\text{R}}}(\tau,\bar \tau) = 0$, one obtains the modular invariant partition function of the level-two WZW model for $\mathfrak{e}_7$ via the GSO projection,
\begin{align}
\begin{split}
Z(\tau, \bar{\tau}) & = \frac{1}{2} \Big[ Z_{\text{NS}}(\tau,\bar \tau) +Z_{\widetilde{\text{NS}}} (\tau,\bar \tau) +  Z_{\text{R}} (\tau,\bar \tau) + Z_{\widetilde{\text{R}}} (\tau,\bar \tau) \Big] 
\nonumber \\ & = 
\Big|\chi_{0}^{ {\mathfrak e}_{7,2}}(\tau)\Big|^2 + \Big|\chi_{\frac{3}{2}}^{ {\mathfrak e}_{7,2}}(\tau)\Big|^2 + \Big|\chi_{\frac{9}{10}}^{ {\mathfrak e}_{7,2}}(\tau)\Big|^2+ \Big|\chi_{\frac{7}{5}}^{ {\mathfrak e}_{7,2}}(\tau)\Big|^2+ \Big|\chi_{\frac{21}{16}}^{ {\mathfrak e}_{7,2}}(\tau)\Big|^2
+ \Big|\chi_{\frac{57}{80}}^{ {\mathfrak e}_{7,2}}(\tau)\Big|^2.
\end{split}
\end{align}
This result suggests that the fermionic description of the level-two WZW model for $\mathfrak{e}_7$ preserves the supersymmetry but has a non-supersymmetric Ramond ground state. We leave this problem of verifying the emergent supersymmetry as a future work.

As a side remark, it turns out that the solutions of $c=\frac{133}{10}$ are paired up with the characters of the first unitary $\mathcal{N}=1$ supersymmetric minimal model to produce the $(\widehat{E}_{7,1})^{\otimes 2}$ WZW model, in the same way that the Ising model pairs up with the $\mathfrak{e}_8$ WZW model at level-two to give $(\widehat{E}_{8,1})^{\otimes 2}$. Precisely the bilinear relations read
\begin{align}
\begin{split}
\left(\chi_0^{ {\mathfrak e}_{7,1}}(\tau)\right)^2 + \left(\chi_{\frac{3}{4}}^{ {\mathfrak e}_{7,1}}(\tau)\right)^2 &= {f}^{\rm{NS}}_{0}(\tau) \tilde{f}^{\rm{NS}}_{0}(\tau) + {f}^{\rm{NS}}_{1}(\tau) \tilde{f}^{\rm{NS}}_{1}(\tau),  \\
\left(\chi_0^{ {\mathfrak e}_{7,1}}(\tau)\right)^2 - \left(\chi_{\frac{3}{4}}^{ {\mathfrak e}_{7,1}}(\tau)\right)^2 &= {f}^{\widetilde{\rm{NS}}}_{0}(\tau) \tilde{f}^{\widetilde{\rm{NS}}}_{0}(\tau) + {f}^{\widetilde{\rm{NS}}}_{1}(\tau) \tilde{f}^{\widetilde{\rm{NS}}}_{1}(\tau),  \\
\left(\chi_0^{ {\mathfrak e}_{7,1}}(\tau) \chi_{\frac{3}{4}}^{ {\mathfrak e}_{7,1}}(\tau)\right)^2 &= {f}^{\rm{R}}_{0}(\tau) \tilde{f}^{\rm{R}}_{0}(\tau) + {f}^{\rm{R}}_{1}(\tau) \tilde{f}^{\rm{R}}_{1}(\tau),
\end{split}
\end{align}
where $\chi_h^{ {\mathfrak e}_{7,1}}(\tau)$ means the characters of the $\mathfrak{e}_7$ WZW model at level-one.

\paragraph{Orbifold of $\mathfrak{d}_7$ WZW model with level three}

We observe that the NS solutions with $c=91/5$ can be expressed  in terms of the characters of the level-three WZW model for $\mathfrak{d}_7$ as follows, 
\begin{align}
\label{D7 level 3 WZW}
\begin{split}
    f^{\rm{NS}}_{0}(\tau) & = \chi^{{\mathfrak d}_{7,3}}_{0}(\tau) + \chi^{{\mathfrak d}_{7,3}}_{2}(\tau) + \chi^{{\mathfrak d}_{7,3}}_{\frac{3}{2}}(\tau) + {\widetilde \chi}^{{\mathfrak d}_{7,3}}_{\frac{3}{2}}(\tau), 
    \\
    f^{\rm{NS}}_{1}(\tau) & =\chi^{{\mathfrak d}_{7,3}}_{\frac{11}{10}}(\tau) + \chi^{{\mathfrak d}_{7,3}}_{\frac{21}{10}}(\tau) +  \chi^{{\mathfrak d}_{7,3}}_{\frac{8}{5}}(\tau) + {\widetilde \chi}^{{\mathfrak d}_{7,3}}_{\frac{8}{5}}(\tau),\\
\end{split}
\end{align}
where $\chi^{{\mathfrak d}_{7,3}}_{\frac{3}{2}}(\tau)$ and  ${\widetilde \chi}^{{\mathfrak d}_{7,3}}_{\frac{3}{2}}(\tau)$ ($\chi^{{\mathfrak d}_{7,3}}_{\frac{8}{5}}(\tau)$ and ${\widetilde \chi}^{{\mathfrak d}_{7,3}}_{\frac{8}{5}}(\tau)$) are characters of the same weight $h=3/2$ ($h=8/5$) but in different representations under ${\mathfrak d}_7$. Performing $S$ and $T$ transformations on them, the solutions for other sectors are given as
\begin{align}
\begin{split}
    f^{\widetilde{\rm{NS}}}_{0}(\tau) = \chi^{{\mathfrak d}_{7,3}}_{0}(\tau) + \chi^{{\mathfrak d}_{7,3}}_{2}(\tau) - \chi^{{\mathfrak d}_{7,3}}_{\frac{3}{2}}(\tau)-{\widetilde \chi}^{{\mathfrak d}_{7,3}}_{\frac{3}{2}}(\tau),
    \\
    f^{\widetilde{\rm{NS}}}_{1}(\tau) =\chi^{{\mathfrak d}_{7,3}}_{\frac{11}{10}}(\tau) + \chi^{{\mathfrak d}_{7,3}}_{\frac{21}{10}}(\tau) -  \chi^{{\mathfrak d}_{7,3}}_{\frac{8}{5}}(\tau)- {\widetilde \chi}^{{\mathfrak d}_{7,3}}_{\frac{8}{5}}(\tau),
\end{split}
\end{align}
and 
\begin{align}
    f^{\rm{R}}_{0}(\tau) & = 2\Big( \chi^{{\mathfrak d}_{7,3}}_{\frac{49}{40}}(\tau) + \chi^{{\mathfrak d}_{7,3}}_{\frac{89}{40}}(\tau)\Big), 
    \\
    f^{\rm{R}}_{1}(\tau) & =  2\Big( \chi^{{\mathfrak d}_{7,3}}_{\frac{13}{8}}(\tau) + \chi^{{\mathfrak d}_{7,3}}_{\frac{21}{8}}(\tau) \Big).    
\end{align}
We propose the vanishing $Z_{\widetilde{\text{R}}}(\tau,\bar \tau)$ so that the bosonization of the fermionic RCFT with $c=91/5$ provides the non-diagonal modular invariant partition function of the $\mathfrak{d}_7$ WZW model with level three,
\begin{align}\label{lim}
\begin{split}
Z(\tau, \bar{\tau}) & = \frac{1}{2} \Big[ Z_{\text{NS}}(\tau,\bar \tau) +Z_{\widetilde{\text{NS}}} (\tau,\bar \tau) +  Z_{\text{R}} (\tau,\bar \tau) + Z_{\widetilde{\text{R}}} (\tau,\bar \tau) \Big] 
\\ & = 
\Big|\chi_{0}^{{\mathfrak d}_{7,3}} + \chi_{2 }^{{\mathfrak d}_{7,3}} \Big|^2 + 
\Big|\chi^{{\mathfrak d}_{7,3}}_{\frac{3}{2}} + {\widetilde \chi}^{{\mathfrak d}_{7,3}}_{\frac{3}{2}}\Big|^2 +
\Big|\chi_{\frac{11}{10}}^{{\mathfrak d}_{7,3}} + \chi_{\frac{21}{10}}^{{\mathfrak d}_{7,3}} \Big|^2 +
\Big|\chi^{{\mathfrak d}_{7,3}}_{\frac{8}{5}} + {\widetilde \chi}^{{\mathfrak d}_{7,3}}_{\frac{8}{5}}\Big|^2 
\\ & 
+ 2 \Big|\chi_{\frac{49}{40}}^{{\mathfrak d}_{7,3}} +\chi_{\frac{89}{40}}^{{\mathfrak d}_{7,3}} \Big|^2 
+ 2 \Big|\chi_{\frac{13}{8}}^{{\mathfrak d}_{7,3}} +\chi_{\frac{21}{8}}^{{\mathfrak d}_{7,3}} \Big|^2. 
\end{split}
\end{align}
Each multiplicity $2$ in \eqref{lim} says that two primaries in different representations under $\mathfrak{d}_7$ are associated with the same character. We expect that a certain orbifold of the $\mathfrak{d}_7$  WZW model with level three has \eqref{lim} as the diagonal modular invariant partition function. From the Verlinde formula, one can verify that the orbifold model has a consistent fusions algebra. This result is tantalizing that the fermionic RCFT of $c=91/5$  has an emergent supersymmetry but no supersymmetric vacuum, which can explain why $Z_{\widetilde{\text{R}}}(\tau,\bar \tau)$ vanishes. 

\paragraph{Orbifold of $\mathfrak{e}_6$ WZW model with level four}

Similarly, one can express the solutions with $c=39/2$ in terms of the characters of the level-four WZW model for $\mathfrak{e}_6$ as follows, 
\begin{align} 
\begin{split}
    f^{\rm{NS}}_{0}(\tau) & = \chi^{{\mathfrak e}_{6,4}}_{0}(\tau) + \chi^{{\mathfrak e}_{6,4}}_{2}(\tau) + \chi^{{\mathfrak e}_{6,4}}_{\frac{3}{2}}(\tau) + \chi^{{\mathfrak e}_{6,4}}_{\frac{5}{2}}(\tau), \\
    f^{\rm{NS}}_{1}(\tau) & =\chi^{{\mathfrak e}_{6,4}}_{\frac{7}{6}}(\tau) 
    + \chi^{{\mathfrak e}_{6,4}}_{\frac{13}{6}}(\tau) + \chi^{{\mathfrak e}_{6,4}}_{\frac{5}{3}}(\tau) + \chi^{{\mathfrak e}_{6,4}}_{\frac{8}{3}}(\tau),\\
    f^{\widetilde{\rm{NS}}}_{0}(\tau) & =   \chi^{{\mathfrak e}_{6,4}}_{0}(\tau) + \chi^{{\mathfrak e}_{6,4}}_{2}(\tau) - \chi^{{\mathfrak e}_{6,4}}_{\frac{3}{2}}(\tau) - \chi^{{\mathfrak e}_{6,4}}_{\frac{5}{2}}(\tau),\\
    f^{\widetilde{\rm{NS}}}_{1}(\tau) & = \chi^{{\mathfrak e}_{6,4}}_{\frac{7}{6}}(\tau) + \chi^{{\mathfrak e}_{6,4}}_{\frac{13}{6}}(\tau) - \chi^{{\mathfrak e}_{6,4}}_{\frac{5}{3}}(\tau) - \chi^{{\mathfrak e}_{6,4}}_{\frac{8}{3}}(\tau),
\end{split}
\end{align} 
and
\begin{align}
\begin{split}
    f^{\rm{R}}_{0}(\tau) & = 2 \Big( \chi^{{\mathfrak e}_{6,4}}_{\frac{65}{48}}(\tau) + \chi^{{\mathfrak e}_{6,4}}_{\frac{113}{48}}(\tau) \Big) , \quad f^{\rm{R}}_{1}(\tau)  = 2 \chi^{{\mathfrak e}_{6,4}}_{\frac{27}{16}}(\tau).  
\end{split}
\end{align}
The modular matrices then determine the torus partition functions for various spin structures, 
\begin{align}
    Z_{\text{NS}} (\tau,\bar \tau) & = \Big| \chi^{{\mathfrak e}_{6,4}}_{0} + \chi^{{\mathfrak e}_{6,4}}_{2} + \chi^{{\mathfrak e}_{6,4}}_{\frac{3}{2}} + \chi^{{\mathfrak e}_{6,4}}_{\frac{5}{2}} \Big|^2 
    +2\Big|\chi^{{\mathfrak e}_{6,4}}_{\frac{7}{6}}  + \chi^{{\mathfrak e}_{6,4}}_{\frac{13}{6}} + \chi^{{\mathfrak e}_{6,4}}_{\frac{5}{3}} + \chi^{{\mathfrak e}_{6,4}}_{\frac{8}{3}}\Big|^2,
    \nonumber \\ 
    Z_{\widetilde{\text{NS}}}(\tau,\bar \tau) & = \Big| \chi^{{\mathfrak e}_{6,4}}_{0} + \chi^{{\mathfrak e}_{6,4}}_{2} - \chi^{{\mathfrak e}_{6,4}}_{\frac{3}{2}} - \chi^{{\mathfrak e}_{6,4}}_{\frac{5}{2}} \Big|^2
    + 2\Big|\chi^{{\mathfrak e}_{6,4}}_{\frac{7}{6}} + \chi^{{\mathfrak e}_{6,4}}_{\frac{13}{6}} - \chi^{{\mathfrak e}_{6,4}}_{\frac{5}{3}} - \chi^{{\mathfrak e}_{6,4}}_{\frac{8}{3}}\Big|^2, 
    \\ 
    Z_{\text{R}} (\tau,\bar \tau) & = 4 \Big|\chi^{{\mathfrak e}_{6,4}}_{\frac{65}{48}}+ \chi^{{\mathfrak e}_{6,4}}_{\frac{113}{48}}\Big|^2 + 8 \Big|\chi^{{\mathfrak e}_{6,4}}_{\frac{27}{16}} \Big|^2.
    \nonumber 
\end{align}
We again suppose that the fermionic model of our interest has vanishing $Z_{\widetilde{\text{R}}}(\tau,\bar \tau)$. As a consequence, the GSO projection leads to 
\begin{align}\label{lim2}
\begin{split}
Z(\tau, \bar{\tau}) & = \frac{1}{2} \Big[ Z_{\text{NS}}(\tau,\bar \tau) +Z_{\widetilde{\text{NS}}} (\tau,\bar \tau) +  Z_{\text{R}} (\tau,\bar \tau) + Z_{\widetilde{\text{R}}} (\tau,\bar \tau) \Big] 
\\ & = 
\Big|\chi^{{\mathfrak e}_{6,4}}_{0} + \chi^{{\mathfrak e}_{6,4}}_{2} \Big|^2 +
\Big|\chi^{{\mathfrak e}_{6,4}}_{\frac{3}{2}} + \chi^{{\mathfrak e}_{6,4}}_{\frac{5}{2}} \Big|^2 +
2\Big|\chi^{{\mathfrak e}_{6,4}}_{\frac{7}{6}}  + \chi^{{\mathfrak e}_{6,4}}_{\frac{13}{6}} \Big|^2 +
2\Big|\chi^{{\mathfrak e}_{6,4}}_{\frac{5}{3}} + \chi^{{\mathfrak e}_{6,4}}_{\frac{8}{3}} \Big|^2 
\\ & 
+ 2 \Big|\chi^{{\mathfrak e}_{6,4}}_{\frac{65}{48}}+ \chi^{{\mathfrak e}_{6,4}}_{\frac{113}{48}} \Big|^2 
+ 4 \Big|\chi^{{\mathfrak e}_{6,4}}_{\frac{27}{16}}  \Big|^2, 
\end{split}
\end{align}
which is a non-diagonal modular invariant partition function of the level-four $\mathfrak{e}_6$ WZW models. In the $\mathfrak{e}_6$ WZW model, some pairs of primaries in different representations share the same character. For instance, this happens for the characters of weight $7/6$, $13/6$, $5/3$, $8/3$, $65/48$, $113/48$ and $27/16$. It is likely that a certain orbifold of the WZW model would have \eqref{lim2} as the diagonal modular invariant partition function. We also propose that the fermionic RCFT with $c=39/2$ preserves the supersymemtry, but its Ramond vacuum breaks the supersymmetry spontaneously.

\paragraph{The solutions with $c=\frac{114}{5}$ and $21$}
Let us consider the solutions with $c=114/5$ and $c=21$. We first make the conjecture that under the assumption of vanishing $Z_{\widetilde{\text{R}}}(\tau,\bar \tau)$, the solutions with $c=114/5$ describes the fermionization of the bosonic RCFT exhibiting the moonshine phenomena for $2.^{2}E_6(2).2$ \cite{Bae:2020pvv}. More precisely, the NS, $\widetilde{\text{NS}}$ and R sector solutions can be expressed as follows 
\begin{align}
\begin{split}
&f^{\rm{NS}}_0(\tau) = \tilde{\chi}_0(\tau) + \tilde{\chi}_2(\tau), \quad f^{\rm{NS}}_1(\tau) = \tilde{\chi}_1(\tau) + \tilde{\chi}_5(\tau),\\
&f^{\widetilde{\rm{NS}}}_0(\tau) = \tilde{\chi}_0(\tau) - \tilde{\chi}_2(\tau), \quad f^{\widetilde{\rm{NS}}}_1(\tau) = \tilde{\chi}_1(\tau) - \tilde{\chi}_5(\tau), \\
&f^{ \rm{R}}_0(\tau) = 2\tilde{\chi}_2(\tau), \quad f^{ \rm{R}}_1(\tau) =  2\tilde{\chi}_5(\tau).
\end{split}
\end{align}
Here $\tilde{\chi}_i(\tau)$ are the characters of putative bosonic RCFT with $c=\frac{114}{5}$ and their explicit forms are given in the equation (3.55) of \cite{Bae:2020pvv}. Let us combine the partition functions for the NS, $\widetilde{\text{NS}}$ and R sector,
\begin{align}
\begin{split}
Z(\tau,\bar{\tau}) & = \frac{1}{2} Z^{\text{NS}}(\tau,\bar{\tau}) + \frac{1}{2} Z^{\widetilde{\text{NS}}}(\tau,\bar{\tau}) + \frac{1}{2} Z^{\text{R}}(\tau,\bar{\tau}) \\
                  &= |\tilde{\chi}_0(\tau)|^2 + |\tilde{\chi}_1(\tau)|^2 + 3 |\tilde{\chi}_2(\tau)|^2 + 3 |\tilde{\chi}_5(\tau)|^2,
\end{split}
\end{align}
thus the GSO projection reproduce the modular invariant partition function of the hypothetical bosonic RCFT with $c=\frac{114}{5}$ with the assumption of $Z^{\widetilde{\text{R}}} = 0$.

We finally comment that the solutions of $c=21$ admits the linear group $2^9.L_3(4)$ as an automorphism group. This can be shown with the help of a rank 21 lattice that constructed in \cite{Hoehn:2015rsa}. We will discuss more details of this theory in an upcoming paper \cite{ONP20}.

\subsubsection*{Non-BPS Type {\bf{II}} : Non-BPS pairs with $h^{\rm NS}=\frac{1}{2}$}

Here we discuss the solutions obtained by imposing the two conditions $h^{\rm{NS}}=\frac{1}{2}$ and $a_1=0$. There are seven solutions in this class which have non-negative integers in $q$-series. The profiles of the seven solutions are listed in Table \ref{type II} and their explicit $q$-series are presented in Table \ref{type II NS} and \ref{type II R}. The R sector conformal weights are given by $h^{\rm R}_{-}= \frac{c-4}{8}$ and $h^{\rm R}_{+}=\frac{c}{8}$. Note that the $\mathcal{N}=1$ unitarity bound is violated for $c<6$, because $h^{\rm R}_{-}=\frac{c-4}{8}<\frac{c}{24}$. Therefore the theories with $c<6$ in this class cannot be unitary $\mathcal{N}=1$ supersymmetric CFTs. The unitarity bound is saturated for the solution of $c=6$ and it agrees with the self-dual solution in BPS type $\bf{I}$. As discussed before, the solution with $c=6$ can be understood as a hexic product of the $su(2)_1$ WZW model and we will not discuss it any further in this article. 

\begin{table}[htp]
\begin{center}
\begin{tabular}{c|ccccccc}
$c$ & $\frac{9}{2}$ & 5  & $\frac{11}{2}$  & 6 & $\frac{13}{2}$ & 7 & $\frac{15}{2}$   \\ [1mm]
\hline
$h^{\rm {NS}} $&  $\frac12$ & $\frac12$  & $\frac12$ & $\frac12$ & $\frac12$ & $\frac12$ & $\frac12$ \\[1mm]
$h^{\rm R}$ & $\frac{1}{16},\frac{9}{16}$ & $\frac18,\frac58$ & $\frac{3}{16},\frac{11}{16}$  & $\frac{1}{4},\frac{3}{4}$ & $\frac{5}{16},\frac{13}{16}$ & $\frac38,\frac78$ & $\frac{7}{16},\frac{15}{16}$ \\ [1mm]
$b_0$ & 1 & 2 & 1 & 4 & 1 & 6 & 1 \\ [1mm] 
$\mathcal{M}$ & 207 & 55 & 231 & 15 & 247 & 7 & 255 \\ [1mm]
$a_0'$ & 1 & 1 & 1 & 2 & 2 & 4 & 4 \\ [1mm]
$b_0'$ & 1 & 8 & 8 & 8 & 16 & 96 & 32
\end{tabular}
\caption{\label{type II} The seven solutions obtained by imposing two conditions $h^{\rm{NS}}=\frac{1}{2}$ and $a_1=0$. The solution of $c=6$ saturate the unitarity bound hence it is also found in the BPS type {\bf{I}}. In this table, $b_0$ denote a coefficient of the leading order term of the NS sector non-degenerate solution.$\mathcal{M}$ denote a degeneracy of the NS sector primary state. $a_0'$ and $b_0'$ stand for the leading order coefficients of the R sector solutions.}
\end{center}
\end{table}

The NS sector characters take the analytic form of
\begin{align}
\label{analytic type II}
\begin{split}
f_0^{\rm{NS}}(\tau) &= (\psi^{\rm NS})^{2c} -  2c (\psi^{\rm NS})^{2c-24} \left( \frac{\vartheta_3(\tau)}{\vartheta_4(\tau)} \right)^{4} = (\psi^{\rm NS})^{2c} \left( 1 -  \frac{c\lambda}{8} \right), \\
f_1^{\rm{NS}}(\tau)=&= b_0(\psi^{\rm NS})^{2c-24} \left( \frac{\vartheta_3(\tau)}{\vartheta_4(\tau)} \right)^{4} =b_0 (\psi^{\rm NS})^{2c}  \frac{\lambda}{16}, 
\end{split}
\end{align}
where the central charge $c$ is constrained between $4 \le c \le \frac{15}{2}$ to have non-negative integer coefficients in the $q$-series. Because \eqref{analytic type II} reduces to the characters of level-one WZW model for ${\mathfrak d}_4 $ when $c=4$, we only focus on the solutions between $\frac{9}{2} \le c \le \frac{15}{2}$ to explore the fermionic RCFT. Now it is straightforward to see that the two characters $f_0^{\rm{NS}}(\tau)$ and $f_1^{\rm{NS}}(\tau)$ satisfy the following identity,
\begin{align}
\begin{split}
f_0^{\rm{NS}}(\tau) + \frac{2c}{b_0} f_1^{\rm{NS}}(\tau) = (\psi^{\rm NS})^{2c},
\end{split}
\end{align}
where $\psi^{\rm NS}$ denote the NS partition function of a free Majorana-Weyl fermion which introduced in \eqref{free fermion}. This identity indicates that two independent solutions in non-BPS type {\bf II} split the partition function of $n=2c$ tensor product of the free Majorana-Weyl fermions. The vacuum solution $f_0^{\rm{NS}}(\tau)$ does not involves free fermion currents because $\lambda \sim 16 q^{\frac{1}{2}}$ thus it satisfies a condition $a_1 = 0$. 

The $S$ transformation of the vacuum solution ought to be written as a linear combination of $f_0^{\rm{NS}}(\tau)$ and $f_1^{\rm{NS}}(\tau)$. Therefore, $f_0^{\rm{NS}}(\tau)/(\psi^{\rm NS})^{2c}$ should be a linear polynomial of $\lambda$. With the help of the analytic expression \eqref{analytic type II}, one can read the $S$-transformation rule for $f_0^{\rm{NS}}(\tau)$ and $f_1^{\rm{NS}}(\tau)$,
\begin{align}
\label{S matrix of type II NS}
\begin{split}
\left(
\begin{array}{c}
f_0^{\rm{NS}}(-\frac{1}{\tau}) \\
f_1^{\rm{NS}}(-\frac{1}{\tau})     
\end{array}
\right)
=
\frac{1}{16}
\left(
\begin{array}{cc}
16-2c & \ \ \frac{1}{b_0}(64c-4c^2)\\
b_0 &  \  \ (2c-16)
\end{array}
\right)
\left(
\begin{array}{c}
f_0^{\rm{NS}}(\tau) \\
f_1^{\rm{NS}}(\tau)     
\end{array}
\right)
\end{split}\,.
\end{align}
We demand $64c-4c^2 >0$ to have a symmetric extended matrix, thus $c$ should be smaller than $16$. This constraint is automatically satisfied since we focus on the solutions between $\frac{9}{2} \le c \le \frac{15}{2}$. For this case, one can show that the NS sector partition function take the form of,
\begin{align}
Z_{\rm{NS}}(\tau,\bar{\tau}) = f_0^{\rm{NS}}(\tau) \bar{f}_0^{\rm{NS}}(\bar{\tau}) + \frac{4c(16-c)}{b_0^2}  f_1^{\rm{NS}}(\tau) \bar{f}_1^{\rm{NS}}(\bar{\tau}),
\end{align}
using the $S$-matrix \eqref{S matrix of type II NS}.

We now discuss the bilinear relation which is satisfied by the NS sector solutions in this class. Let us take below three pairs of the non-BPS type $\mathbf{II}$ solutions,
\begin{align}
\label{three pairs}
(c, \tilde{c}=12-c) = \left(\frac{9}{2}, \frac{15}{2}\right), \quad (5, 7), \quad \left(\frac{11}{2}, \frac{13}{2}\right).
\end{align}
Using the analytic expression of the solutions \eqref{analytic type II}, we arrive at the following bilinear relation.
\begin{align}
\label{bilinear of type II}
\begin{split}
f_0^{\rm{NS}} \widetilde{f}_0^{\rm{NS}} + m f_1^{\rm{NS}} \widetilde{f}_1^{\rm{NS}} &= (\psi^{\rm{NS}})^{24}-24 + \left( \frac{4c(12-c)+m b_0 \widetilde{b_0}}{16} - 24 \right) \frac{\lambda}{1-\lambda} \\ 
&= K(\tau) + \left( \frac{4c(12-c)+m b_0 \widetilde{b_0}}{16} - 24 \right) \frac{\lambda}{1-\lambda}\,.
\end{split}
\end{align}
In the above relation, $(\widetilde{f}_0^{\rm{NS}},\widetilde{f}_1^{\rm{NS}})$ denotes the dual pair of $(f_0^{\rm{NS}},f_1^{\rm{NS}})$. Also $b_0$ and $\widetilde{b_0}$ stand for the normalization constant of the solutions $f_1^{\rm{NS}}$ and $\widetilde{f}_1^{\rm{NS}}$, respectively. When the constant $m$ satisfies the constraint $m = \frac{4(96-12c+c^2)}{b_0 \widetilde{b_0}} $, the right hand side of \eqref{bilinear of type II} simply becomes $K(\tau)$. Assuming this constraint is satisfied, the duality relations for three pairs \eqref{three pairs} are given by
\begin{align}
\label{bilinear type II}
\begin{split}
K(\tau) &= f_{0}^{\text{NS}}(q) \tilde f_{0}^{\text{NS}}(q) + 249 f_{1}^{\text{NS}}(q) \tilde f_{1}^{\text{NS}}(q), \quad \text{for} \quad c=\frac{9}{2}, \\
K(\tau) &= f_{0}^{\text{NS}}(q) \tilde f_{0}^{\text{NS}}(q) + \frac{61}{3} f_{1}^{\text{NS}}(q) \tilde f_{1}^{\text{NS}}(q), \quad \text{for} \quad c=5, \\
K(\tau) &= f_{0}^{\text{NS}}(q) \tilde f_{0}^{\text{NS}}(q) + 241 f_{1}^{\text{NS}}(q) \tilde f_{1}^{\text{NS}}(q), \quad \text{for} \quad c=\frac{11}{2}. 
\end{split}
\end{align}

Let us turn our attention to the $\widetilde{\rm{NS}}$ and R sector solutions. To this end, we apply $T$ and $ST$ transformation to \eqref{analytic type II}. First we note that the character of a single free fermion for $\widetilde{\rm{NS}}$ and R sector can be obtained as follows,
\begin{align}
\label{modular transform of the free fermion}
\psi^{\widetilde{\text{NS}}}(\tau) \equiv \psi^{\text{NS}}(\tau + 1) &= \frac{\eta(\frac{\tau}{2})}{\eta(\tau)}, \quad \psi^{\text{R}}(\tau) \equiv \psi^{\text{NS}}\left( -\frac{1}{\tau} + 1\right) = \sqrt{2} \frac{\eta(2\tau)}{\eta(\tau)}\,.
\end{align}
Since the $\lambda$ variable is mapped to $\frac{\lambda}{\lambda-1}$ and  $\frac{\lambda-1}{\lambda}$ under the $T$ and $ST$ transformation respectively, the $\widetilde{\rm{NS}}$ and R sector characters can be expressed as
\begin{align}
\label{NSR characters in lambda}
\begin{split}
f_0^{\widetilde{\rm{NS}}}(\tau) &=  \left( 1 + \frac{c}{8}  \frac{\lambda}{1-\lambda} \right) (\psi^{\widetilde{\rm{NS}}})^{2c}, \qquad f_1^{\widetilde{\rm{NS}}}(\tau) =  b_0 \frac{1}{16} \frac{\lambda}{1-\lambda} (\psi^{\widetilde{\rm{NS}}})^{2c}, \\
f_0^{\rm{R}}(\tau) &= a_0' \frac{2^{3-c}(2-\lambda)}{\lambda}(\psi^{\rm R})^{2c}, \qquad f_1^{\rm{R}}(\tau) = 2^{-c}  b_0' (\psi^{\rm R})^{2c}, 
\end{split}
\end{align}
and their $q$-series agree with that of the Table \ref{type II R} with suitable choice of $a_0'$ and $b_0'$. From \eqref{NSR characters in lambda}, the modular transformation rules for the $\widetilde{\rm{NS}}$ and R sector characters read
\begin{align}
\begin{split}
\left(
\begin{array}{c}
f_0^{\widetilde{\rm{NS}}}(-\frac{1}{\tau})  \\
f_1^{\widetilde{\rm{NS}}}(-\frac{1}{\tau})
\end{array}
\right)
&= 2^{c-8}
\left(
\begin{array}{cc}
   2c/{a_0'}  & \ \ 16(16-c)/{b_0'}  \\
   b_0/{a_0'}  & \ \  -8b_0/{b_0'}
\end{array}
\right)
\left(
\begin{array}{c}
f_0^{{\rm{R}}}({\tau})  \\
f_1^{{\rm{R}}}({\tau})
\end{array}
\right), \\
\left(
\begin{array}{c}
f_0^{{\rm{R}}}(-\frac{1}{\tau})  \\
f_1^{{\rm{R}}}(-\frac{1}{\tau})
\end{array}
\right)
&= 2^{-c}
\left(
\begin{array}{cc}
   8 a_0'  & \ \ 16 (16- c) {a_0'}/{b_0} \\
   {b_0'}  & \ \ -2c  {b_0'}/{b_0}
\end{array}
\right)
\left(
\begin{array}{c}
f_0^{\widetilde{\rm{NS}}}({\tau})  \\
f_1^{\widetilde{\rm{NS}}}({\tau})
\end{array}
\right).
\end{split}
\end{align}
Because the $\widetilde{\text{NS}}$ sector partition function has a form of
\begin{align}
Z_{\widetilde{\rm{NS}}}(\tau,\bar{\tau}) = |f_0^{\widetilde{\rm{NS}}}(\tau)|^2 + \frac{4c(16-c)}{b_0'}  |f_1^{\widetilde{\rm{NS}}}(\tau)|^2,
\end{align}
the R sector partition function is given by
\begin{align}
\begin{split}
Z_{\rm{R}}(\tau,\bar{\tau}) &= Z_{\widetilde{\rm{NS}}}\left(-\frac{1}{\tau},-\frac{1}{\bar{\tau}}\right) = \Big|f_0^{\widetilde{\rm{NS}}}(-\frac{1}{\tau})\Big|^2 + \frac{4c(16-c)}{{b_0}^2}  \Big|f_1^{\widetilde{\rm{NS}}}(-\frac{1}{\tau})\Big|^2 \\
&= \frac{2^{2c-10} c}{{a_0'}^2} |f_0^{\rm{R}}(\tau)|^2 + \frac{2^{2c-4}(16-c)}{{b_0'}^2} |f_1^{\rm{R}}(\tau)|^2. 
\end{split}
\end{align}
Finally, let us comment on the normalization constants $a_0'$ and $b_0'$. We choose integer $a_0'$ and $b_0'$ such that satisfy $\sqrt{2^{2c-10} c} = {a_0'} \sqrt{\mathcal{M}_1}$ and $\sqrt{2^{2c-4}(16-c)} = {b_0'} \sqrt{\mathcal{M}_2}$, where $\frac{\mathcal{M}_1}{n_1^2}, \frac{\mathcal{M}_2}{n_2^2} \notin \mathbbm{Z}$ for any integer $n_1,n_2>1$. Then we absorb $a_0'$ and $b_0'$ into the R sector characters and consider $\mathcal{M}_1$ and $\mathcal{M}_2$ are related to the degeneracy of two characters in the R sector.

\subsubsection*{Non-BPS Type {\bf{III}} : One-parameter family with $c=16$}
The second order MLDE exhibits infinitely many solutions of $c=16$. Here we discuss a relation between these infinite solutions and the characters of the $c=16$ bosonic RCFT without Kac-Moody symmetry that discussed in \cite{Bae:2017kcl}.  The $c=16$ bosonic theory of our interest involves the vacuum and two primaries of weight $h=\frac{1}{2}, 1$. The three characters for the vacuum and primaries are known to have below $q$-expansion.
\begin{align}
\label{c=16 character}
\begin{split}
f_{h=0}(q) &= q^{-2/3} \left( 1 + 2296 q^2 + 65536 q^3 + 1085468 q^4 + \cdots \right),  \\
f_{h=1}(q) &= q^{1/3} \left( 1+ 136 q + 4132 q^2 + 67712 q^3 + \cdots \right),  \\
f_{h=\frac{3}{2}}(q) &= q^{5/6} \left( 1+ 52 q + 1106 q^2 + 14808 q^3 + \cdots \right).
\end{split}
\end{align}

We find that the infinitely many solutions with $c=16$ are  one-parameter family. More precisely, the  solutions of non-BPS type {\bf{III}} can be expressed as follows.
\begin{align}
\label{fermionic characters c=16}
\begin{split}
f^{\rm NS}_0(q) &= \left( f_{h=0}(q) + 256 f_{h=\frac{3}{2}}(q) \right) + n \left( f_{h=1}(q) + 16 f_{h=\frac{3}{2}}(q)  \right), \quad n \in \mathbbm{Z}_{\ge 0}, \\
f^{\rm NS}_1(q) &= f_{h=1}(q) + 16 f_{h=\frac{3}{2}}(q), \\
f^{\widetilde{\rm NS}}_0(q) &= \left( f_{h=0}(q) - 256 f_{h=\frac{3}{2}}(q) \right) + n \left( f_{h=1}(q) - 16 f_{h=\frac{3}{2}}(q)  \right), \quad n \in \mathbbm{Z}_{\ge 0}, \\
f^{\widetilde{\rm NS}}_1(q) &= f_{h=1}(q) - 16 f_{h=\frac{3}{2}}(q), \qquad f^{\rm R}_0(q) = f_{h=1}(q), \qquad f^{\rm R}_1(\tau) = f_{h=\frac{3}{2}}(q).
\end{split}
\end{align}
Let us see how the characters \eqref{fermionic characters c=16} form $\Gamma_\theta$ invariant partition functions as well as the consistent fusion rule algebra. To this end, it is necessary to find the $S$-matrices of \eqref{fermionic characters c=16}. Note that the characters \eqref{c=16 character} ought to solve the third order differential equation with $\ell=0$ \footnote{We thank Sunil Mukhi for letting us know the analytic solution of the third order MLDE for $\text{SL}_2(\mathbbm{Z})$.} \cite{Mathur:1988gt},
\begin{align}
\label{MDE lambda}
\begin{split}
&\lambda^2 (1-\lambda)^2 \partial_\lambda^3 \tilde{f}(\lambda) - (6\alpha_0+2) \lambda (1-\lambda) (2\lambda-1) \partial_\lambda^2 \tilde{f}(\lambda) \\
&+ \left[ 12\alpha_0^2 + 2\alpha_0 + \nu_1 - 2 - \lambda(1-\lambda)(48\alpha_0^2+20\alpha_0+\nu_1) \right] \partial_\lambda \tilde{f}(\lambda) \\
&+ \left[24\alpha_0^2+6(2-\nu_1)\alpha_0-9\nu_2 \right](2\lambda-1) \tilde{f}(\lambda) = 0,
\end{split}
\end{align}
where $\tilde{f}(\lambda) \equiv \left( \frac{\lambda(1-\lambda)}{16}\right)^{\frac{c}{12}} f(\lambda)$ and $\alpha_i = h_i - \frac{c}{24}$. The parameters $\nu_i$ are expressed in terms of $\alpha_i$ as follows,
\begin{align}
\nu_1  = 2 + 4\alpha_0 \alpha_1 + 4\alpha_0 \alpha_2  + 4\alpha_1 \alpha_2 , \quad \nu_2  = - 4\alpha_0 \alpha_1 \alpha_2 .
\end{align}
We find that the NS, $\widetilde{\text{NS}}$ and R sector characters can be expressed as linear combination of the three independent solutions of \eqref{MDE lambda} with $\alpha_0 = -\frac{2}{3}, \alpha_1 = -\frac{1}{3}, \alpha_2 = \frac{1}{6}$. Explicitly,
\begin{align}
\begin{split}
f_{h=0}(q) &= \left( 1 + \frac{1}{32} \lambda (\lambda^3 -2 \lambda^2 + 34 \lambda -64)\right) \left( \frac{\lambda(1-\lambda)}{16}\right)^{-\frac{4}{3}},\\
f_{h=1}(q) &= \frac{1}{512}\lambda^2(\lambda^2-2\lambda+2)\left( \frac{\lambda(1-\lambda)}{16}\right)^{-\frac{4}{3}},\\
f_{h=\frac{3}{2}}(q) &= - \frac{\lambda^3(\lambda-2)}{8192}  \left( \frac{\lambda(1-\lambda)}{16}\right)^{-\frac{4}{3}},
\end{split}
\end{align}
thus the fermionic characters \eqref{fermionic characters c=16} can be written in terms of the $\lambda$ variable.

Using the $S$ transformation rule of the $\lambda$ variable, i.e., $\lambda \rightarrow 1-\lambda$, it is straightforward to read the $S$-matrices for the characters in the NS, $\widetilde{\text{NS}}$ and R sector. We find that the NS sector characters are transformed as
\begin{align}
\begin{split}
\left(
\begin{array}{c}
 f_0^{\rm NS}(-\frac{1}{\tau})   \\
 f_1^{\rm NS}(-\frac{1}{\tau}) 
\end{array}
\right) = \frac{1}{256}
\left(
\begin{array}{cc}
 (n+16)  & \ \ -(n+272)(n-240) \\
   1  & \ \ -(n+16)
\end{array}
\right)
\left(
\begin{array}{c}
 f_0^{\rm NS}(\tau)   \\
 f_1^{\rm NS}(\tau) 
\end{array}
\right),
\end{split}
\end{align}
while the transformation rules for the $\widetilde{NS}$ and $R$ sector characters read
\begin{align}
\begin{split}
\left(
\begin{array}{c}
 f_0^{\widetilde{\rm NS}}(-\frac{1}{\tau})   \\
 f_1^{\widetilde{\rm NS}}(-\frac{1}{\tau}) 
\end{array}
\right) = 
\left(
\begin{array}{cc}
 (n+272)  & \ \ -16(n-240) \\
  1  & \ \ -16
\end{array}
\right)
\left(
\begin{array}{c}
 f_0^{\rm R}(\tau)   \\
 f_1^{\rm R}(\tau) 
\end{array}
\right),\\
\left(
\begin{array}{c}
 f_0^{\rm R}(-\frac{1}{\tau})   \\
 f_1^{\rm R}(-\frac{1}{\tau}) 
\end{array}
\right) = \frac{1}{8192}
\left(
\begin{array}{cc}
 16  & \ \ -16(n-240) \\
  1  & \ \ -(n+272)
\end{array}
\right)
\left(
\begin{array}{c}
 f_0^{\widetilde{\rm NS}}(\tau)   \\
 f_1^{\widetilde{\rm NS}}(\tau) 
\end{array}
\right).
\end{split}
\end{align}
There should be a symmetric extended matrix to have the consistent fusion rule algebra. If $n\ge 240$, the off-diagonal component becomes 0 or negative thus one cannot find an extended matrix. In other words, the existence of the consistent fusion rule algebra constrains $n < 240$. This indicate that only finite number of solutions have the consistent fusion rule algebra and can be considered as the characters of certain RCFT.

\subsubsection*{Non-BPS Type {\bf{IV}} : Single-character theories}

We now focus on the class of the single-character solutions. For the solutions in this class, the vacuum solutions are allowed to have the non-negative integer coefficients. In contrast, the other solutions exhibit rational coefficients \textit{with unbounded denominators}, thus we do not pay much attention to these solutions. We will show that the vacuum solutions of the non-BPS type {$\bf{IV}$} are invariant under $S$ transformation by themselves.

In principle, the single-character of our interest  could appear as the solution of  the first order MLDE with higher $\ell$. In some cases, the first order MLDE can be lifted to the higher order MLDE without poles and the single-character can be identified to the vacuum solution of the higher order MLDE with $\ell =0$. To illustrate this point, we revisit the character of $(\mathfrak{e}_8)_1$ WZW model. It has been argued that the character of this theory appears as a solution with non-negative integer coefficients of the second order MLDE \cite{Mathur:1988gt}
\begin{equation}
\label{eq:MLDE_E8}
    \left(\mathcal D^2-\frac{E_4}{6}\right)f=0\,.
\end{equation}
On the other hand, the other independent solution of the above MLDE is characterized by rational coefficients in the $q$-expansion, hence cannot be interpreted as a character of certain RCFT. Because the vacuum solution is modular invariant by itself, $(\mathfrak{e}_8)_1$ WZW model can be understood as a single-character theory with $f_0(\tau)=j^{1/3}=E_4/\eta^8$. 

As a single character theory, the character $f_0(\tau)$ has to be solution to a first order MLDE given in \cite{Mathur:1988na}, 
\begin{equation}
\label{eq:1MLDE_E8}
    \left(\mathcal D+\frac{1}{3}\frac{E_6}{E_4}\right)f=0\,,
\end{equation}
where the Wronskian have a zero of order two. Given the equation (\ref{eq:1MLDE_E8}), the following identity is true, 
\begin{equation}
\label{eq:newMLDE_E8}
    \left(\mathcal D+H\right)\left(\mathcal D+\frac{1}{3}\frac{E_6}{E_4}\right)f=0,
\end{equation}
for any meromorphic modular form $H$. For a generic choice of $H$, the above second order MLDE will be characterized by an arbitrary complicated pole structure in the Wronskian. Note that the choice of $H$ is sensitive to the data of the second independent solution of (\ref{eq:newMLDE_E8}). Once we choose $H=-\tfrac{E_6}{3E_4}$, then \eqref{eq:newMLDE_E8} is reduced to the second order MLDE with $\ell=0$, namely \eqref{eq:MLDE_E8}. In conclusion, the unphysical solution of \eqref{eq:MLDE_E8} can be considered as a remnant of lifting the first order MLDE to the second order MLDE without poles. For the solutions in the non-BPS type $\bf{IV}$ class, we conjecture that a pole structure in the first order MLDE can be trivialized by lifting it to a higher order MLDE.

Now let us describe the NS sector solutions of the non-BPS type {\bf{IV}} in detail. To have non-negative integer coefficients in the vacuum solution, the central charges are restricted to 32 values listed below,
\begin{align}
\label{c for class II}
c = 8, \frac{17}{2}, 9, \frac{19}{2}, 10, \cdots, 23, \frac{47}{2}.
\end{align}
Among them, eight solutions with $c = \frac{17}{2}, 9, \cdots, 12$ are presented in \cite{hoehn2007selbstduale}. The explicit form of the solutions in $q$-series are summarized in Table \ref{type V solution NS} and \ref{type V solution R}. The NS sector vacuum character can be expressed in terms of the NS partition function of a free Majorana-Weyl fermion $\psi^{\text{NS}}$ as follow. 
\begin{align}\label{qwe}
\begin{split}
&f^{\rm NS}_0(\tau) = (\psi^{\text{NS}})^{2c} - 2c (\psi^{\text{NS}})^{2c-24}, \quad \text{for} \ \ c = 8, \frac{17}{2}, \cdots, \frac{47}{2}.
\end{split}
\end{align}
Because $\psi^{\text{NS}}$ is invariant under the $\Gamma_\theta$, it is straightforward to see that $f^{\text{NS}}_0(\tau)$ is a weight-zero modular form of $\Gamma_\theta$. Therefore $f^{\text{NS}}_0(\tau)$ can be interpreted as the left-moving $NS$ sector partition function of single-character theories. This character does not have any free fermion as
\begin{equation}\label{NS32}
    f^{\rm NS}_0(\tau)= q^{-\frac{c}{24}}\Big[1+ c(47-2c)q+ \frac{8c}{3}(100-(c-18)^2)q^\frac32+\cdots\Big].
\end{equation}
The coefficients of $f_0^{\rm NS}$ being non-negative integer leads to the allowed value of   the central charge  to be $8, \frac{17}{2},...   \frac{47}{2}$. Note that for $c=8$, the coefficient of $q$ to half integer power vanishes and so the theory is not fermionic. The theory is not new one. One can show easily that $f^{\rm NS}_0$ for $c=8$ is identical to the character $j^{1/3}$ of the $({\mathfrak e}_8)_1$ WZW model. For $c=12$, the character $f^{\rm NS}_0(\tau)$ becomes the NS character $K(\tau)$ of the $c=12$ ${\rm Co}_0$ SCFT.

To explore the $\widetilde{\rm{NS}}$ and R sector partition function, we take modular transformation to $\psi^{\text{NS}}$. After all, the analytic structure of $\widetilde{\rm{NS}}$ sector and R sector vacuum solutions are given by
\begin{align}
\label{qwe2}
\begin{split}
f^{\widetilde{\rm NS}}_0(\tau) &= (\psi^{\widetilde{\text{NS}}}(\tau))^{2c} - 2c (\psi^{\widetilde{\text{NS}}}(\tau))^{2c-24}, \\
f^{\rm R}_0(\tau) &= (\psi^{\text{R}}(\tau))^{2c} + 2c (\psi^{\text{R}}(\tau))^{2c-24}.
\end{split}
\end{align}
In the $q$-series expansion, $f^{\rm R}_0$ becomes 
\begin{equation}
    f^{\rm R}_0(\tau)= q^{\frac{c}{12}-1}\Big[2^{-11+c}c + 2^{-10+c}(1024-12c+c^2)q+ 2^{-11+c}c(4372-47c+2c^2)q^2 \cdots\Big].
\end{equation}
As recently discussed in \cite{Benjamin:2020zbs}, the R sector constraint could be stronger than that of the NS sector. To demonstrate this aspect, we present the $q$-expansion of $f^{R}_0(\tau)$ for  $c=\frac{19}{2}$, $c=10$ and $c=11$ as the illustrative examples.
\begin{align}
\begin{split}
f^{\rm R}_0(\tau)&=  \sqrt{2} q^{-\frac{5}{24}} \left(\frac{19}{8}  + \frac{4001}{8} q + \frac{78014}{8} q^2 + \cdots \right), \quad \text{for} \ \ c=\frac{19}{2}, \\
f^{\rm R}_0(\tau)&= q^{-\frac{1}{6}} \left(5  + 1004 q + 20510 q^2 + \cdots \right),  \quad \text{for} \ \ c=10, \\
f^{\rm R}_0(\tau)&= q^{-\frac{1}{12}} \left(11  + 2026 q + 45067 q^2 + \cdots \right),  \quad \text{for} \ \ c=11.
\end{split}
\end{align}
For $c=\frac{19}{2}$, the coefficients in $q$-series cannot be the integer values  up to $\sqrt{2}$.
For the identical reason,  we exclude the theories $c= \frac{17}{2},  \frac{19}{2},\frac{21}{2},\frac{23}{2}$.
The R sector character for $c=9$ is excluded as the coefficients are fractional. Only acceptable R sector solutions for $c<12$ are the $c=10,11$ cases. 
  Note that the unitarity bound $h \ge \frac{c}{24}$ is violated when $c<12$,  thus the solutions for $c= 10, 11$   cannot be considered as the vacuum character of an unitary supersymmetric RCFT. For $c\ge 12$, there is no constraint from the R-sector. For all surviving theories with $c=12,12+\frac12,13,13+\frac12,...,23,23+\frac12$, the coefficient of  $q^\frac32$ in\eqref{NS32} is positive and so these fermionic theories are expected to have a ${\mathcal N}=1$ supersymmetric  with the broken supersymmetric vacuum in the R-sector. Thus we can  assume $f_0^{\tilde R}=0$. Especially one with $c=47/2$ is the fermionic CFT for the baby Monster CFT \cite{Lin:2019hks}. 

We argue that the GSO projection of the characters \eqref{qwe} and \eqref{qwe2} leads to a three-character bosonic CFT with $h=0, \frac{3}{2}, \frac{c}{8}-1$ for $c\ge12$ with $\ell=0$.\footnote{To have unitary bosonic CFTs, the central charge should be larger than eight.}. Some of these solutions with $\ell=0$ appeared in  \cite{Gaberdiel:2016zke}. One can imagine their dual bosonic theories of central charge $24-c$ and conformal weights $h=0,\frac12,3-\frac{c}{8}$ were found to be  $SO(48-2c)_1$ WZW model whose fermionic version made of $48-2c$ free fermions has    characters $(\psi^{\rm NS})^{48-c}, (\psi^{\widetilde{\rm NS}})^{48-c}$, and $ (\psi^{\rm R})^{48-c} $.   The bilinear relation between the pair of these theories become 
\begin{equation} \frac12 \Big[ (\psi^{\rm NS})^{48-2c}f_0^{\rm NS}    + (\psi^{\widetilde{\rm NS}})^{48-2c}  f_0^{\widetilde{\rm NS}}    + (\psi^{\rm R})^{48-2c}   f_0^{\rm R}(\tau)\Big]  = J(\tau)+ 48 (\frac{47}{2}-c). \end{equation}
For $c=23+\frac12$, the above relation corresponds  exactly the  fermionization of $c=24$ Monster CFT with 2A ${\mathbb Z}_2$ symmetry in terms of the fermionic Baby Monster CFT and a single Majorana fermion \cite{Lin:2019hks}.

Let us explore the bosonic theory a bit more.
The three characters should be realized as the independent solutions of the third order MLDE with $\ell = 0$, namely \eqref{MDE lambda}. With the parametrization
\begin{align}
\alpha_0 = -\frac{c}{24} , \quad \alpha_1 = -\frac{c}{24} + \frac{3}{2}, \quad  \alpha_2 = \frac{c}{12}-1,
\end{align}
we find that the analytic structure of three independent solutions are given by
\begin{align}
\label{solutions of bosonic 3rd MLDE}
\begin{split}
g_0(\lambda)&=\frac{1}{8} \left( \frac{\lambda(1-\lambda)}{16} \right)^{-\frac{c}{12}}(8-c\lambda+c \lambda^2), \\
g_1(\lambda)&=\frac{1}{8} \left( \frac{\lambda(1-\lambda)}{16} \right)^{-\frac{c}{12}}(1-\lambda)^{\frac{c}{4}-2}(8+(c-16)\lambda+8 \lambda^2), \\
g_2(\lambda)&=\frac{1}{8} \left( \frac{\lambda(1-\lambda)}{16} \right)^{-\frac{c}{12}}\lambda^{\frac{c}{4}-2}(c-c\lambda+8 \lambda^2),
\end{split}
\end{align}
and they agree with the $f_0^{\rm NS}(\tau)$, $f_0^{\widetilde{\rm NS}}(\tau)$ and $f_0^{\rm R}(\tau)$  respectively. Furthermore, the three characters of the bosonic CFT can be expressed as
\begin{align}
\label{character relation}
f_0(\lambda) = \frac{1}{2}(g_0(\lambda) + g_1(\lambda)), \quad f_1(\lambda) = \frac{1}{2}(g_0(\lambda) - g_1(\lambda)), \quad f_2(\lambda) = g_2(\lambda).
\end{align}
For $c=12$, the above three characters reproduce the characters of self-dual RCFT discussed in  \cite{Bae:2018qfh}. The modular transformation rule of $f_0(\lambda), f_1(\lambda), f_2(\lambda)$ for general central charge is given by
\begin{align}
\begin{split}
\left(
\begin{array}{c}
f_0(1-\lambda)   \\
f_1(1-\lambda)   \\
f_2(1-\lambda)
\end{array}
\right)
=
\left(
\begin{array}{ccc}
\frac{1}{2} & \frac{1}{2} &  \frac{1}{2} \\
\frac{1}{2} & \frac{1}{2} & -\frac{1}{2} \\
1 & -1 &  0
\end{array}
\right)
\left(
\begin{array}{c}
f_0(\lambda)   \\
f_1(\lambda)   \\
f_2(\lambda)
\end{array}
\right),
\end{split}
\end{align}
therefore the modular invariant partition function reads
\begin{align}
\label{single character partition function}
Z(\tau, \bar{\tau}) = |f_0(\tau)|^2 + |f_1(\tau)|^2 + \frac{1}{2} |f_2(\tau)|^2.
\end{align}

Substituting \eqref{character relation} into \eqref{single character partition function}, one can recast the partition function as follows,
\begin{align}
\label{GSO for single character}
\begin{split}
Z(\tau, \bar{\tau}) &= \frac{1}{2} f_0^{\rm NS}(\tau) \bar{f}_0^{\rm NS}(\bar{\tau}) +  \frac{1}{2} f_0^{\widetilde{\rm NS}}(\tau) \bar{f}_0^{\widetilde{\rm NS}}(\bar{\tau}) +  \frac{1}{2} f_0^{\rm R}(\tau) \bar{f}_0^{\rm R}(\bar{\tau}) \\
                    &= \frac{1}{2} Z_{\text{NS}} + \frac{1}{2}  Z_{\widetilde{\text{NS}}} + \frac{1}{2} Z_{\text{R}}.
\end{split}
\end{align}
In the second line, we use the fact of that the partition functions of each spin structure can be factorized.
Under the assumption of vanishing $Z_{\widetilde{\text{R}}}$,  the GSO projection \eqref{GSO for single character} of single-character fermionic theory leads to the three-character bosonic CFT with $h=0, \frac{3}{2}, \frac{c}{8}-1$ only when $c\ge12$. 

\section*{Acknowledgment} We would like to thank  Changrim Ahn, Gil Young Cho, Dongmin Gang, Hee-Cheol Kim and Sunil Mukhi for discussions. 
The work of J.B. is supported by the European Research Council(ERC) under the European Union's Horizon 2020 research and innovation programme (Grant No. 787185). J.B. also acknowledge the hospitality of the Korea Institute for Advanced Study where this work was completed.  Z.D., K.L., S.L., and M.S. are supported by KIAS Individual Grant PG076901, PG006904, PG056502, and PG064201. K.L. is also supported in part by National Research Foundation Individual Grant NRF-2017R1D1A1B06034369. 

\appendix


\section{${\rm SL}_2({\mathbb Z})$ Modular Group  and Forms}
\label{sec:conventions}

In this appendix, we present the definitions and properties of various modular forms that appeared in the main text. In what follows, we denote $\tau$ as a point in the Poincar\'e upper half plane $\mathbb H$, and $q=\exp(2i\pi\tau)$.  
\paragraph{The Eisenstein series} 

The Eisenstein series of weight $2k$, $k>1$, is defined as
\begin{align}
E_{2k}(\tau) = \sum_{(m,n) \in \mathbbm{Z}^2 \backslash \{(0,0)\}} \frac{1}{(m + n\tau)^{2k}},
\end{align}
and it satisfies the property
\begin{align}
\begin{split}
E_{2k}(\gamma \tau) &= (c\tau + d)^{2k} E_{2k}(\tau), \qquad \text{for} \ \ k \ge 2 \ \ \text{and} \ \ \gamma \in \text{SL}_2(\mathbbm{Z)},
\end{split}
\end{align}
In particular, the Eisenstein series of weight two, four and six are given by
\begin{align}
\begin{split}
E_2(\tau)&=  1-24\sum_{n=1}^\infty \frac{nq^n}{1-q^n} = 1 - 24 q - 72 q^2 - 96 q^3 - 168 q^4 - 144 q^5 + \cdots , \\
E_4(\tau)&=  1+240\sum_{n=1}^\infty \frac{n^3q^n}{1-q^n} = 1 + 240 q + 2160 q^2 + 6720 q^3 + 17520 q^4 + \cdots , \\
E_6(\tau)&= 1-504\sum_{n=1}^\infty \frac{n^5q^n}{1-q^n} = 1 - 504 q - 16632 q^2 - 122976 q^3 - 532728 q^4 + \cdots .
\end{split}
\end{align}
The Eisenstein series $E_4(\tau)$ and $E_6(\tau)$ are modular forms of weight four and six, respectively. On the other hand, the modular transformation rule of the weight two Eisenstein series is given by 
\begin{align} 
E_{2}(\gamma \tau) = (c \tau + d)^2 E_{2}(\tau) + \frac{\pi i}{6} c (c\tau+d),
\end{align}
thus it is referred to as the quasi-modular form. It enters in the definition of the Ramanujan-Serre covariant derivative to be defined below.

\paragraph{The Dedekind eta function} 

The Dedekind eta function is a modular form of weight 1/2 and defined by the infinite product
	\begin{equation}
	\eta(\tau)=q^{\frac{1}{24}}\prod_{n=1}^{\infty}\left(1-q^{n}\right)\,.
	\end{equation}
Under $T$ and $S$, the $\eta$ function transforms as follows:
\begin{align}
    \eta(\tau+1)=e^{\frac{\pi i}{12}}\eta(\tau), \ \ \eta(-1/\tau) = \sqrt{-i\tau}\eta(\tau).
\end{align}

\paragraph{The Jacobi theta functions} 

The theta functions with vanishing elliptic argument are given by the following infinite products:
\begin{align}
\begin{split}
  \vartheta_2(\tau) &=  \theta_2(\tau,0)  =2 q^\frac18 \prod^\infty_{n=1}(1-q^n)(1+q^n)^2  = 2 q^{\frac{1}{8}} + 2 q^{\frac{9}{8}} + 2 q^{\frac{25}{8}} + 2 q^{\frac{49}{8}} + \cdots, \\
\vartheta_3(\tau)&=\theta_3(\tau,0) = 
\prod^\infty_{n=1}(1-q^n)(1+q^{n-\frac12}  )^2 = 1 + 2q^{\frac{1}{2}} + 2q^2 + 2 q^{\frac{9}{2}} + 2q^8 + \cdots, \\
\vartheta_4(\tau)& =\theta_4(\tau,0) =
\prod^\infty_{n=1}(1-q^n)(1-q^{n-\frac12}  )^2= 1 - 2q^{\frac{1}{2}} + 2q^2 - 2 q^{\frac{9}{2}} + 2q^8 + \cdots.
\end{split}
\end{align}
Under  $S$ and  $T$ transformations, the theta functions transform as,
\begin{align}
    &   \vartheta_2(\tau+1)= \sqrt{i}\vartheta_2(\tau), \ \vartheta_3(\tau+1)= \vartheta_4(\tau), \ \vartheta_4
    (\tau+1)=    \vartheta_3(\tau), \nonumber \\
  \vartheta_2(-1/\tau)&= \sqrt{-i\tau}\vartheta_4(\tau), \ \vartheta_3(-1/\tau)= \sqrt{-i\tau} \vartheta_3(\tau), \ \vartheta_4(-1/\tau  )=  \sqrt{-i\tau}\vartheta_2(\tau).
\end{align}

\paragraph{Useful identities} 

We summarize some useful identities satisfied by the above defined modular forms:
\begin{align}
\begin{split}
\vartheta_3^4(\tau)&=\vartheta_2^4(\tau)+\vartheta_4^4(\tau)\,, \quad  \eta^3(\tau)=\frac12 \vartheta_2(\tau)\vartheta_3(\tau)\vartheta_4(\tau)\,, \\
   E_4(\tau) &=\frac12 (\vartheta_2^8+\vartheta_3^8+\vartheta_4^8)\,, \quad 
   E_6(\tau) =\frac12 (\vartheta_2^4+\vartheta_3^4) (\vartheta_3^4+\vartheta_4^4)(\vartheta_4^4-\vartheta_2^4)\,.
\end{split}
\end{align}
The derivative of eta function with respect to  $\tau$ is given by 
\begin{align}\label{eq:derivative_eta} \frac{1}{2\pi i} \frac{\text{d}}{\text{d}\tau} \eta(\tau) = \frac{1}{24} E_2(\tau)\eta(\tau). \end{align}
The Ramanujan-Serre covariant derivative is then defined as
\begin{align}
    {\mathcal{D}}_k=\frac{1}{2\pi i}\frac{\text{d}}{\text{d}\tau} -\frac{k}{12} E_2(\tau),
\end{align} 
and it sends a weight $k$ modular form to a modular form of weight $k+2$. By (\ref{eq:derivative_eta}), one can see that the covariant derivative of the eta function vanishes. The Eisenstein series $E_2,E_4,E_6$ satisfy the Ramanujan identities
\begin{align}
    \mathcal{D}_2 E_2= -\frac{1}{12}E_4,\ \ \mathcal{D}_4 E_4=-\frac13 E_6, \ \ \mathcal{D}_6 E_6=-\frac12 E_4^2.
\end{align}
The Klein $j$-invariant is the unique holomorphic modular function on $\mathbb{H}$ with a simple pole at $\tau = i\infty$. It can be expressed in terms of the Eisenstein series as follows:
\begin{align}
\begin{split}
j(\tau) &=\frac{E_4(\tau)^3}{\eta(\tau)^{24}} =  \frac{1728E_4(\tau)^3}{E_4(\tau)^3-E_6(\tau)^2 }\,. \\
        &= q^{-1}+ 744+196884 q   +21493760 q^2 +864299970 q^3+   20245856256 q^4 + \cdots
\end{split}
\end{align}
%

\section{Modular Matrix of the Fermionic Second Order MLDE} \label{App:A}

In this appendix, we provide a closed-from expression for the NS sector solutions of the second order MLDE \eqref{MDEs by lambda} and their $S$-matrix. Since the differential equation for $f^{\rm{NS}}(\lambda)$
\begin{align}
\label{Riemann}
\begin{split}
    \Big[ \frac{\text{d}^2}{\text{d}\lambda^2} +   \frac{2(1+3\mu_1)(1 -2\lambda)}{3\lambda(1-\lambda)} \frac{\text{d}}{\text{d}\lambda} +  \frac{4(\mu_2 + \mu_3)  -4\mu_3\lambda(1-\lambda)   }{\lambda^2(1-\lambda)^2}  \Big] f^{\rm{NS}}(\lambda) &=0,
\end{split}
\end{align}
takes the form of Riemann's differential equation, the solutions of \eqref{Riemann} can be written in terms of the Riemann's $P$-symbol \cite{Naculich:1988xv}
\begin{align}
\begin{split}
f^{\rm{NS}}(\lambda) = P
   \left\{
   \begin{array}{ccc|c}
   0 & 1 & \infty & \\[1em]
   -\frac{c}{12} & -\frac{c}{12} & \alpha^{+} & \lambda  \\[1em]
   2h^{\rm{NS}}-\frac{c}{12} & 2h^{\rm{NS}}-\frac{c}{12} & \alpha^{-} & \\
   \end{array}
   \right\},
\end{split}
\end{align}
where
\begin{align}
\alpha^{\pm} = \frac{1}{12} \left( 6 + 2c - 24 h^{\rm{NS}} \pm \sqrt{4-a_1 + 2c-32 h^{\rm{NS}} + 2 a_1 h^{\rm{NS}} - 4 c h^{\rm{NS}} + 64 (h^{\rm{NS}})^2}\right).
\end{align}
To cast the above solution into a form involving the hypergeometric function $\left._2F_1\right.$, we use the following identity satisfied by Riemann's $P$-symbol: 
\begin{align}
\label{P-symbol identity}
\begin{split} 
   \lambda^{\sigma} (1-\lambda)^{\rho}P
   \left\{
   \begin{array}{ccc|c}
   0 & 1 & \infty & \\
   \alpha_1 & \beta_1 & \gamma_1 & \lambda  \\
   \alpha_2 & \beta_2 & \gamma_2 & \\
   \end{array}
   \right\}
   =
   P
   \left\{
   \begin{array}{ccc|c}
   0 & 1 & \infty & \\
   \alpha_1 + \sigma & \beta_1 + \rho & \gamma_1 - \sigma - \rho & \lambda  \\
   \alpha_2 + \sigma & \beta_2 + \rho& \gamma_2 - \sigma - \rho & \\
   \end{array}
   \right\}
\end{split}
\end{align}
After redefining the NS sector character by
\begin{align}
f^{\rm{NS}}(\lambda) = \lambda^{-\frac{c}{12}} (1-\lambda)^{-\frac{c}{12}} \widetilde{f}^{\rm{NS}}(\lambda),
\end{align}
and using the identity \eqref{P-symbol identity}, one can find
\begin{align}
\label{P symbol redefine}
\begin{split}
\widetilde{f}^{\rm{NS}}(\lambda) = P
   \left\{
   \begin{array}{ccc|c}
   0 & 1 & \infty & \\[0.5em]
   0 & 0 & \beta^{+} & \lambda  \\[0.5em]
   2h^{\rm{NS}} & 2h^{\rm{NS}} & \beta^{-} & \\
   \end{array}
   \right\},
\end{split}
\end{align}
where $\beta^{\pm} = \alpha^{\pm} - \frac{c}{6}$. It is known that the Riemann's $P$-symbol of the form \eqref{P symbol redefine} solves the ordinary hypergeometric differential equation. Therefore, we obtain the closed-form expression of NS sector characters in terms of the  hypergeometric function:
\begin{align}
\begin{split}
f^{\rm{NS}}_0(\lambda) &= 2^{\frac{c}{3}}\lambda^{-\frac{c}{12}} (1-\lambda)^{-\frac{c}{12}} {_2F_1}\left(\beta^+,\beta^-;1-2h^{\rm{NS}} ; \lambda \right) \\
f^{\rm{NS}}_1(\lambda) &= b_0 2^{\frac{c}{3}-8h^{\rm{NS}}} \lambda^{2h^{\rm{NS}}-\frac{c}{12}} (1-\lambda)^{-\frac{c}{12}} {_2F_1}\left(\beta^+ + 2h^{\rm{NS}},\beta^- + 2h^{\rm{NS}};1+2h^{\rm{NS}} ; \lambda \right)
\end{split}
\end{align}
Here $b_0$ is overall normalization constant.\footnote{We fix the overall factor $2^{\frac{c}{3}}$ in the vacuum character by $\chi^{\rm{NS}}_0(\lambda) = q^{-\frac{c}{24}} + \cdots $.}

The modular matrix of vector-valued modular form $(f^{\rm{NS}}_0(\lambda), f^{\rm{NS}}_1(\lambda))$ can be found by Gauss identity of hypergeometric function
\begin{align}
\begin{split}\label{appB:gauss1}
{_2F_1}&\left(\alpha_1,\alpha_2;\alpha_3;1-\lambda \right) = \frac{\Gamma(\alpha_3)\Gamma(\alpha_3-\alpha_1-\alpha_2)}{\Gamma(\alpha_3-\alpha_1)\Gamma(\alpha_3-\alpha_2)} {_2F_1}\left(\alpha_1,\alpha_2;\alpha_1+\alpha_2-\alpha_3+1;\lambda \right) \\
& + \frac{\Gamma(\alpha_3)\Gamma(\alpha_1+\alpha_2-\alpha_3)}{\Gamma(\alpha_1) \Gamma(\alpha_2)} \lambda^{\alpha_3-\alpha_1-\alpha_2} {_2F_1}\left(\alpha_3-\alpha_1,\alpha_3-\alpha_2;\alpha_3-\alpha_1-\alpha_2+1;\lambda \right),
\end{split}
\end{align}
and Euler transformation formula of hypergeometric function
\begin{align}\label{appB:gauss2}
{_2F_1}(\alpha,\beta; \gamma;\lambda) = (1-\lambda)^{\gamma-\alpha-\beta} {_2F_1}\left(\gamma-\alpha,\gamma-\beta;\gamma;\lambda\right).
\end{align}
After some algebra, one can read the transformation rule of NS sector characters.
\begin{align}
\label{S-matrix of 2nd SMDE}
\begin{split}
\left(
\begin{array}{c}
f^{\rm{NS}}_0(1-\lambda)    \\[0.5em]
f^{\rm{NS}}_1(1-\lambda)
\end{array}
\right)
=
\left(
\begin{array}{cc}
\frac{\Gamma(1-2h^{\rm{NS}}) \Gamma(2h^{\rm{NS}})}{\Gamma(1-2h^{\rm{NS}}-\beta^+)\Gamma(1-2h^{\rm{NS}}-\beta^-)} & \frac{2^{8h^{\rm{NS}}}}{b_0} \frac{\Gamma(1-2h^{\rm{NS}})\Gamma(-2h^{\rm{NS}})}{\Gamma(\beta^+) \Gamma(\beta^-)}    \\[0.5em]
\frac{b_0}{2^{8h^{\rm{NS}}}} \frac{\Gamma(1+2h^{\rm{NS}})\Gamma(2h^{\rm{NS}})}{\Gamma(1-\beta^{+})\Gamma(1-\beta^{-})} & \frac{\Gamma(1+2h^{\rm{NS}})\Gamma(-2h^{\rm{NS}})}{\Gamma(\beta^+ +2h^{\rm{NS}})\Gamma(\beta^- + 2h^{\rm{NS}})}
\end{array}
\right)
\left(
\begin{array}{c}
f^{\rm{NS}}_0(\lambda)    \\[0.5em]
f^{\rm{NS}}_1(\lambda)
\end{array}
\right) \nonumber
\end{split}
\end{align}

\section{Solutions of the Fermionic Second Order MLDE in $q$-series}
\label{App:B}

\begin{table}[ht]
\centering
\scalebox{1.0}{
\begin{tabular}{c | c | c | c}
\hline
 $c$  & $h^{\rm{NS}}$ & ($\mu_1, \mu_2, \mu_3$) & NS sector character  \\
\hline
\multirow{2}{*}{1} &  $0$ & \multirow{2}{*}{($\frac{1}{12}, -\frac{1}{192}, 0$)} & $ q^{-\frac{1}{24}} \left(1 + q + 2 q^{3/2} +2 q^2 +2 q^{5/2} +3 q^3  + \cdots \right)$   \\[0.3em]
    &  $\frac{1}{6}$ &   & $ q^{\frac{1}{8}}  \left(1+q^{1/2}+q+q^{3/2}+2 q^2+3 q^{5/2}+3 q^3 + \cdots \right) $  \\[0.3em]
    \hline
\multirow{2}{*}{$\frac{9}{4}$} & $0$ & \multirow{2}{*}{($\frac{5}{48},-\frac{15}{1024},0$)} & $q^{-\frac{3}{32}} \left(1+3 q+7 q^{3/2}+9 q^2+12 q^{5/2}+22 q^3+\cdots\right)$  \\[0.3em]
             & $\frac{1}{4}$ &  & $ q^{\frac{5}{32}}  \left(3+5q^{1/2}+9 q+15 q^{3/2}+27 q^2+45 q^{5/2}+\cdots\right) $  \\[0.3em]
    \hline
\multirow{2}{*}{$6$} & $0$ & $\multirow{2}{*}{$(\frac{1}{6}, -\frac{1}{16}, 0)$}$ &  $q^{-\frac{1}{4}} \left(1+18 q+64 q^{3/2}+159 q^2+384 q^{5/2}+ \cdots \right)$ \\[0.3em]
    & $\frac{1}{2}$ &   & $q^{\frac{1}{4}}  \left(4+16 q^{1/2}+56 q+160 q^{3/2}+404 q^2+944 q^{5/2}+ \cdots \right)$   \\[0.3em]
    \hline
\multirow{2}{*}{$\frac{39}{4}$} & $0$ & \multirow{2}{*}{$(\frac{11}{48},-\frac{143}{1024},0)$} & $q^{-\frac{13}{32}} \left(1+78 q+429 q^{3/2}+1794 q^2+6435 q^{5/2}+\cdots\right)$  \\[0.3em]
     & $\frac{3}{4}$ & $ $ & $ q^{\frac{11}{32}}  \left(65+429q^{1/2}+2145 q+8437q^{3/2}+28236q^2+\cdots \right)$ \\[0.3em]
    \hline
\multirow{2}{*}{$11$} & $0$ & \multirow{2}{*}{$(\frac{1}{4}, -\frac{11}{64}, 0)$} & $q^{-\frac{11}{24}} \left(1+143 q+924 q^{3/2}+4499 q^2+18084 q^{5/2} + \cdots \right)$  \\[0.3em]
     & $\frac{5}{6}$ &  & $ q^{\frac{3}{8}}  \left(66+495q^{1/2}+2718q+11649q^{3/2}+42174 q^2+ \cdots \right)$  \\[0.3em]
\hline
\end{tabular}
}
\caption{\label{NS type I} NS sector solutions of BPS type \bf{I}}
\end{table}

\begin{table}[ht]
\centering
\scalebox{1.0}{
\begin{tabular}{c | c | c | c}
\hline
 $c$  & $h^{\rm{R}}$ & ($\mu_1, \mu_2, \mu_3$) & R sector character  \\
\hline
\multirow{2}{*}{1} &  $\frac{1}{24}$ & \multirow{2}{*}{($\frac{1}{12}, -\frac{1}{192}, 0$)} & $ \sqrt{2} \left(1 + 2 q + 4 q^2 + 6 q^3 + 10 q^4 + 16 q^5 + 24 q^6 + \cdots \right)$   \\[0.3em]
    &  $\frac{3}{8}$ &   & $q^{\frac{1}{3}} \left( 2+2q+4 q^2+8 q^3 + 12 q^4 + 18 q^5 + 28 q^6  + \cdots \right) $  \\[0.3em]
    \hline
\multirow{2}{*}{$\frac{9}{4}$} & $\frac{3}{32}$ & \multirow{2}{*}{($\frac{5}{48},-\frac{15}{1024},0$)} & $ 2 + 12q + 36q^2 + 88q^3 + 204 q^4 + 432 q^5 + 856 q^6 + \cdots $  \\[0.3em]
             & $\frac{15}{32}$ &  & $\sqrt{2}q^{\frac{3}{8}}  \left(4+12 q+36 q^2+88 q^{3} + 192 q^4 + 396 q^5 + 776 q^6 + \cdots\right) $  \\[0.3em]
    \hline
\multirow{2}{*}{$6$} & $\frac{1}{4}$ & $\multirow{2}{*}{$(\frac{1}{6}, -\frac{1}{16}, 0)$}$ &  $2 + 64 q +512 q^2 + 2816 q^3 + 12288 q^4  + 45952 q^5 + \cdots $ \\[0.3em]
    & $\frac{3}{4}$ &   & $q^{\frac{1}{2}}  \left(8+96 q+624 q^2+3008 q^3 + 12072 q^4 + 42528 q^5 +  \cdots \right)$   \\[0.3em]
    \hline
\multirow{2}{*}{$\frac{39}{4}$} & $\frac{13}{32}$ & \multirow{2}{*}{$(\frac{11}{48},-\frac{143}{1024},0)$} & $12 + 1144 q + 19032 q^2 + 180336 q^3 + 1247688 q^4  +\cdots$  \\[0.3em]
     & $\frac{33}{32}$ & $ $ & $\sqrt{2} q^{\frac{5}{8}}  \left(208 +5136 q+57408 q^2+439504q^{3}+2647632 q^4 +\cdots \right)$ \\[0.3em]
    \hline
\multirow{2}{*}{$11$} & $\frac{11}{24}$ & \multirow{2}{*}{$(\frac{1}{4}, -\frac{11}{64}, 0)$} & $\sqrt{2}\left(12 + 1584 q + 32472 q^2 + 360096 q^3 + 2846448 q^4 + \cdots\right)$  \\[0.3em]
     & $\frac{9}{8}$ &  & $q^{\frac{2}{3}}  \left(440 + 13024 q+ 169048 q^2+1470944 q^{3}+9929392 q^4+\cdots \right)$  \\[0.3em]
\hline
\end{tabular}
}
\caption{\label{R type I}R sector solutions of BPS type {\bf{I}}}
\end{table}

\begin{table}[ht]
\centering
\scalebox{1.0}{
\begin{tabular}{c|L|c|c}
\hline
$c$  & h^{\textsc{ns}} & $(\mu_1,\mu_2,\mu_3)$ & \text{NS sector characters}  \\[0.3em]
\hline \multirow{2}{*}{$\frac{3}{4}$} & 0 & \multirow{2}{*}{($-\frac{1}{48}, -\frac{7}{1024}, 0$)} &  $q^{-\frac{1}{32}}\left(1+q^{\frac{1}{2}}+q+q^{\frac{3}{2}}+2 q^2+2 q^{\frac{5}{2}}+2 q^3 + 3q^{\frac{7}{2}} + \cdots \right)$ \\[0.3em]
& \frac{1}{4} & & $q^{\frac{7}{32}}\left(1+q^{\frac{3}{2}}+q^2+q^{\frac{5}{2}}+ q^3 + q^{\frac{7}{2}} + 2q^4 + 2q^{\frac{9}{2}}\cdots \right)$ \\[0.3em]
 \hline \multirow{2}{*}{$\frac{3}{2}$} &0 & \multirow{2}{*}{($\frac{1}{24}, -\frac{3}{256}, 0$)} &  $q^{-\frac{1}{16}}\left(1+q^{\frac{1}{2}}+3 q+4 q^{\frac{3}{2}}+5 q^2+8 q^{\frac{5}{2}}+11 q^3+ \cdots \right)$ \\[0.3em]
& \frac{1}{4} & & $q^{\frac{3}{16}}\left(2+2q^{\frac{1}{2}}+2q+4 q^{\frac{3}{2}}+8 q^2+10 q^{\frac{5}{2}}+ 12 q^3 + \cdots\right)$ \\[0.3em]
 \hline \multirow{2}{*}{$3$} &0 & \multirow{2}{*}{($\frac{1}{12}, -\frac{5}{192}, 0$)} &  $q^{-\frac{1}{8}}\left(1+2 q^{\frac{1}{2}}+9 q+18 q^{
 \frac{3}{2}}+29 q^2+54 q^{\frac{5}{2}}+100 q^3+\cdots \right)$ \\[0.3em]
& \frac{1}{3} & & $q^{\frac{5}{24}}\left(3+6q^{\frac{1}{2}}+12 q+24 q^{\frac{3}{2}}+48 q^2+84 q^{\frac{5}{2}}+135 q^3 +\cdots\right)$ \\[0.3em]
 \hline \multirow{2}{*}{$6$} &0  & \multirow{2}{*}{($\frac{1}{6}, -\frac{1}{16}, 0$)} &  $q^{-\frac{1}{4}}\left( 1 + 4 q^{\frac{1}{2}} + 34 q + 120 q^{\frac{3}{2}} + 319 q^2 + 788 q^{\frac{5}{2}} + \cdots \right)$ \\[0.3em]
& \frac{1}{2} & & $q^{\frac{1}{4}} \left( 8 + 32 q^{\frac{1}{2}} + 112 q + 320 q^{\frac{3}{2}} + 808 q^2 + 1888 q^{\frac{5}{2}} + \cdots \right)$ \\[0.3em]
\hline \multirow{2}{*}{$9$} & 0 & \multirow{2}{*}{($\frac{1}{4}, -\frac{7}{64}, 0$)} &  $q^{-\frac{3}{8}}\left(1+6 q^{\frac{1}{2}}+93 q+494 q^{\frac{3}{2}}+1950 q^2+6504 q^{\frac{5}{2}}+\cdots \right)$ \\[0.3em]
& \frac{2}{3} & & $q^{\frac{7}{24}}\left(27+162 q^{\frac{1}{2}}+783 q+2970 q^{\frac{3}{2}}+9531 q^2+\cdots \right)$ \\[0.3em]
\hline \multirow{2}{*}{$\frac{21}{2}$} & 0 & \multirow{2}{*}{($\frac{7}{24}, -\frac{35}{256}, 0$)} &  $q^{-\frac{7}{16}}\left(1+7 q^{\frac{1}{2}}+154 q+973 q^{\frac{3}{2}}+4550 q^2+17472 q^{\frac{5}{2}}+\cdots\right)$ \\[0.3em]
& \frac{3}{4} & & $q^{\frac{5}{16}}\left(56+392 q^{\frac{1}{2}}+2144 q+9128 q^{\frac{3}{2}}+32536 q^2+\cdots \right)$ \\[0.3em]
\hline \multirow{2}{*}{$\frac{45}{4}$} & 0 & \multirow{2}{*}{($\frac{17}{48}, -\frac{135}{1024}, 0$)} &  $q^{-\frac{15}{32}}\left(1+15 q^{\frac{1}{2}}+225 q+1555 q^{\frac{3}{2}}+7920 q^2+\cdots\right)$ \\[0.3em]
 &\frac{3}{4} & & $q^{\frac{9}{32}}\left(35+252q^{\frac{1}{2}}+1485q+6805 q^{\frac{3}{2}}+25845q^2+\cdots \right)$ \\[0.3em]
\hline
\end{tabular}
}
\caption{\label{NS type VI} NS sector solutions of BPS type \bf{II}}
\end{table}

\begin{table}[ht]
\centering
\scalebox{1.0}{
\begin{tabular}{c | L | c | c}
\hline
$c$  & h^{\textsc{r}}  & $(\mu_1,\mu_2,\mu_3)$ & \text{R sector characters}  \\[0.3em]
\hline \multirow{2}{*}{$\frac{3}{4}$}& \frac{1}{32} & \multirow{2}{*}{($-\frac{1}{48}, -\frac{7}{1024}, 0$)} &$1+2 q+2 q^2+4 q^3+6 q^4+ 8 q^5 + 12 q^6 + 16 q^7 +\cdots$ \\[0.3em]
 &\frac{5}{32} &  & $\sqrt{2} q^{\frac{1}{8}}\left(1+q+2 q^2+3 q^3+ 4 q^4 + 6 q^5 + 9 q^6 + 12 q^7 + \cdots \right)$ \\[0.3em]
 \hline \multirow{2}{*}{$\frac{3}{2}$}&\frac{1}{16}  & \multirow{2}{*}{($\frac{1}{24}, -\frac{3}{256}, 0$)} & $\sqrt{2} \left( 1+4 q+8 q^2+16 q^3+32 q^4+ 56q^5 + 96 q^6 + \cdots \right)$ \\[0.3em]
 &\frac{5}{16} & & $\sqrt{2} q^{\frac{1}{4}}\left(2+4 q+10 q^2+20 q^3 + 36 q^4 + 64 q^5 + 110 q^6 + \cdots \right)$ \\[0.3em]
\hline \multirow{2}{*}{$3$} &\frac{1}{8} & \multirow{2}{*}{($\frac{1}{12}, -\frac{5}{192}, 0$)} & $2+20 q+72 q^2+220 q^3+596 q^4 + 1440q^5 +\cdots$ \\[0.3em]
 &\frac{11}{24} & & $\sqrt{2} q^{\frac{1}{3}}\left(6+30 q+108 q^2+312 q^3+ 804 q^4 + 1902 q^5 + \cdots \right)$ \\[0.3em]
\hline \multirow{2}{*}{$6$} &0 & $\multirow{2}{*}{$(\frac{1}{6}, -\frac{1}{16}, 0)$}$ & $2+64 q+512 q^2+2816 q^3+12288 q^4 + 45952 q^5 +\cdots$ \\[0.3em]
 &\frac{1}{2} & & $ q^{\frac{1}{2}}\left(32+384 q+2496 q^2+12032 q^3 + 48288q^4 +\cdots \right)$ \\[0.3em]
 \hline \multirow{2}{*}{$9$} &\frac{3}{8} & \multirow{2}{*}{($\frac{1}{4}, -\frac{7}{64}, 0$)} &$8+672 q+9744 q^2+83648 q^3+532128 q^4 +\cdots$ \\[0.3em]
 &\frac{25}{24} & & $\sqrt{2}q^{\frac{2}{3}}\left(216+4320 q+42552 q^2+294624 q^3+  \cdots \right)$ \\[0.3em]
\hline \multirow{2}{*}{$\frac{21}{2}$} &\frac{7}{16} & \multirow{2}{*}{($\frac{7}{24}, -\frac{35}{256}, 0$)} & $\sqrt{2}\left(8+1120 q+21056 q^2+219520 q^3+\cdots \right)$ \\[0.3em]
 &\frac{19}{16}& & $\sqrt{2}q^{\frac{3}{4}}\left(448+10880q+126784 q^2+1018752 q^3+\cdots \right)$ \\[0.3em]
\hline \multirow{2}{*}{$\frac{45}{4}$} &\frac{15}{32} & \multirow{2}{*}{($\frac{17}{48}, -\frac{135}{1024}, 0$)} &$8+2160 q+45840 q^2+524000 q^3+4250160 q^4+\cdots$\\[0.3em]
 &\frac{43}{32} & & $\sqrt{2}q^{\frac{7}{8}}\left(960+23104q+279360 q^2+2346240 q^3+\cdots \right)$ \\[0.3em]
\hline
\end{tabular}
}
\caption{\label{R type VI} R sector solutions of BPS type \bf{II}}
\end{table}

\begin{table}[ht]
\centering
\scalebox{1.0}{
\begin{tabular}{c | c | c | c}
\hline
 $c$  & $h^{\rm{NS}}$ & ($\mu_1, \mu_2, \mu_3$) & NS sector character  \\
\hline
\multirow{2}{*}{$\frac{7}{10}$} &  $0$ & \multirow{2}{*}{($\frac{1}{8}, -\frac{7}{1280}, \frac{49}{14400}$)} & $  q^{-\frac{7}{240}} \left(1+ q^{3/2}+q^2+q^{5/2}+q^3+2q^{7/2}+\cdots \right) $   \\[0.3em]
    &  $\frac{1}{10}$ &   & $q^{\frac{17}{240}}  \left(1+ q^{1/2}+ q+q^{3/2}+q^2 + 2q^{5/2}+2q^3+\cdots \right)$  \\[0.3em]
    \hline
\multirow{2}{*}{$\frac{133}{10}$} & $0$ & \multirow{2}{*}{($\frac{3}{8},-\frac{399}{1280},\frac{1729}{14400}$)} & $ q^{-\frac{133}{240}} \left(1+133 q+1463 q^{3/2}+9044 q^2+\cdots \right)$  \\[0.3em]
             & $\frac{9}{10}$ &  & $q^{\frac{83}{240}}  \left(133+1539 q^{1/2}+10318 q+52535 q^{3/2}+\cdots \right) $  \\[0.3em]
    \hline
\multirow{2}{*}{$\frac{91}{5}$} & $0$ & $\multirow{2}{*}{$(\frac{7}{12}, -\frac{637}{960}, \frac{91}{225})$}$ &  $q^{-\frac{91}{120}} \left(1+91 q+2548 q^{3/2}+28301 q^2+ \cdots \right)$ \\[0.3em]
    & $\frac{11}{10}$ &   & $q^{\frac{41}{120}}  \left(364+7007 q^{1/2}+69732 q+487109 q^{3/2}+\cdots \right)$   \\[0.3em]
    \hline
\multirow{2}{*}{$\frac{39}{2}$} & $0$ & \multirow{2}{*}{$(\frac{5}{8},-\frac{195}{256},\frac{91}{192})$} & $q^{-\frac{13}{16}} \left(1+78 q+2925 q^{3/2}+37908 q^2 + \cdots \right) $  \\[0.3em]
     & $\frac{7}{6}$ & $ $ & $q^{\frac{17}{48}}  \left(351+7371 q^{1/2}+79353 q+594243 q^{3/2} + \cdots \right)$ \\[0.3em]
    \hline
\multirow{2}{*}{$\frac{102}{5}^*$} & $0$ & \multirow{2}{*}{$(\frac{2}{3}, -\frac{17}{20}, \frac{221}{400})$} & $q^{-\frac{17}{20}} \left(1+102 q+4352 q^{3/2}+62577 q^2 + \cdots \right)$  \\[0.3em]
     & $\frac{6}{5}$ &  & $q^{\frac{7}{20}}  \left(374+8448 q^{1/2}+96900 q+765952 q^{3/2}+\cdots \right)$  \\[0.3em]
\hline
\multirow{2}{*}{$21$} & $0$ & \multirow{2}{*}{$(\frac{2}{3},-\frac{7}{8},\frac{35}{64})$} & $ q^{-\frac{7}{8}} \left(1+63 q+3584 q^{3/2}+55734 q^2+ \cdots \right)$  \\[0.3em]
     & $\frac{5}{4}$ & $ $ & $q^{\frac{3}{8}}  \left(672+ 15360 q^{1/2}+178976 q+1440768 q^{3/2}+ \cdots \right)$ \\[0.3em]
    \hline
\multirow{2}{*}{$\frac{85}{4}^*$} & $0$ & \multirow{2}{*}{$(\frac{11}{16},-\frac{935}{1024},\frac{85}{144})$} & $ q^{-\frac{85}{96}} \left(1+85 q+4675 q^{3/2}+74630 q^2+\cdots \right)$  \\[0.3em]
     & $\frac{5}{4}$ & $ $ & $q^{\frac{35}{96}}  \left(357+8415 q^{1/2}+100555 q+825945 q^{3/2}+ \cdots \right)$ \\[0.3em]
    \hline
\multirow{2}{*}{$22^*$} & $0$ & \multirow{2}{*}{$(\frac{2}{3},-\frac{11}{12},\frac{77}{144})$} & $ q^{-\frac{11}{12}} \left(1+22 q+2816 q^{3/2}+50171 q^2+\cdots \right)$  \\[0.3em]
     & $\frac{4}{3}$ & $ $ & $q^{\frac{5}{12}}  \left(891+20736 q^{1/2}+247698 q+2052864 q^{3/2}+\cdots \right)$ \\[0.3em]
    \hline
\multirow{2}{*}{$\frac{114}{5}$} & $0$ & \multirow{2}{*}{$(\frac{2}{3},-\frac{19}{20},\frac{209}{400})$} & $ q^{-\frac{19}{20}} \left(1+2432 q^{3/2}+48621 q^2+\cdots \right)$  \\[0.3em]
     & $\frac{7}{5}$ & $ $ & $q^{\frac{9}{20}}  \left(1938+45696 q^{1/2}+556206 q+4713216 q^{3/2}+ \cdots \right)$ \\[0.3em]
    \hline
\end{tabular}
}
\caption{\label{NS type III} NS sector solutions of non-BPS type {\bf{I}}. The solutions with asterisk ($c=\frac{102}{5},\frac{85}{4},22$) do not have a consistent fusion rule algebra.}
\end{table}

\begin{table}[ht]
\centering
\scalebox{1.0}{
\begin{tabular}{c | c | c | c}
\hline
 $c$  & $h^{\rm{R}}$ & ($\mu_1, \mu_2, \mu_3$) & R sector character  \\
\hline
\multirow{2}{*}{$\frac{7}{10}$} &  $\frac{3}{80}$ & \multirow{2}{*}{($\frac{1}{8}, -\frac{7}{1280}, \frac{49}{14400}$)} & $\sqrt{2}q^{\frac{1}{120}}\left(1 + q + 2 q^2 + 3 q^3 + 4 q^4 + 6 q^5 + \cdots \right) $   \\[0.3em]
    &  $\frac{7}{16}$ &   & $ \sqrt{2}q^{\frac{49}{120}}  \left(1 + q + q^2 + 2 q^3 + 3 q^4 + 4 q^5 + \cdots \right) $ \\[0.3em]
    \hline
\multirow{2}{*}{$\frac{133}{10}$} & $\frac{57}{80}$ & \multirow{2}{*}{($\frac{3}{8},-\frac{399}{1280},\frac{1729}{14400}$)} & $\sqrt{2}q^{\frac{19}{120}}\left(56 + 7448 q + 186352 q^2 + 2512104 q^3 + \cdots \right)$  \\[0.3em]
             & $\frac{21}{16}$ &  & $\sqrt{2} q^{\frac{91}{120}}  \left(912 + 35112 q +577752 q^2 + 6183968 q^3 + \cdots \right) $  \\[0.3em]
\hline
\multirow{2}{*}{$\frac{91}{5}$} & $\frac{49}{40}$ & $\multirow{2}{*}{$(\frac{7}{12}, -\frac{637}{960}, \frac{91}{225})$}$ &  $q^{\frac{7}{15}}\left(1664 + 232960 q + 8118656 q^2 + 153033216 q^3 + \cdots \right)$ \\[0.3em]
   & $\frac{13}{8}$ &   & $q^{\frac{13}{15}}  \left(9856 + 658944 q +17169152 q^2 + 273681408 q^3 + \cdots \right) $   \\[0.3em]
    \hline
\multirow{2}{*}{$\frac{39}{2}$} & $\frac{65}{48}$ & \multirow{2}{*}{$(\frac{5}{8},-\frac{195}{256},\frac{91}{192})$} & $q^{\frac{13}{24}}\left(3456 + 494208 q + 18555264 q^2 + 377574912 q^3 + \cdots \right) $  \\[0.3em]
     & $\frac{27}{16}$ & $ $ & $ q^{\frac{7}{8}}  \left(11648 + 908544 q + 26687232 q^2 +  470863744 q^3 + \cdots \right)$ \\[0.3em]
   \hline
\multirow{2}{*}{$\frac{102}{5}^*$} & $\frac{3}{2}$ & \multirow{2}{*}{$(\frac{2}{3}, -\frac{17}{20}, \frac{221}{400})$} & $q^{\frac{13}{20}}\left(8704 + 1122816 q + 42275328 q^2 + 881340928 q^3 + \cdots \right)$  \\[0.3em]
     & $\frac{17}{10}$ &  & $q^{\frac{17}{20}}  \left(16896 + 1531904q + 49795584 q^2 + 952435200 q^3 + \cdots \right)$  \\[0.3em]
\hline
\multirow{2}{*}{$21$} & $\frac{3}{2}$ & \multirow{2}{*}{$(\frac{2}{3},-\frac{7}{8},\frac{35}{64})$} & $ q^{\frac{5}{8}}\left(7168 + 1053696q + 42897408 q^2 + 948921344 q^3 + \cdots \right) $  \\[0.3em]
     & $\frac{7}{4}$ & $ $ & $q^{\frac{7}{8}}  \left(30720 + 2881536q + 97298432 q^2 + 1928994816 q^3 + \cdots \right)$ \\[0.3em]
    \hline
\multirow{2}{*}{$\frac{85}{4}^*$} & $\frac{51}{32}$ & \multirow{2}{*}{$(\frac{11}{16},-\frac{935}{1024},\frac{85}{144})$} & $  q^{\frac{17}{24}}\left(8960 + 1153280q + 45012736 q^2 + 977589760 q^3 + \cdots \right)$  \\[0.3em]
     & $\frac{55}{32}$ & $ $ & $q^{\frac{5}{6}}  \left(26112 + 2698240 q + 95979520 q^2 + 1973719040 q^3 + \cdots \right)$ \\[0.3em]
    \hline
\multirow{2}{*}{$22^*$} & $\frac{3}{2}$ & \multirow{2}{*}{$(\frac{2}{3},-\frac{11}{12},\frac{77}{144})$} & $ q^{\frac{7}{12}}\left(5632 + 1036288q + 48080384 q^2 + 1173607424 q^3 + \cdots \right)$  \\[0.3em]
     & $\frac{11}{6}$ & $ $ & $q^{\frac{11}{12}}  \left(41472 + 4105728 q + 147350016 q^2 + 3095262720 q^3 +\cdots \right)$ \\[0.3em]
    \hline
\multirow{2}{*}{$\frac{114}{5}$} & $\frac{3}{2}$ & \multirow{2}{*}{$(\frac{2}{3},-\frac{19}{20},\frac{209}{400})$} & $  q^{\frac{11}{20}}\left(4864 + 1079808q + 55653888 q^2 + 1469453312 q^3 + \cdots \right)$  \\[0.3em]
     & $\frac{19}{10}$ & $ $ & $q^{\frac{19}{20}}  \left(91392 + 9426432q +354483456 q^2 + 7786145280 q^3 +\cdots \right)$ \\[0.3em]
    \hline
\end{tabular}
}
\caption{\label{R type III}R sector solutions of non-BPS type \bf{I}}
\end{table}

\begin{table}[ht]
\centering
\scalebox{1.0}{
\begin{tabular}{c | c | c | c}
\hline
 $c$  & $h^{\rm{NS}}$ & ($\mu_1, \mu_2, \mu_3$) & NS sector character  \\
\hline
\multirow{2}{*}{$\frac{9}{2}$} &  $0$ & \multirow{2}{*}{($\frac{1}{24}, -\frac{3}{256}, -\frac{3}{64}$)} & $ q^{-\frac{3}{16}} \left(1+27 q+21 q^{3/2}+126 q^2+126 q^{5/2}+\cdots \right) $   \\[0.3em]
    &  $\frac{1}{2}$ &   & $q^{\frac{5}{16}}  \left(1+q^{1/2}+8 q+9 q^{3/2}+37 q^2+45 q^{5/2}+ \cdots \right) $  \\[0.3em]
    \hline
\multirow{2}{*}{$5$} & $0$ & \multirow{2}{*}{($\frac{1}{12},-\frac{5}{192},-\frac{5}{144}$)} & $q^{-\frac{5}{24}} \left(1+25 q+40 q^{3/2}+130 q^2+232 q^{5/2}+\cdots \right)$  \\[0.3em]
             & $\frac{1}{2}$ &  & $q^{\frac{7}{24}}  \left(2+4 q^{1/2}+18 q+36 q^{3/2}+96 q^2+184 q^{5/2}+\cdots \right) $  \\[0.3em]
    \hline
\multirow{2}{*}{$\frac{11}{2}$} & $0$ & $\multirow{2}{*}{$(\frac{1}{8}, -\frac{11}{256}, -\frac{11}{576})$}$ &  $q^{-\frac{11}{48}} \left(1+22 q+55 q^{3/2}+143 q^2+319 q^{5/2}+ \cdots \right)$ \\[0.3em]
    & $\frac{1}{2}$ &   & $q^{\frac{13}{48}}  \left(1+3 q^{1/2}+11 q+28 q^{3/2}+69 q^2+ \cdots \right)$   \\[0.3em]
    \hline
\multirow{2}{*}{$\frac{13}{2}$} & $0$ & \multirow{2}{*}{$(\frac{5}{24},-\frac{65}{768},\frac{13}{576})$} & $q^{-\frac{13}{48}} \left(1+13 q+65 q^{3/2}+169 q^2+416 q^{5/2}+ \cdots \right) $  \\[0.3em]
     & $\frac{1}{2}$ & $ $ & $q^{\frac{11}{48}}  \left(1+5 q^{1/2}+18 q+55 q^{3/2}+146 q^2+\cdots \right)$ \\[0.3em]
    \hline
\multirow{2}{*}{$7$} & $0$ & \multirow{2}{*}{$(\frac{1}{4}, -\frac{7}{64}, \frac{7}{144})$} & $q^{-\frac{7}{24}} \left(1+7 q+56 q^{3/2}+161 q^2+392 q^{5/2}+\cdots \right)$  \\[0.3em]
     & $\frac{1}{2}$ &  & $q^{\frac{5}{24}}  \left(6+36 q^{1/2}+138 q+444 q^{3/2}+1242 q^2+ \cdots \right)$  \\[0.3em]
\hline
\multirow{2}{*}{$\frac{15}{2}$} & $0$ & \multirow{2}{*}{$(\frac{287}{1560},-\frac{287}{3328},-\frac{5}{832})$} & $ q^{-\frac{5}{16}} \left(1+35 q^{3/2}+120 q^2+273 q^{5/2}+ \cdots \right)$  \\[0.3em]
     & $\frac{1}{2}$ & $ $ & $q^{\frac{3}{16}}  \left(1+7 q^{1/2}+29 q+98 q^{3/2}+288 q^2+ \cdots \right)$ \\[0.3em]
    \hline
\end{tabular}
}
\caption{\label{type II NS}NS sector characters of non-BPS type {\bf{II}}}
\end{table}

\begin{table}[ht]
\centering
\scalebox{1.0}{
\begin{tabular}{c | c | c | c}
\hline
 $c$  & $h^{\rm{R}}$ & ($\mu_1, \mu_2, \mu_3$) & R sector character  \\
\hline
\multirow{2}{*}{$\frac{9}{2}$} &  $\frac{1}{16}$ & \multirow{2}{*}{($\frac{1}{24}, -\frac{3}{256}, -\frac{3}{64}$)} & $ q^{-\frac{1}{8}}\left( 1 + 29 q + 163 q^2 + 732 q^3 + 2569 q^4 + \cdots \right)$   \\[0.3em]
    &  $\frac{9}{16}$ &   & $q^{\frac{3}{8}}  \left(1 + 9 q + 45 q^2 + 174 q^3 + 576 q^4 +\cdots \right) $  \\[0.3em]
    \hline
\multirow{2}{*}{$5$} & $\frac{1}{8}$ & \multirow{2}{*}{($\frac{1}{12},-\frac{5}{192},-\frac{5}{144}$)} & $q^{-\frac{1}{12}}\left( 1 + 30 q + 193 q^2 + 926 q^3 + 3524 q^4 + \cdots \right) $  \\[0.3em]
             & $\frac{5}{8}$ &  & $q^{\frac{5}{12}}  \left(8 + 80 q + 440 q^2 + 1840 q^3 + 6520 q^4 + \cdots \right) $  \\[0.3em]
    \hline
\multirow{2}{*}{$\frac{11}{2}$} & $\frac{3}{16}$ & $\multirow{2}{*}{$(\frac{1}{8}, -\frac{11}{256}, -\frac{11}{576})$}$ &  $q^{-\frac{1}{24}}\left( 1 + 31 q + 224 q^2 + 1151 q^3 + 4705 q^4 + \cdots \right)$ \\[0.3em]
    & $\frac{11}{16}$ &   & $q^{\frac{11}{24}}  \left(8 + 88 q + 528 q^2 + 2376 q^3 + 8976 q^4 + \cdots \right)$   \\[0.3em]
    \hline
\multirow{2}{*}{$\frac{13}{2}$} & $\frac{5}{16}$ & \multirow{2}{*}{$(\frac{5}{24},-\frac{65}{768},\frac{13}{576})$} & $q^{\frac{1}{24}}\left(2 + 66 q + 578 q^2 + 3396 q^3 + 15748 q^4 + \cdots \right) $  \\[0.3em]
     & $\frac{13}{16}$ & $ $ & $q^{\frac{13}{24}}  \left(16 +208 q + 1456 q^2 + 7488 q^3 + 31824 q^4 + \cdots \right)$  \\[0.3em]
    \hline
\multirow{2}{*}{$7$} & $\frac{3}{8}$ & \multirow{2}{*}{$(\frac{1}{4}, -\frac{7}{64}, \frac{7}{144})$} & $q^{\frac{1}{12}}\left(4 + 136 q + 1292 q^2 + 8088 q^3 + 39716 q^4 + \cdots \right)$  \\[0.3em]
     & $\frac{7}{8}$ &  & $q^{\frac{7}{12}}  \left(96 + 1344 q + 10080 q^2 + 55104 q^3 + 247296 q^4 + \cdots \right)$  \\[0.3em]
\hline
\multirow{2}{*}{$\frac{15}{2}$} & $\frac{7}{16}$ & \multirow{2}{*}{$(\frac{287}{1560},-\frac{287}{3328},-\frac{5}{832})$} & $ q^{\frac{1}{8}}\left(4 + 140 q + 1432 q^2 + 9524 q^3 + 49376 q^4 + \cdots \right) $  \\[0.3em]
     & $\frac{15}{16}$ & $ $ & $q^{\frac{5}{8}}  \left(32 + 480 q + 3840 q^2 + 22240 q^3 + 105120 q^4 + \cdots \right)$ \\[0.3em]
    \hline
\end{tabular}
}
\caption{\label{type II R}R sector characters of non-BPS type {\bf{II}}}
\end{table}

\begin{table}[h!]
\centering
\scalebox{1.0}{
\begin{tabular}{c | c | c}
\hline
 $c$  & ($\mu_1, \mu_2, \mu_3$) & NS sector character  \\
\hline
$10$  &  $(\frac{3}{22}, -\frac{15}{176}, -\frac{10}{99})$ & $ q^{-\frac{5}{12}} \left(1 + 270 q + 960 q^{3/2} + 5725 q^2 + 18304 q^{5/2} + \cdots \right)$   \\[0.3em]
$11$  &  $(\frac{17}{84}, -\frac{187}{1344}, -\frac{55}{1008})$ & $ q^{-\frac{11}{24}} \left(1 + 275 q + 1496 q^{3/2} + 7931 q^2 + 31240 q^{5/2}  + \cdots \right)$   \\[0.3em]
$12$  &  $(\frac{4}{15},-\frac{1}{5},0)$ & $ q^{-\frac{1}{2}} \left(1 + 276 q + 2048 q^{3/2} + 11202 q^2 + 49152 q^{5/2} + \cdots \right)$   \\[0.3em]
$\frac{25}{2}$  &  $(\frac{31}{104},-\frac{775}{3328}, \frac{25}{832})$ & $ q^{-\frac{25}{48}} \left(1 + 275 q + 2325 q^{3/2} + 13250 q^2 + 60630 q^{5/2} + \cdots \right)$   \\[0.3em]
$13$  &  $(\frac{25}{76}, -\frac{325}{1216}, \frac{169}{2736})$ & $ q^{-\frac{13}{24}} \left(1 + 273 q + 2600 q^{3/2} + 15574 q^2 + 74152 q^{5/2}  + \cdots \right)$   \\[0.3em]
$\frac{27}{2}$  &  $(\frac{319}{888}, -\frac{2871}{9472}, \frac{225}{2368})$ & $ q^{-\frac{9}{16}} \left(1 + 270 q + 2871 q^{3/2} + 18171 q^2 + 89991 q^{5/2}  + \cdots \right)$   \\[0.3em]
$14$  &  $(\frac{7}{18},-\frac{49}{144},\frac{7}{54})$ & $ q^{-\frac{7}{12}} \left(1 + 266 q + 3136 q^{3/2} + 21035 q^2 + 108416 q^{5/2} + \cdots \right)$   \\[0.3em]
$\frac{29}{2}$  &  $(\frac{117}{280},-\frac{3393}{8960},\frac{667}{4032})$ & $ q^{-\frac{29}{48}} \left(1 + 261 q + 3393 q^{3/2} + 24157 q^2 + 129688 q^{5/2}  + \cdots \right)$   \\[0.3em]
$15$  &  $(\frac{91}{204}, -\frac{455}{1088}, \frac{55}{272})$ & $ q^{-\frac{5}{8}} \left(1 + 255 q + 3640 q^{3/2} + 27525 q^2 + 154056 q^{5/2}  + \cdots \right)$   \\[0.3em]
$\frac{31}{2}$  &  $(\frac{125}{264}, -\frac{3875}{8448}, \frac{1519}{6336})$ & $ q^{-\frac{31}{48}} \left(1 + 248 q + 3875 q^{3/2} + 31124 q^2 + 181753 q^{5/2}  + \cdots \right)$   \\[0.3em]
$16$  &  $(\frac{1}{2}, -\frac{1}{2}, \frac{5}{18})$ & $ q^{-\frac{2}{3}} \left(1 + 240 q + 4096 q^{3/2} + 34936 q^2 + 212992 q^{5/2}  + \cdots \right)$  \\[0.3em]
$\frac{33}{2}$  &  $(\frac{391}{744}, -\frac{4301}{7936}, \frac{627}{1984})$ & $ q^{-\frac{11}{16}} \left(1 + 231 q + 4301 q^{3/2} + 38940 q^2 + 247962 q^{5/2}  + \cdots \right)$   \\[0.3em]
$17$  &  $(\frac{11}{20},-\frac{187}{320}, \frac{17}{48})$ & $ q^{-\frac{17}{24}} \left(1 + 221 q + 4488 q^{3/2} + 43112 q^2 + 286824 q^{5/2} + \cdots \right)$   \\[0.3em]
$\frac{35}{2}$  &  $(\frac{133}{232}, -\frac{4655}{7424}, \frac{6545}{16704})$ & $ q^{-\frac{35}{48}} \left(1 + 210 q + 4655 q^{3/2} + 47425 q^2 + 329707 q^{5/2} + \cdots \right)$   \\[0.3em]
$18$  &  $(\frac{25}{42},-\frac{75}{112},\frac{3}{7})$ & $ q^{-\frac{3}{4}} \left(1 + 198 q + 4800 q^{3/2} + 51849 q^2 + 376704 q^{5/2}  + \cdots \right)$   \\[0.3em]
$\frac{37}{2}$  &  $(\frac{133}{216},-\frac{4921}{6912},\frac{2405}{5184})$ & $ q^{-\frac{37}{48}} \left(1 + 185 q + 4921 q^{3/2} + 56351 q^2 + 427868 q^{5/2} + \cdots \right)$   \\[0.3em]
$19$  &  $(\frac{33}{52},-\frac{627}{832},\frac{931}{1872})$ & $ q^{-\frac{19}{24}} \left(1 + 171 q + 5016 q^{3/2} + 60895 q^2 + 483208 q^{5/2} + \cdots \right)$   \\[0.3em]
$\frac{39}{2}$  &  $(\frac{391}{600}, -\frac{5083}{6400},\frac{169}{320})$ & $ q^{-\frac{13}{16}} \left(1 + 156 q + 5083 q^{3/2} + 65442 q^2 + 542685 q^{5/2}  + \cdots \right)$   \\[0.3em]
$20$  &  $(\frac{2}{3}, -\frac{5}{6}, \frac{5}{9})$ & $ q^{-\frac{5}{6}} \left(1 + 140 q + 5120 q^{3/2} + 69950 q^2 + 606208 q^{5/2}  + \cdots \right)$   \\[0.3em]
$\frac{41}{2}$  &  $(\frac{125}{184}, -\frac{5125}{5888},\frac{7667}{13248})$ & $ q^{-\frac{41}{48}} \left(1 + 123 q + 5125 q^{3/2} + 74374 q^2 + 673630 q^{5/2} + \cdots \right)$   \\[0.3em]
$21$  &  $(\frac{91}{132}, -\frac{637}{704}, \frac{105}{176})$ & $ q^{-\frac{7}{8}} \left(1 + 105 q + 5096 q^{3/2} + 78666 q^2 + 744744 q^{5/2}  + \cdots \right)$   \\[0.3em]
$\frac{43}{2}$  &  $(\frac{39}{56}, -\frac{1677}{1792},\frac{817}{1344})$ & $ q^{-\frac{43}{48}} \left(1 + 86 q + 5031 q^{3/2} + 82775 q^2 + 819279 q^{5/2}  + \cdots \right)$   \\[0.3em]
$22$  &  $(\frac{7}{10},-\frac{77}{80},\frac{11}{18})$ & $ q^{-\frac{11}{12}} \left(1 + 66 q + 4928 q^{3/2} + 86647 q^2 + 896896 q^{5/2}  + \cdots \right)$   \\[0.3em]
$\frac{45}{2}$  &  $(\frac{319}{456},-\frac{4785}{4864},\frac{735}{1216})$ & $ q^{-\frac{15}{16}} \left(1 + 45 q + 4785 q^{3/2} + 90225 q^2 + 977184 q^{5/2} + \cdots \right)$   \\[0.3em]
$23$  &  $(\frac{25}{36},-\frac{575}{576},\frac{253}{432})$ & $ q^{-\frac{23}{24}} \left(1 + 23 q + 4600 q^{3/2} + 93449 q^2 + 1059656 q^{5/2} + \cdots \right)$   \\[0.3em]
$\frac{47}{2}$  &  $(\frac{93}{136},-\frac{4371}{4352}, \frac{5405}{9792})$ & $ q^{-\frac{47}{48}} \left(1 + 4371 q^{3/2} + 96256 q^2 + 1143745 q^{5/2}  + \cdots \right)$   \\[0.3em]
\hline
\end{tabular}
}
\caption{\label{type V solution NS}NS sector solutions of non-BPS type \bf{IV}}
\end{table}

\begin{table}[ht]
\centering
\scalebox{1.0}{
\begin{tabular}{c | c | c}
\hline
 $c$  & ($\mu_1, \mu_2, \mu_3$) & R sector character  \\
\hline
$10$  &  $(\frac{3}{22}, -\frac{15}{176}, -\frac{10}{99})$ & $ q^{-\frac{1}{6}} \left(5 + 1004 q + 20510 q^2 + 215000 q^3  + \cdots \right)$   \\[0.3em]
$11$  &  $(\frac{17}{84}, -\frac{187}{1344}, -\frac{55}{1008})$ & $ q^{-\frac{1}{12}} \left(11 + 2026 q + 45067 q^2 + 518122 q^3  + \cdots \right)$   \\[0.3em]
$12$  &  $(\frac{4}{15},-\frac{1}{5},0)$ & $  24 + 4096 q + 98304 q^2 + 1228800 q^3  + \cdots $   \\[0.3em]
$\frac{25}{2}$  &  $(\frac{31}{104},-\frac{775}{3328}, \frac{25}{832})$ & $ q^{\frac{1}{24}} \left(25 + 4121 q + 102425 q^2 + 1331250 q^3 + \cdots \right)$   \\[0.3em]
$13$  &  $(\frac{25}{76}, -\frac{325}{1216}, \frac{169}{2736})$ & $ q^{\frac{1}{12}} \left(52 + 8296 q + 213148 q^2 + 2875704 q^3 + \cdots \right)$   \\[0.3em]
$\frac{27}{2}$  &  $(\frac{319}{888}, -\frac{2871}{9472}, \frac{225}{2368})$ & $ q^{\frac{1}{8}} \left(54 + 8354 q + 221508 q^2 + 3097278 q^3 + \cdots \right)$   \\[0.3em]
$14$  &  $(\frac{7}{18},-\frac{49}{144},\frac{7}{54})$ & $ q^{\frac{1}{6}} \left(112 + 16832 q + 459872 q^2 + 6654592 q^3 + \cdots \right)$   \\[0.3em]
$\frac{29}{2}$  &  $(\frac{117}{280},-\frac{3393}{8960},\frac{667}{4032})$ & $ q^{\frac{5}{24}} \left(116 + 16964 q + 476876 q^2 + 7131680 q^3 + \cdots \right)$   \\[0.3em]
$15$  &  $(\frac{91}{204}, -\frac{455}{1088}, \frac{55}{272})$ & $ q^{\frac{1}{4}} \left(240 + 34208 q + 988080 q^2 + 15252000 q^3 + \cdots \right)$   \\[0.3em]
$\frac{31}{2}$  &  $(\frac{125}{264}, -\frac{3875}{8448}, \frac{1519}{6336})$ & $ q^{\frac{7}{24}} \left(248 + 34504 q + 1022752 q^2 + 16275496 q^3 + \cdots \right)$   \\[0.3em]
$16$  &  $(\frac{1}{2}, -\frac{1}{2}, \frac{5}{18})$ & $ q^{\frac{1}{3}} \left(512 + 69632 q + 2115584 q^2 + 34668544 q^3 + \cdots \right)$  \\[0.3em]
$\frac{33}{2}$  &  $(\frac{391}{744}, -\frac{4301}{7936}, \frac{627}{1984})$ & $ q^{\frac{3}{8}} \left(528 + 70288 q + 2186448 q^2 + 36857568 q^3 + \cdots \right)$   \\[0.3em]
$17$  &  $(\frac{11}{20},-\frac{187}{320}, \frac{17}{48})$ & $ q^{\frac{5}{12}} \left(1088 + 141952 q + 4516288 q^2 + 78238080 q^3 + \cdots \right)$   \\[0.3em]
$\frac{35}{2}$  &  $(\frac{133}{232}, -\frac{4655}{7424}, \frac{6545}{16704})$ & $ q^{\frac{11}{24}} \left(1120 + 143392 q + 4661440 q^2 + 82908000 q^3 + \cdots \right)$   \\[0.3em]
$18$  &  $(\frac{25}{42},-\frac{75}{112},\frac{3}{7})$ & $ q^{\frac{1}{2}} \left(2304 + 289792 q + 9616896 q^2 + 175454208 q^3 + \cdots \right)$   \\[0.3em]
$\frac{37}{2}$  &  $(\frac{133}{216},-\frac{4921}{6912},\frac{2405}{5184})$ & $ q^{\frac{13}{24}} \left(2368 + 292928 q + 9914816 q^2 + 185395456 q^3 + \cdots \right)$   \\[0.3em]
$19$  &  $(\frac{33}{52},-\frac{627}{832},\frac{931}{1872})$ & $ q^{\frac{7}{12}} \left(4864 + 592384 q + 20433664 q^2 +  \cdots \right)$   \\[0.3em]
$\frac{39}{2}$  &  $(\frac{391}{600}, -\frac{5083}{6400},\frac{169}{320})$ & $ q^{\frac{5}{8}} \left(4992 + 599168 q + 21046272 q^2 +  \cdots \right)$   \\[0.3em]
$20$  &  $(\frac{2}{3}, -\frac{5}{6}, \frac{5}{9})$ & $ q^{\frac{2}{3}} \left(10240 + 1212416 q + 43335680 q^2 + \cdots \right)$   \\[0.3em]
$\frac{41}{2}$  &  $(\frac{125}{184}, -\frac{5125}{5888},\frac{7667}{13248})$ & $ q^{\frac{17}{24}} \left(10496 + 1227008 q + 44597504 q^2 + \cdots \right)$   \\[0.3em]
$21$  &  $(\frac{91}{132}, -\frac{637}{704}, \frac{105}{176})$ & $ q^{\frac{3}{4}} \left(21504 + 2484224 q + 91757568 q^2 + \cdots \right)$   \\[0.3em]
$\frac{43}{2}$  &  $(\frac{39}{56}, -\frac{1677}{1792},\frac{817}{1344})$ & $ q^{\frac{19}{24}} \left(22016 + 2515456 q + 94360576 q^2  + \cdots \right)$   \\[0.3em]
$22$  &  $(\frac{7}{10},-\frac{77}{80},\frac{11}{18})$ & $ q^{\frac{5}{6}} \left(45056 + 5095424 q + 194011136 q^2 + \cdots \right)$   \\[0.3em]
$\frac{45}{2}$  &  $(\frac{319}{456},-\frac{4785}{4864},\frac{735}{1216})$ & $ q^{\frac{7}{8}} \left(46080 + 5161984 q + 199388160 q^2  + \cdots \right)$   \\[0.3em]
$23$  &  $(\frac{25}{36},-\frac{575}{576},\frac{253}{432})$ & $ q^{\frac{11}{12}} \left(94208 + 10461184 q + 409710592 q^2  + \cdots \right)$   \\[0.3em]
$\frac{47}{2}$  &  $(\frac{93}{136},-\frac{4371}{4352}, \frac{5405}{9792})$ & $ q^{\frac{23}{24}} \left(96256 + 10602496 q + 420831232 q^2 + \cdots \right)$   \\[0.3em]
\hline
\end{tabular}
}
\caption{\label{type V solution R}R sector solutions of non-BPS type \bf{IV}}
\end{table}


    \clearpage

 
\bibliographystyle{IEEEtran}
\bibliography{main}

\end{document}